\documentclass[twocolumn,trackchanges]{aastex62}
\usepackage[section]{placeins}
\usepackage{longtable}
\usepackage{hyperref}
\usepackage{url}
\usepackage{amsmath}
\graphicspath{{./}{figures/}}
\newcommand{\pslsn}{$P({\rm SLSN-I})$}

\usepackage{enumitem}

\usepackage[hang,flushmargin]{footmisc} 
\usepackage{url}

\newcommand{\CfA}{\affiliation{Center for Astrophysics \textbar{} Harvard \& Smithsonian, 60 Garden Street, Cambridge, MA 02138-1516, USA}}

\newcommand{\Edinburgh}{\affiliation{Institute for Astronomy, University of Edinburgh, Royal Observatory, Blackford Hill EH9 3HJ, UK}}
\newcommand{\Birmingham}{\affiliation{Birmingham Institute for Gravitational Wave Astronomy and School of Physics and Astronomy, University of Birmingham, Birmingham B15 2TT, UK}}
\newcommand{\CIERA}{\affiliation{Center for Interdisciplinary Exploration and Research in Astrophysics and Department of Physics and Astronomy, \\Northwestern University, 2145 Sheridan Road, Evanston, IL 60208-3112, USA}}

\shorttitle{FLEET}
\shortauthors{Gomez et al.}

\begin{document}

\title{FLEET: A Redshift-Agnostic Machine Learning Pipeline to Rapidly Identify Hydrogen-Poor Superluminous Supernovae}

\correspondingauthor{Sebastian Gomez}
\email{sgomez@cfa.harvard.edu}

\author[0000-0001-6395-6702]{Sebastian Gomez}
\CfA

\author[0000-0002-9392-9681]{Edo Berger}
\CfA

\author[0000-0003-0526-2248]{Peter K. Blanchard}
\CIERA

\author[0000-0002-0832-2974]{Griffin Hosseinzadeh}
\CfA

\author[0000-0002-2555-3192]{Matt Nicholl}
\Birmingham\Edinburgh

\author[0000-0002-5814-4061]{V. Ashley Villar}
\CfA

\author[0000-0002-5723-8023]{Yao Yin}
\CfA

\begin{abstract}

Over the past decade wide-field optical time-domain surveys have increased the discovery rate of transients to the point that $\lesssim 10\%$ are being spectroscopically classified. Despite this, these surveys have enabled the discovery of new and rare types of transients, most notably the class of hydrogen-poor superluminous supernovae (SLSN-I), with about 150 events confirmed to date. Here we present a machine-learning classification algorithm targeted at rapid identification of a pure sample of SLSN-I to enable spectroscopic and multi-wavelength follow-up. This algorithm is part of the FLEET (Finding Luminous and Exotic Extragalactic Transients) observational strategy. It utilizes both light curve and contextual information, but without the need for a redshift, to assign each newly-discovered transient a probability of being a SLSN-I. This classifier can achieve a maximum purity of about 85\% (with 20\% completeness) when observing a selection of SLSN-I candidates. Additionally, we present two alternative classifiers that use either redshifts or complete light curves and can achieve an even higher purity and completeness. At the current discovery rate, the FLEET algorithm can provide about $20$ SLSN-I candidates per year for spectroscopic follow-up with 85\% purity; with the Legacy Survey of Space and Time we anticipate this will rise to more than $\sim 10^3$ events per year.

\end{abstract}


\section{Introduction}\label{sec:intro}

Type I Superluminous Supernovae (hereafter, \mbox{SLSN-I}) are a class of astrophysical transients that exceed the luminosity of normal SNe by up to two orders of magnitude. They were originally classified based on their luminosity, since most have typical peak absolute magnitudes of $\lesssim -21$ \citep{Chomiuk11,Quimby11}. However, events with spectroscopic signatures that match those of SLSN-I have been discovered at lower luminosities (e.g., \citealt{Lunnan13}) and they are now classified based on their hydrogen-free spectra, strong \ion{O}{2} absorption lines at early time, and a blue continuum \citep{Angus19}. At present, about 150 SLSN-I have been spectroscopically classified; see Table~A.\ref{tab:SLSNe} for a listing and references.

While the energy source of SLSN-I was intensely debated for a few years following their discovery, it now appears that radioactive decay of $^{56}$Ni (as in normal Type I SNe) and circumstellar interaction (as in Type IIn SNe) cannot explain the bulk of the population. Instead, the most likely energy source appears to be the spin-down of a millisecond magnetar produced in the explosion \citep{Kasen10, Metzger15}. This model can explain the diverse light curve behavior \citep{Nicholl17_mosfit}, the early-time UV spectra \citep{Mazzali16}, the late-time light curve flattening \citep{Blanchard18, Nicholl18_1000days}, and the nebular spectra \citep{Dessart12, Nicholl19_nebular} of SLSN-I. Still, the nature of SLSN-I progenitors, their environments, and their relation to those of other stripped-envelope explosions remain areas of active investigation (e.g., \citealt{Blanchard20}). Similarly, the ubiquity and origin of unusual light curve and spectroscopic features seen in some SLSN-I, such as late time ``bumps'' \citep{Nicholl16_15bn,Inserra17,Blanchard18,Lunnan19_four}, double-peaked light curves \citep{Nicholl15_LSQ14bdq}, or potential helium lines \citep{Yan20} remain unclear.

Making progress on these open questions requires a substantial increase in the identification rate of SLSN-I, preferably at early times to enable spectroscopic follow-up. A significant challenge is that SLSN-I are intrinsically rare, at a volumetric rate of $\sim 90$ SNe yr$^{-1}$ Gpc$^{-3}$ at a weighted redshift of $z = 1.13$, they represent $< 0.1\%$ of the core-collapse SN rate \citep{Prajs17}. Even accounting for their larger discovery volume they represent only $\sim 1.5\%$ of the detection rate in magnitude-limited surveys \citep{Villar19,Fremling20}. Currently, only $\sim 10\%$ of all optical transients are classified spectroscopically, and with the Legacy Survey of Space and Time (LSST) on the Vera C.~Rubin Observatory, this will decline to $\lesssim 0.1\%$. Thus, efficient and rapid selection of SLSN-I candidates is essential.

One approach to identifying SLSN-I candidates is to use general purpose machine learning (ML) classification algorithms that attempt to sort optical transients into various spectroscopic classes. Some of these (e.g., {\tt RAPID}: \citealt{Muthukrishna19}, {\tt Avocado}: \citealt{Boone19}) have been trained on synthetic data, such as the Photometric LSST Astronomical Time-series Classification project (PLAsTiCC; \citealt{Kessler19}), but their performance with real data remains untested. Other classifiers such as {\tt SuperRAENN} \citep{Villar20} or {\tt Superphot} \citep{Villar19,Hosseinzadeh20} have been trained on real survey data from the Pan-STARRS1 Medium Deep Survey (PS1/MDS). Overall, these classifiers have a fairly high success rate and recover $\sim 80$\% of SLSN-I, but only when using redshift information and fairly complete light curves. Additionally, the Automatic Learning for the Rapid Classification of Events (ALeRCE) broker, which is currently providing real-time classifications for transients from ZTF \citep{Sanchez20}, is able to recover up to $100\%$ of the SLSN-I in their training sample, but with a large standard deviation of $\sim 26\%$ for the predicted classification, which they estimate by running 20 versions of their classifier.

An alternative approach, which we develop and use in this paper, is to devise a classification algorithm that is optimized specifically for SLSN-I. In \citet{Blanchard_thesis} we introduced an initial simple algorithm that improved SLSN-I selection from the random $\sim 1.5\%$ to $\sim 20\%$, using the brightness contrast between a transient and its host galaxy. This approach yielded other unusual transients as well \citep{Blanchard17,Gomez19,Nicholl20_nature}. Here, we describe a more sophisticated machine learning algorithm that utilizes light curve and contextual information to enable efficient real-time SLSN-I selection without the need for redshift information. This classifier is the core of our FLEET (Finding Luminous and Exotic Extragalactic Transients) observational program. We find that this targeted approach achieves an overall higher success rate than all-encompassing classifiers.

The structure of the paper is as follows. In \S\ref{sec:philo} we introduce and motivate the philosophy behind our approach. In \S\ref{sec:training} we present the data set used to train our algorithm. In \S\ref{sec:context} and \S\ref{sec:model} we outline the contextual and light curve information used for classification, respectively. In \S\ref{sec:algorithm} we describe the ML algorithm and the classification results. In \S\ref{sec:alternatives} we present our alternative classifiers that use redshifts and full light curves as additional information. Finally, we summarize our conclusions in \S\ref{sec:conclusions}. FLEET is provided as a Python package on Github\footnote{\url{https://github.com/gmzsebastian/FLEET}} and Zenodo \citep{Gomez20}, as well as included in the Python Package Index with the name {\tt fleet-pipe}.

\section{Guiding Principles}\label{sec:philo}

As discussed above, there are several efforts aimed at ML classification of astronomical transients, mainly based on light curve information from wide-field surveys. By design, some classifiers make choices that tend to optimize their overall classification success rate across a range of astronomical transients (e.g. \citealt{Boone19,Muthukrishna19,Gagliano20,Hosseinzadeh20,Villar20}). Here, we take a distinct approach by focusing on optimized classification of a single class of transients. Our algorithm is based on the following guiding principles:
\begin{enumerate}[topsep=5pt,partopsep=8pt,itemsep=4pt,parsep=5pt]
  \item Classifying only SLSN-I with no regard for the classification success of other transients.
  \item Obtaining the purest possible sample of SLSN-I, at the expense of sample completeness.
  \item Prioritizing speed and computational resources over model complexity to allow for rapid classification.
  \item Finding SLSN-I at early times to enable real-time follow-up.
\end{enumerate}
This approach enables us to make efficient use of large-aperture telescopes for spectroscopic classification, as well as perform later follow-up studies.

At the present, most transients are reported to the Transient Name Server (TNS), a repository for transient discoveries and classifications. FLEET is designed to assign any transient reported to the TNS a classification probability of being a SLSN-I. The current rate of $\sim 1500$ transients per month reported to the TNS (and $\gtrsim 10^5$ per month expected from LSST) motivates our emphasis on computational speed, as well as purity at the expense of completeness. In particular, even if we manage to identify less than half of the  SLSN-I in the data stream, but with a high success rate, then we can double the existing sample of SLSN-I by the time LSST commences.

We provide a main rapid version of the classifier in addition to two additional classifiers with somewhat different motivations. First, a full light curve classifier that can more confidently classify SLSN-I, but at the expense of early discovery, mainly aimed at constructing large samples with only photometric data. And second, a classifier that uses redshift information for higher purity classification, mainly in anticipation of robust photometric redshifts that will be provided by LSST.

\section{Test Set}
\label{sec:training}

To train our classifier we obtained all spectroscopically classified transients from the TNS: SNe, tidal disruption events (TDEs), active galactic nuclei (AGN) flares, and Galactic transients (e.g., cataclysmic variables and variable stars). In addition to those, we included the TDEs published in \cite{Velzen20}, which are not yet reported to the TNS, and every unambiguous SLSN-I from the literature; see Table~A.\ref{tab:SLSNe}. We also obtained all of the available photometry for each transient, from the Open Supernova Catalog (OSC; \citealt{guillochon17}) or the Zwicky Transient Facility (ZTF; \citealt{Bellm19}). We require each transient to have at least 2 $g$-band and 2 $r$-band measurements to model their light curves. We restrict the list to transients within the footprint of the Pan-STARRS1 $3\pi$ (PS1/$3\pi$) survey \citep{Chambers18} for the purpose of identifying host galaxies. Finally, we removed from the training set $44$ transients with ambiguous host galaxy identifications or spurious data in order to have the cleanest data set possible; however, we kept these events in our test set to prevent any resulting biases. The resulting sample is composed of 1,813 transients, with the following distinct labels from the TNS: 800 SN\,Ia, 381 SN\,II, 156 SLSN-I, 95 CV, 71 SN\,IIn, 63 SN\,IIP, 59 SN\,Ic, 43 SLSN-II, 37 SN\,Ib, 33 SN\,IIb, 19 TDE, 16 SN\,Ic-BL, 13 SN\,Ibc, 12 AGN, 8 SN\,Ibn, and 7 Varstar (variable stars).

Since the number of events per class varies substantially, making the training set unbalanced, the classification would be biased towards the more common classes. To mitigate this bias we over-sample each class to have a total of 800 events, using the Synthetic Minority Over-sampling Technique (SMOTE; \citealt{Chawla02}). This algorithm draws random samples along vectors joining every pair of objects in feature space until all classes have the same number of events. We tested an alternative multivariate-Gaussian (MVG) oversampling technique, as implemented in \citet{Villar19}, but find that when sampling features that are close to zero and constrained to be positive (e.g., redshift), SMOTE performs significantly better; even when imposing a $>0$ threshold for the samples, or sampling in log-space.

Since some of the classes in our sample are too small to be properly over-sampled, we experiment by grouping different sets of transients together, not only to allow for over-sampling but to attempt to optimize the success of the classifier at finding SLSN-I and to improve computational efficiency. We find that the best performing grouping is: Varstar+CV, TDE+AGN, SN\,II+SN\,IIP, SN\,Ib+SN\,Ibn+SN\,Ibc+SN\,Ic+SN\,Ic-BL, SLSN-I, SLSN-II, SN\,IIb, SN\,IIn, and SN\,Ia. We stress that since our interest is in classifying SLSN-I with high purity, the grouping and classification success of the other classes are not critical. Still, it is interesting to note that the optimized groupings are indeed related in terms of underlying physics.

\begin{deluxetable}{cccc}
	\tablecaption{Observational Rates of Transients \label{tab:rates}}
	\tablehead{\colhead{Transient} & \colhead{Fremling}  & \colhead{TNS} & \colhead{Target $f$}}
	\startdata
    SNI     & 587 (77.1\%) & 6500 (70.8\%) & 73.9\% \\ 
    SNII    & 155 (20.4\%) & 2109 (23.0\%) & 19.6\% \\  
    SLSN-I  & 12   (1.6\%) & 123   (1.3\%) &  1.5\% \\  
    SLSN-II & 7    (0.9\%) & 45    (0.5\%) &  0.9\% \\  
    Nuclear & --           & 58    (0.6\%) &  0.6\% \\  
    Star    & --           & 340   (3.7\%) &  3.5\%     
    \enddata
	\tablecomments{Observational rates for the relevant types of transients considered here. We normalize the rate of events in our test set to an expected Target rate $f$ calculated from the \citet{Fremling20} sample and the TNS sample, used for Equation~\ref{eq:eta}.}
\end{deluxetable}

\subsection{Test Set}
\label{sec:testing}

We test the efficacy of our classifier on all of the events from the training set. In addition to all the events from the training set, we include in the test set the $44$ transients that were removed in \S\ref{sec:training} to avoid introducing possible biases. We implement a leave-one-out cross-validation method, allowing us to train the classifier on every event except for one, and then predict the classification of that one event, cycling through all events. This allows us to robustly test our classifier without having to divide the data set into a training and test set, which would compromise the sample size.

We define {\it completeness}, {\it classifier purity}, and {\it observed purity} as useful metrics to test the efficacy of our algorithm:
$$
\mathrm{Completeness} = \frac{\rm SN_T}{\rm N_{SLSN}}
$$
$$
\mathrm{Classifier\ Purity} = \frac{\rm SN_T}{\rm SN_T + \rm SN_F}
$$
$$
\mathrm{Observed\ Purity} = \frac{\rm SN_T}{\rm SN_T + \sum_{i} \eta_i\rm SN_{F,i}}
$$

\begin{equation}\label{eq:eta}
\eta_i = \frac{{\rm N_{SLSN}} \times f_i}{N_{i} \times f_{\rm SLSN}},
\end{equation}

where ${\rm N_{SLSN}}$ is the total number of SLSN-I in the test set, ${\rm SN_T}$ is the total number of true positive SLSN-I recovered, and ${\rm SN_F}$ is the total number of false positive SLSN-I. The relative fractions of each transient class in our test set, which we obtained directly from the TNS, does not reflect the true fractions of these transients in a magnitude-limited survey. To determine a purity that is representative of on-going and future surveys, we re-normalize the {\it classifier purity} into an {\it observed purity}, which more accurately represents the outcome of our pipeline in a real survey. Here, $\rm SN_{F,i}$ is the false positive rate for an individual transient class $i$, and $f_i$ is the corresponding true observational rate for that class, listed in Table~\ref{tab:rates}. We use the observational rates of SNe from \citet{Fremling20} to estimate the expected Target Rate, $f$, for any magnitude-limited survey. We then include nuclear transients (TDEs + AGN) and Galactic transients (CVs + variable stars) from the TNS, normalizing by the total number of classified transients from the TNS to the total number of SNe in the \citet{Fremling20} sample.

Given that SLSN-I are over-represented in our test set compared to the rate they would have in a magnitude limited survey, observed purity will be lower than the classifier purity. For example, our test set has 800 SN Ia and 156 SLSN-I, or 0.20 SLSN-I for each SN Ia. But in a magnitude limited survey, there is typically only 0.02 SLSN-I for each SN Ia. Therefore, if we wanted to predict how many SLSN-I we would be able to find in a real survey, we need to normalize the classifier purity, in this example by multiplying the false positive rate by a factor of $\sim 10$.

\section{Contextual Information}
\label{sec:context}

\begin{figure}
    \begin{center}
    \centering
    {\includegraphics[width=\columnwidth]{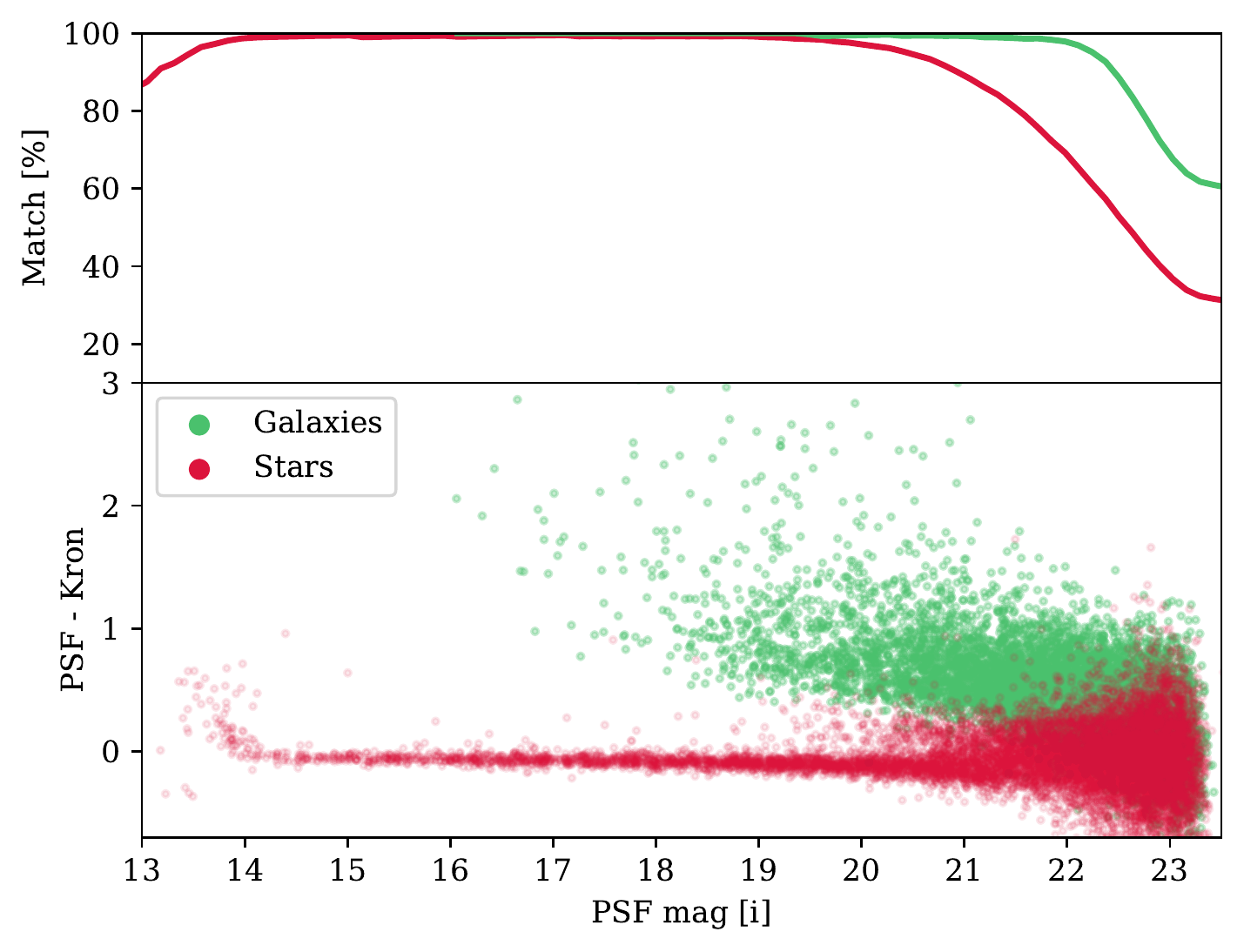}}
    \caption{Galaxies (green) and stars (red) classified by the CFHTLS survey (D1 field) plotted in terms of the difference between their PSF and Kron magnitude as a function of apparent $i$-band magnitude in PS1/3$\pi$. Using this calibration, we assign a probability of being a galaxy to all objects in the field of a transient based on their location in this diagram. The top panel shows the percent of objects for which our classification matches that of the CFHTLS as a function of apparent magnitude, a 90\% match occurs at a magnitude of 22.5. \label{fig:separator}}
    \end{center}
\end{figure}

SLSN-I are known to prefer low-luminosity galaxies \citep{Lunnan14}, and it is therefore advantageous to use contextual information in their classification. Here we describe our method of assigning a host galaxy to each transient, while in \S\ref{sec:features} we explore which host galaxy properties are the most useful features in the SLSN-I classification. For each transient in our training set we obtain PS1/$3\pi$ $grizy$ \citep{Chambers18} and SDSS $ugriz$ \citep{Alam15,Ahumada19} PSF and Kron magnitudes of every cataloged source in a $1'$ radius region around the transient location. We use this information both to separate galaxies from stars, and to identify the most likely host galaxy.

\begin{figure}
    \begin{center}
    \centering
    {\includegraphics[width=\columnwidth]{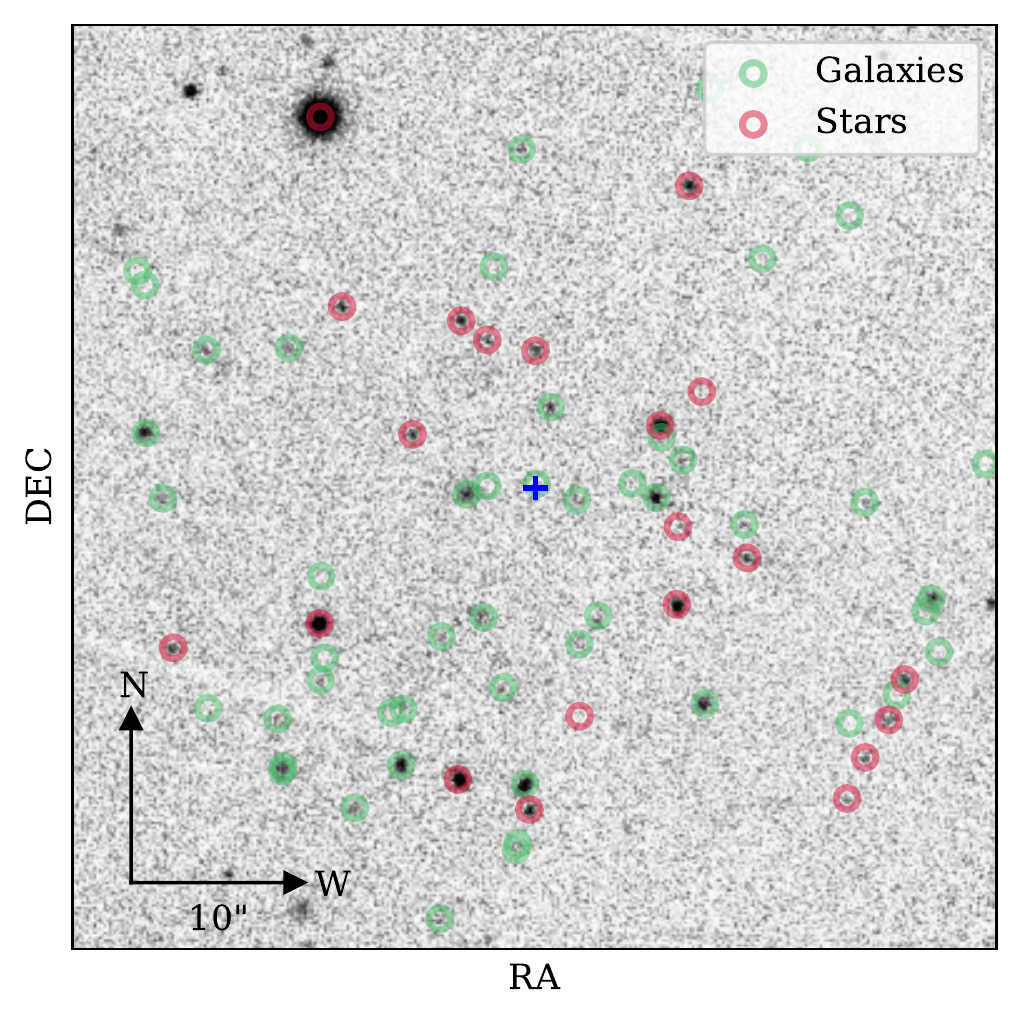}}
    \caption{PS1/3$\pi$ $i$-band image of a $1'\times 1'$ field centered on the position of the SLSN-I SN\,2013hy, indicating objects classified as galaxies (green) and stars (red) based on our star-galaxy separation algorithm (\S\ref{sec:separator}). The blue cross marks the location of the SN and its associated host galaxy with $P_{\rm cc}\approx 0.03$ as determined by the algorithm described in \S\ref{sec:host}. 
    \label{fig:image}}
    \end{center}
\end{figure}

\subsection{Star-Galaxy Separation}
\label{sec:separator}

The first step to identifying the host galaxy of each transient is to separate stars from galaxies. SDSS provides a classification for every object in their catalog, but since SDSS is shallower than PS1/$3\pi$ and has a smaller footprint, this is not sufficient for our purposes. Instead, we develop a method to assign a probabilistic value (between 0 and 1) of how likely every object in SDSS and PS1/$3\pi$ is to be a galaxy.

To train our star-galaxy separation algorithm we use data from the Canada-France-Hawaii Telescope Legacy Survey (CFHTLS; \citealt{Hudelot12}), which provides magnitudes and star-galaxy classifications down to $\approx 26$ mag, significantly deeper than SDSS and PS1/$3\pi$. We specifically use the D1 field (1 deg$^2$) and cross-match with every overlapping object in SDSS and PS1/3$\pi$, for a total of $\sim 23,000$ objects. Galaxies tend to have a larger difference between their PSF and Kron magnitudes than stars, so we use this specific feature (${\rm PSF-Kron}$) to separate them; see Figure~\ref{fig:separator} for an example in the $i$-band. The CFHTLS uses the {\tt CLASS\_STAR} classifier flag in {\tt SExtractor} to separate stars from galaxies, which relies on a multi-layer feed-forward neural network \citep{Bertin96}. 

In our galaxy-star separator we assign a probability of being a galaxy to any object in SDSS or PS1/3$\pi$ by using a custom k-nearest-neighbors algorithm. Given an object's PSF and Kron magnitude, we find the 20 nearest objects in the PSF versus ${\rm PSF-Kron}$ phase-space (Figure~\ref{fig:separator}) to calculate its probability of being a galaxy based on the fraction of those 20 neighbors from the CFHTLS training set that are galaxies. Experimenting with different number of neighbors, we find that at least 10 neighbors are required to produce robust estimates, with only marginal improvement in accuracy beyond 20 neighbors. For every object we calculate its probability of being a galaxy in every available filter, and adopt the average probability among all filters. 

An alternative star-galaxy separator for objects in PS1/3$\pi$ is presented in \citet{Tachibana18}. Although this latest one has a very high accuracy, it does not include objects from SDSS, for which we also require a classification when they are not in the PS1/3$\pi$ catalog. We note that if we label objects with a probability of being a galaxy of $P_G \leq 10\%$ as stars, our classifier agrees with the classification from \citet{Tachibana18} at the $90\%$ level. In Figure~\ref{fig:image} we show an example of our star-galaxy separator applied on a field from PS1/3$\pi$ centered on the location of the SLSN-I SN\,2013hy.

We opt to only label objects with a galaxy probability of $P_G <10\%$ as stars to avoid missing a possible host galaxy identification. While this conservative cut retains more stars in the sample, these are rarely predicted to be the most likely host galaxy of a SN due to the small size of their PSF. We find that a more strict threshold results in a large number of host galaxies being rejected as stars. In the top panel of Figure~\ref{fig:separator} we show that using the classification from the CFHTLS as a reference, our threshold for labeling stars yields a successful galaxy classification for essentially all objects with $\gtrsim 22$ mag and $\approx 65\%$ down to 23 mag.

\subsection{Host Identification}\label{sec:host}

\begin{figure}
    \begin{center}
    \centering
    {\includegraphics[width=\columnwidth]{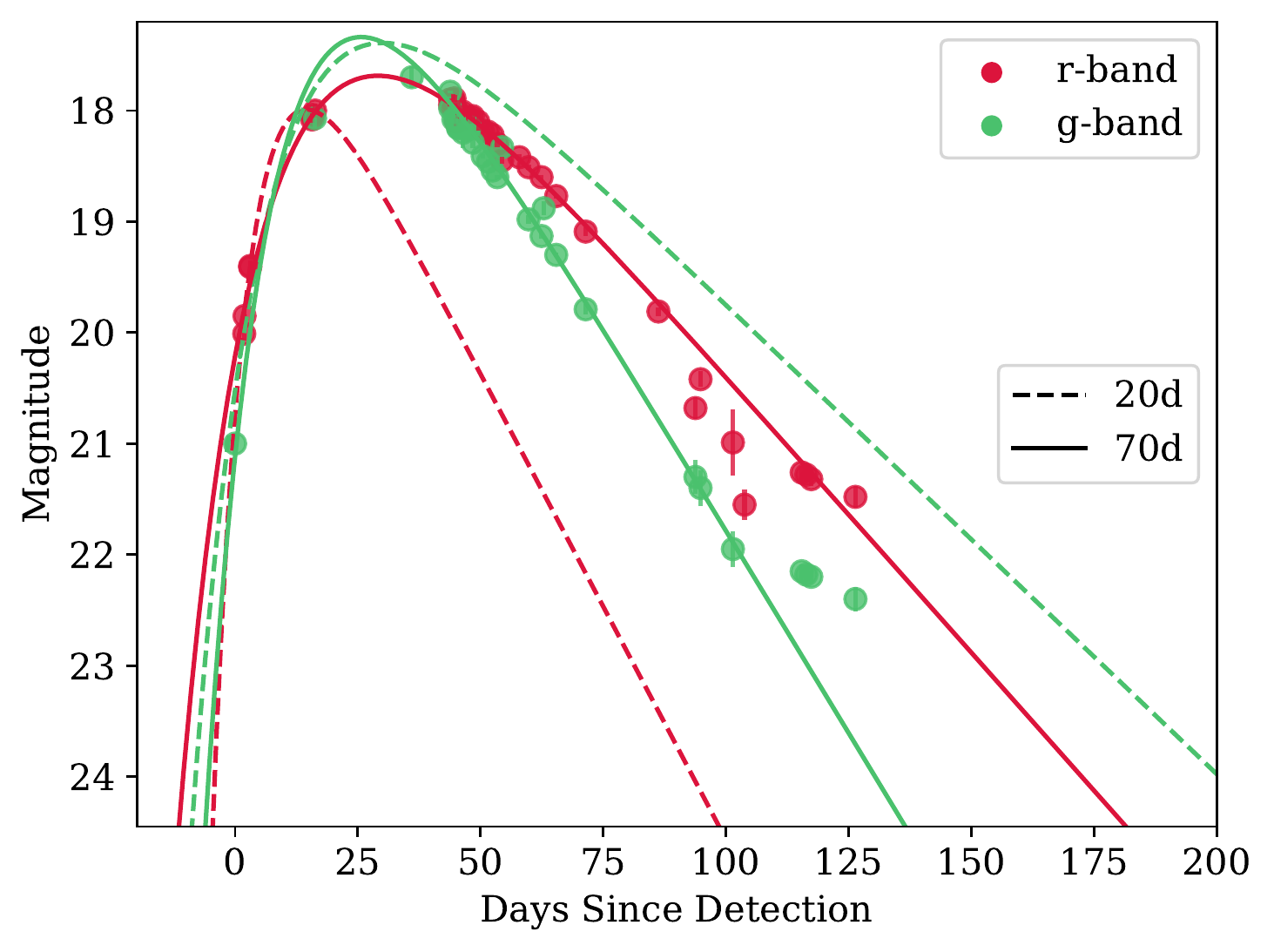}}
    \caption{Light curves of the SLSN-I SN\,2011ke fit with the model described in Equation~\ref{eq:lc}. The dashed lines show the fit using only data up to 20 days after detection (with a fixed value of $A=0.6$), while the solid lines are the result of fitting the data up to 70 days after detection (with $A$ as a free parameter). The former is part of our main rapid classifier, while the latter is part of an alternative classifier that uses full light curves (\S\ref{sec:full}).
    \label{fig:2011ke}}
    \end{center}
\end{figure}

\begin{deluxetable*}{ccccc}
	\tablecaption{Feature Sets \label{tab:features}}
	\tablehead{\colhead{\#} & \colhead{Features}  & \colhead{Peak Classifier Purity} & \colhead{Completeness} & \colhead{\pslsn}}
	\startdata
    1 & $W$+$\Delta t$+$R_n$+ $\Delta m$         & $83.0\pm2.0$\%  & $16.3\pm1.8$\% & 0.73 \\
    2 & $W$+$\Delta t$+ $\Delta m$               & $73.3\pm40.8$\% & $2.6\pm1.7$\%  & 0.81 \\
    3 & $W$+$\Delta t$+$R_n$                     & $81.5\pm2.5$\%  & $20.4\pm1.3$\% & 0.84 \\
    4 & $R_n$+$\Delta m$                         & $98.7\pm6.5$\%  & $4.1\pm1.4$\%  & 0.89 \\
    5 & $W$+$R_n$                                & $75.5\pm4.9$\%  & $16.1\pm1.8$\% & 0.74 \\
    6 & $W$+$\Delta t$+$R_n$+$(g-r)$             & $91.2\pm2.4$\%  & $17.9\pm3.3$\% & 0.80 \\
    7 & $W$+$\Delta t$+$R_n$+ $\Delta m$+$(g-r)$ & $82.0\pm25.4$\% & $2.2\pm0.9$\%  & 0.88 \\
    8 & $W$+$R_n$+ $\Delta m$+$(g-r)$            & $94.7\pm12.2$\% & $3.0\pm0.7$\%  & 0.87 \\
    \enddata
	\tablecomments{Different sets of light curve and contextual features used to train our classifier. We list the highest classifier purity that each set of features achieves, as well as the corresponding completeness and classification probability \pslsn\ that correspond to that peak purity. $W$ is the width of the light curve, $R_n$ is the normalized host separation, $\Delta m$ is the peak transient magnitude minus the host magnitude, $\Delta t$ is the time of peak magnitude minus the time of discovery, and $(g-r)$ is the light curve color at peak.}
\end{deluxetable*}

Once we have identified which objects in the field are likely to be galaxies we can determine which galaxy is the most likely host for a given transient. First, we label stellar transients, using the criterion of a star (i.e., $P_G < 10$\%) being located  $<1''$ from a transient's position. Then, for the non-stellar transients we determine the probability of chance coincidence for each galaxy in the field relative to the transient's position. We follow the method of \citet{Bloom02} and \citet{Berger10} using the measured number density of galaxies, $\Sigma(\le m)$, brighter than a magnitude $m$, to calculate the probability of chance coincidence:
\begin{equation}\label{eq:pcc}
\begin{split}
    P_{\rm cc} & = 1 - e^{-\pi\left(d^2 + 4 R^2\right) \Sigma(\le m)} \\
    \Sigma(\le m) & = \dfrac{10^{0.33 (m-24)-2.44}} {0.33 \, {\rm ln}(10)},
\end{split}
\end{equation}
where $d$ is the angular separation between the center of a galaxy and the transient, and $R$ is the half-light radius of the galaxy obtained from the SDSS catalog, or from the PS1/$3\pi$ catalog if the object is not in the SDSS catalog. We consider the galaxy with the lowest value of $P_{\rm cc}$ to be the host, as long as $P_{cc}\le 0.1$. Otherwise, we designate the transient as ``host-less'' given the more likely situation that its host galaxy is fainter than the magnitude limit of SDSS and PS1/$3\pi$.

\section{Light Curve Model}
\label{sec:model}

In addition to the contextual information, we use the light curves of each transient to predict which transients are most likely SLSN-I. We obtain photometric data from the OSC, as well as from ZTF using the Make Alerts Really Simple (MARS)\footnote{\url{https://mars.lco.global/}} broker. We correct all the photometry for Galactic extinction using the \citet{Schlafly11} dust maps assuming $R_V=3.1$.

Since we are interested in gross features of the light curves, rapid classification, and identifying only SLSN-I (rather than robustly classifying all transient classes), we use a simple exponential light curve model:
\begin{equation}
\label{eq:lc}
m = e^{W (t - \phi)} - A \times W (t - \phi) + m_0,
\end{equation}
where $W$ is the effective width of the light curve, $A$ modifies the decline time relative to the rise time, $m_0$ is the peak magnitude, and $\phi$ is a phase offset relative to the time of the first observation. An example of this function fit to a SLSN-I (SN\,2011ke) is shown in Figure~\ref{fig:2011ke}. We fit this model independently to the $g$- and $r$-band light curves using the {\tt emcee} implementation of the Goodman and Weare \citep{Goodman10} Markov chain Monte Carlo algorithm \citep{foreman13} and adopt the median of the posterior as the best estimate for each parameter. We use flat uninformative priors for all parameters, but initiate the walkers' position at a value of $m_0$ equal to the brightest observed magnitude, and a value of $\phi$ that corresponds to the time of that measurement. We find that a model with 50 walkers and 500 steps converges and provides good results for the majority of transients, with a typical auto-correlation time of $\sim 30$ steps.

We use two versions of Equation~\ref{eq:lc} to test and evaluate the classifier. One version has a fixed value of $A = 0.6$ (the mean value from fitting all of the SLSN-I light curves up to a timescale of 70 days post-discovery), and is used for the rapid version of FLEET, which only uses the first 20 days of data (described in \S\ref{sec:algorithm}). We note that the actual choice of $A$ has only a marginal effect on the results, since this model only uses data up to 20 days after detection, which do not encompass a decline phase. The second version of the model uses data up to 70 days after discovery and has $A$ as a free parameter to fit the light curve decline. This model is used for the full light curve classifier described in \S\ref{sec:full}. In Figure~\ref{fig:2011ke} we show both versions of the model, using only the first 20 days of data (fixed $A$) and 70 days of data ($A$ as a free parameter).

\section{Classification Algorithm}
\label{sec:algorithm}

To classify the transients we use the contextual and light curve information described in \S\ref{sec:context} and \S\ref{sec:model}, respectively, with an implementation of the random forest (RF) algorithm in the {\tt scikit-learn} Python package \citep{Pedregosa12}. In this manner, we assign to each transient a classification probability of being a SLSN-I. This algorithm takes various sub-samples of the training set and forms a number of decision tree classifiers to classify each object. The output classification probability is the result of averaging the output of all the trees in the forest. We run the classifier with 100 estimators to mitigate over-fitting and improve predictive accuracy. We also run each version of the model 25 times using different initial random seeds to estimate the classifier's uncertainties. We run the classifier using the Gini index as the criterion that minimizes the probability of misclassification. We optimize the depth of the trees in each RF by running a grid of models from a depth of 3 to 12 in steps of 1 and find a depth of $7$ performs best (a depth of 6 and 8 performed similarly well, within a $1\sigma$ uncertainty derived from the different random seed iterations).

Additionally, we optimize the grouping of transient classes into different sets, described in \S\ref{sec:testing}. We find that for the most part grouping different classes of SNe together (e.g. SN\,IIn and SN\,II) or separating SNe into distinct classes (e.g. SN\,Ib and SN\,Ic) provides very similar results, with the one exception of grouping all SNe that are not SLSNe into one group, which produces a much lower purity.

\subsection{Feature Selection}
\label{sec:features}

\begin{figure}
\begin{center}
\scalebox{1.}
\centering
{\includegraphics[width=\columnwidth]{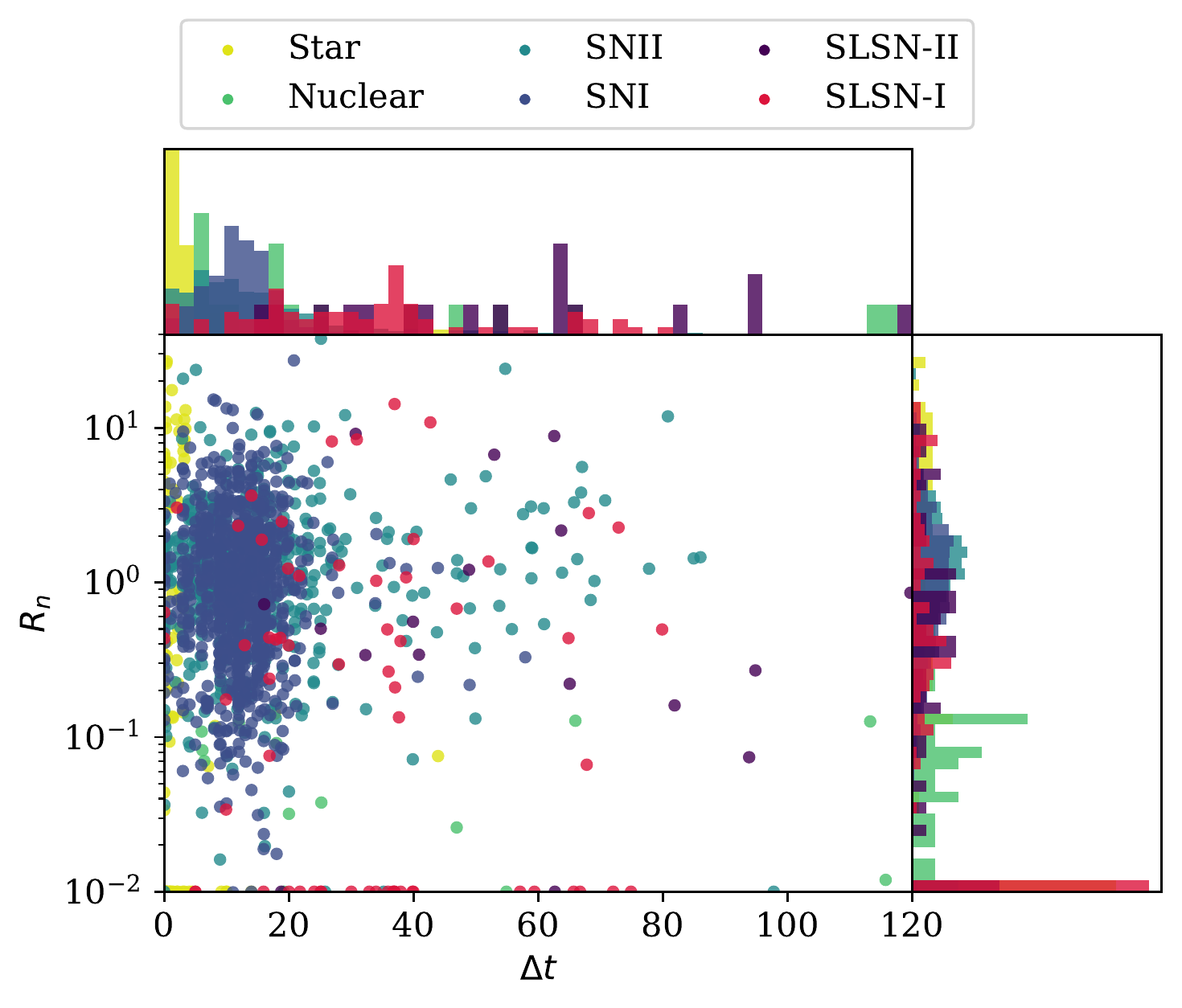}
\includegraphics[width=\columnwidth]{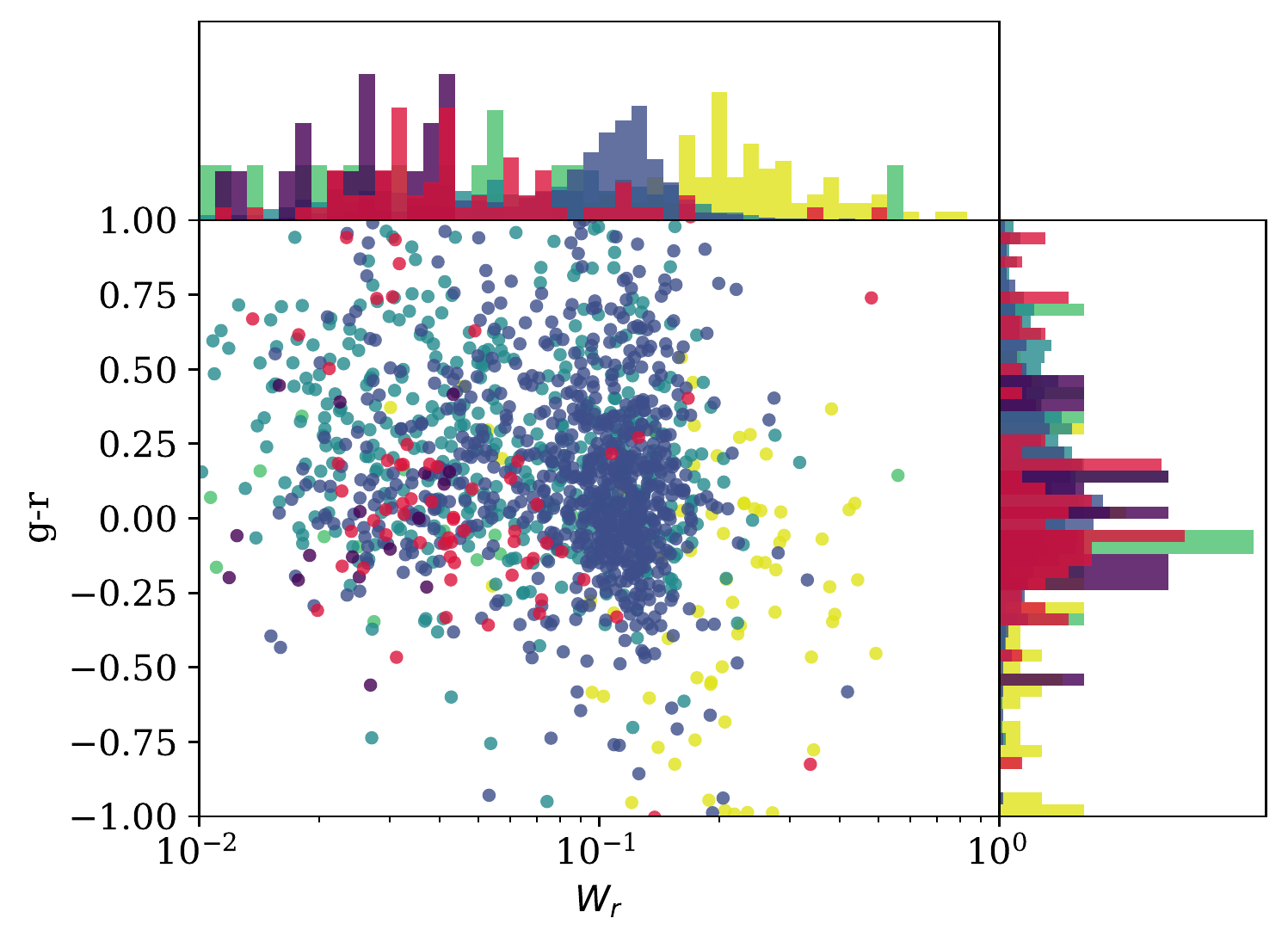}}
\vspace{0.0in}
\caption{Phase-spaces of features selected for the classifier, plotted for the various classes of transients. \textit{Top}: The normalized host separation ($R_n$) versus the time difference between the light curve peak and the first detection ($\Delta t$). For host-less transients we set $R_n=0$ (Shown here at $R_n = 0.01$ for visualization purposes). \textit{Bottom} : Light curve width in $r-$band $W_r$, compared to the color of the transient during peak $(g-r)$. \label{fig:histogram}}
\end{center}
\end{figure}

\begin{figure}
\begin{center}
\scalebox{1.}
\centering
{\includegraphics[width=0.8\columnwidth]{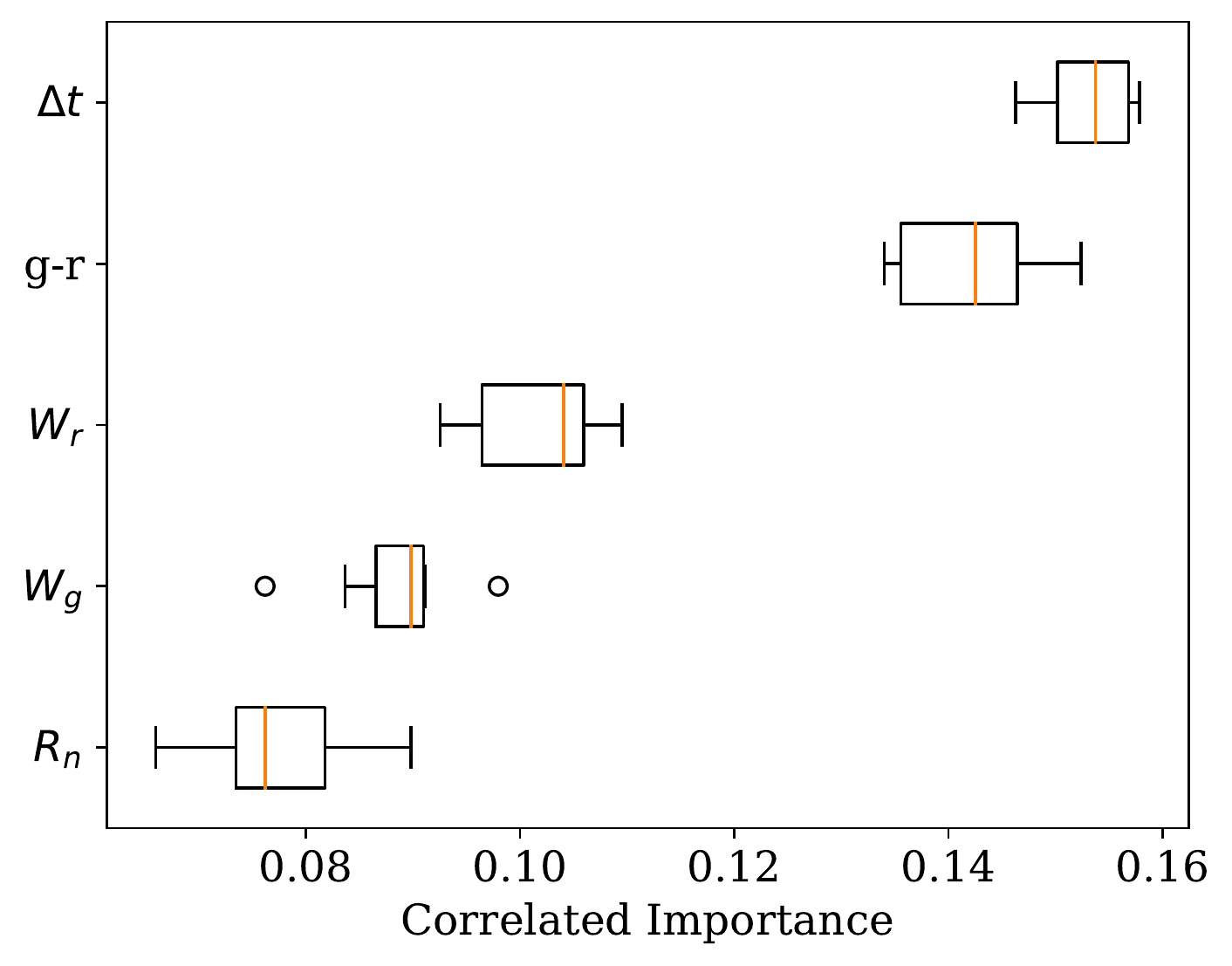}
\includegraphics[width=0.8\columnwidth]{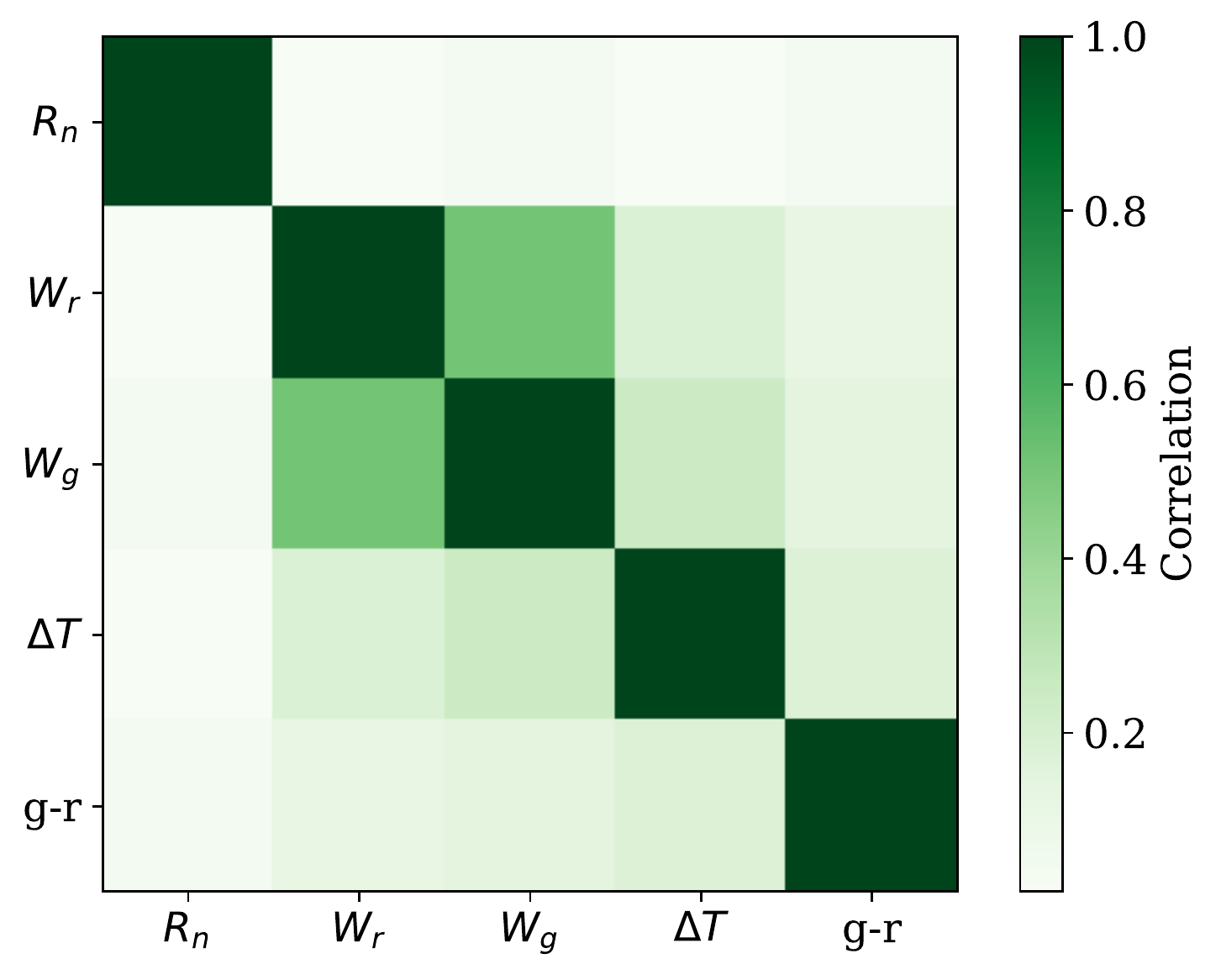}}
\vspace{0.0in}
\caption{\textit{Top}: Correlated importance for the features used in the rapid version of our classifier. \textit{Bottom} : Correlation matrix for the same features. \label{fig:correlation}}
\end{center}
\end{figure}

Unlike newly-discovered transients, the transients in our training set have full light curves. Since a goal of FLEET is to find SLSN-I in real-time we test the algorithm using a varying cutoff time for the light curve data. Naturally, with more data the light curve models are better constrained, but this delays the identification and spectroscopic follow-up into a later phase when the SN is fainter. For our rapid classifier we find optimal results when using the first 20 days of data for each light curve, by which time most SLSN-I have not reached their peak luminosity.

\begin{figure}
    \begin{center}
    \centering
    {\includegraphics[width=\columnwidth]{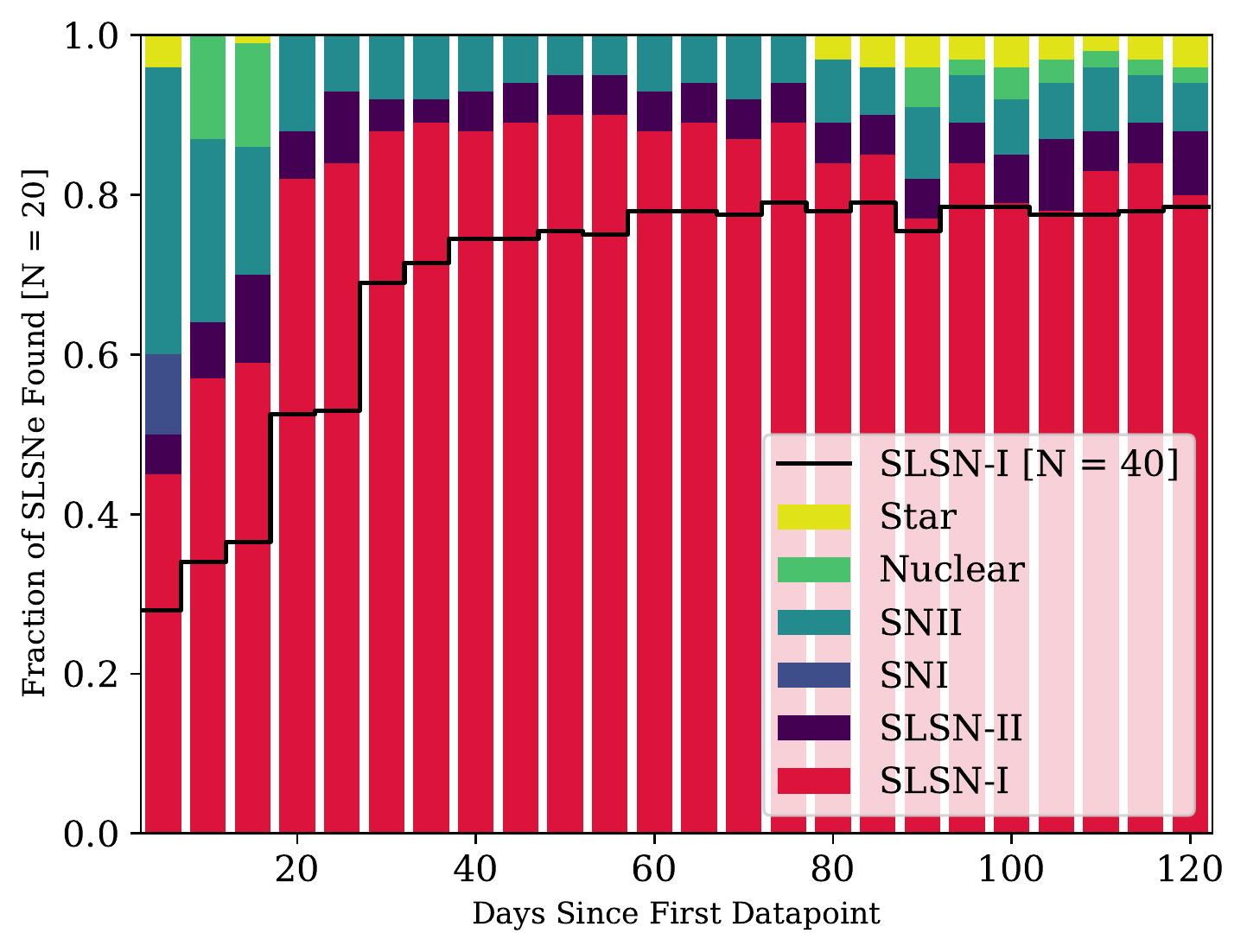}}
    \caption{Fraction of SLSN-I correctly identified by the rapid version of our classifier amongst the top 20 objects predicted to be SLSN-I as a function of days of light curve data used. The peak purity is about 90\% when using $\gtrsim 20$ days of data. This purity is relevant for the training set, before normalizing to observational rates in a magnitude-limited survey (\S\ref{sec:testing}). The black line is the equivalent fraction of SLSN-I found when using the top 40 objects as opposed to the top 20. The small contamination for the Star class at $\gtrsim 70$ days comes from a single CV with a 50-day long outburst that was classified as a SLSN-I due to its long light curve and lack of detected ``host''. \label{fig:topbreakdown}}
    \end{center}
\end{figure}

For the rapid classifier we have 6 light curve parameters (3 in each filter) that can be used as input features: the widths of the light curve, $W$, the phase offsets $\phi$, and the peak magnitudes $m_0$. In addition to these we explore the use of two additional features: (i) $\Delta t$, which is the time difference from first detection of a transient to its observed light curve peak in either $g$- or $r$-band, whichever one is brightest; and (ii) the $g-r$ color at peak, using the model fits, where the time of peak is the one with the brightest observed magnitude in either $g$- or $r$-band.

For the contextual information features we test the use of several host galaxy parameters: the apparent magnitude of the host, $m_h$, its half light radius in $r$-band, $R$, the projected angular separation between the transient and its host center $D$, the projected angular separation normalized by the galaxy radius $R_n$, and the difference between $m_0$ and $m_h$ in $r$-band, $\Delta m$. For host-less transients we use the limiting magnitude of PS1/$3\pi$ of $r=23.2$ as an upper limit on $m_h$, and set all other galaxy parameters to 0 (since those cannot be measured for a non-detected host).

We tested several combinations of the available light curve and contextual features in order to determine which combination set yields the highest purity of SLSN-I, while maintaining reasonable completeness; listed in Table~\ref{tab:features}. We find that the most relevant features that help separate SLSN-I from other transients are $W_g$ and $W_r$, $\Delta t$, $R_n$, $\Delta m$, and $(g-r)$. In Figure~\ref{fig:histogram} we show how the different classes of transients lie in feature-space. In Table~\ref{tab:features} we list the highest purity, and associated uncertainty, achieved for each feature set, as well as the corresponding completeness and classification confidence, \pslsn, at which this highest purity is achieved.

We find that for the rapid classifier, feature set \#6 performs best in terms of purity, while retaining a reasonable completeness. This set contains the width of the light curve $W$ in $g$- and $r$-band, the normalized host separation $R_n$, the time of peak magnitude minus the time of discovery in either band, $\Delta t$, and the light curve color at peak, $(g-r)$.

The importance of each feature used is not defined independently of other features, if two features are correlated then their relative importance might be affected. In the bottom panel of Figure~\ref{fig:correlation} we show the correlation between features, and find that with the exception of $W_g$ and $W_r$, the features are mostly independent. In order to calculate the correlated importance we use the permutation importance method described in \cite{Breiman01}. The correlated importance of each feature is shown on the top panel of Figure~\ref{fig:correlation}.

In Figure~\ref{fig:topbreakdown} we show how the rapid version of the classifier (trained on the first 20 days of light curve data) performs as a function of days of light curve data used, and include the contaminating classes of transients. When considering the top 20 transients with the highest predicted confidence \pslsn, we find that the classifier performance rises for the first $\sim 20$ days, and then plateaus to a peak classifier purity of about $90\%$ (i.e., we correctly identify about 18 of the top 20 transients classified as SLSN-I). This purity is relevant for the training set, without normalizing for the observational rates described in \S\ref{sec:testing}. The remaining $10\%$ of misclassified events are SLSN-II and SNII. We are generally less concerned about misclassifying SLSN-II as SLSN-I since the former are still of scientific interest. The performance of the classifier degrades slightly beyond 70 days, since it is only trained on the rising part of the light curve. If we instead consider the top 40 events predicted to be SLSN-I, we find that the fraction of correctly identified SLSN-I goes down to about $75\%$ (Figure~\ref{fig:topbreakdown}).

\subsection{Model Validation}

\begin{figure}
    \begin{center}
    \centering
    {\includegraphics[width=\columnwidth]{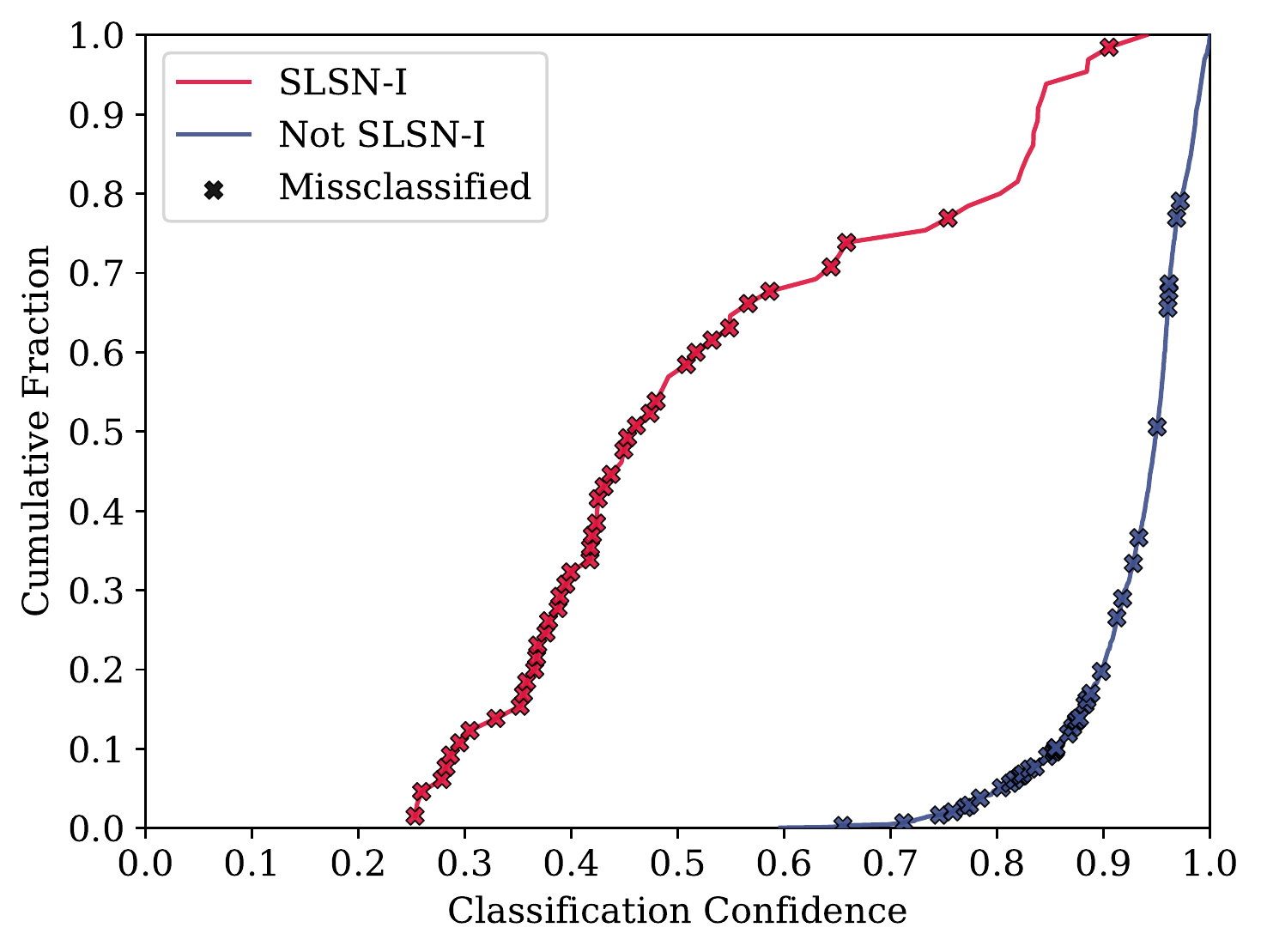}}
    \caption{Cumulative distribution as a function of classification confidence ($P$) for transients classified as SLSN-I (red) and non-SLSN-I (blue). The crosses mark events that are misclassified. We find that for the SLSN-I sample, the misclassified events are mainly concentrated at $P({\rm SLSN-I})\lesssim 0.6$.
    \label{fig:crossy}}
    \end{center}
\end{figure}

\begin{figure}
    \begin{center}
    \centering
    {\includegraphics[width=\columnwidth]{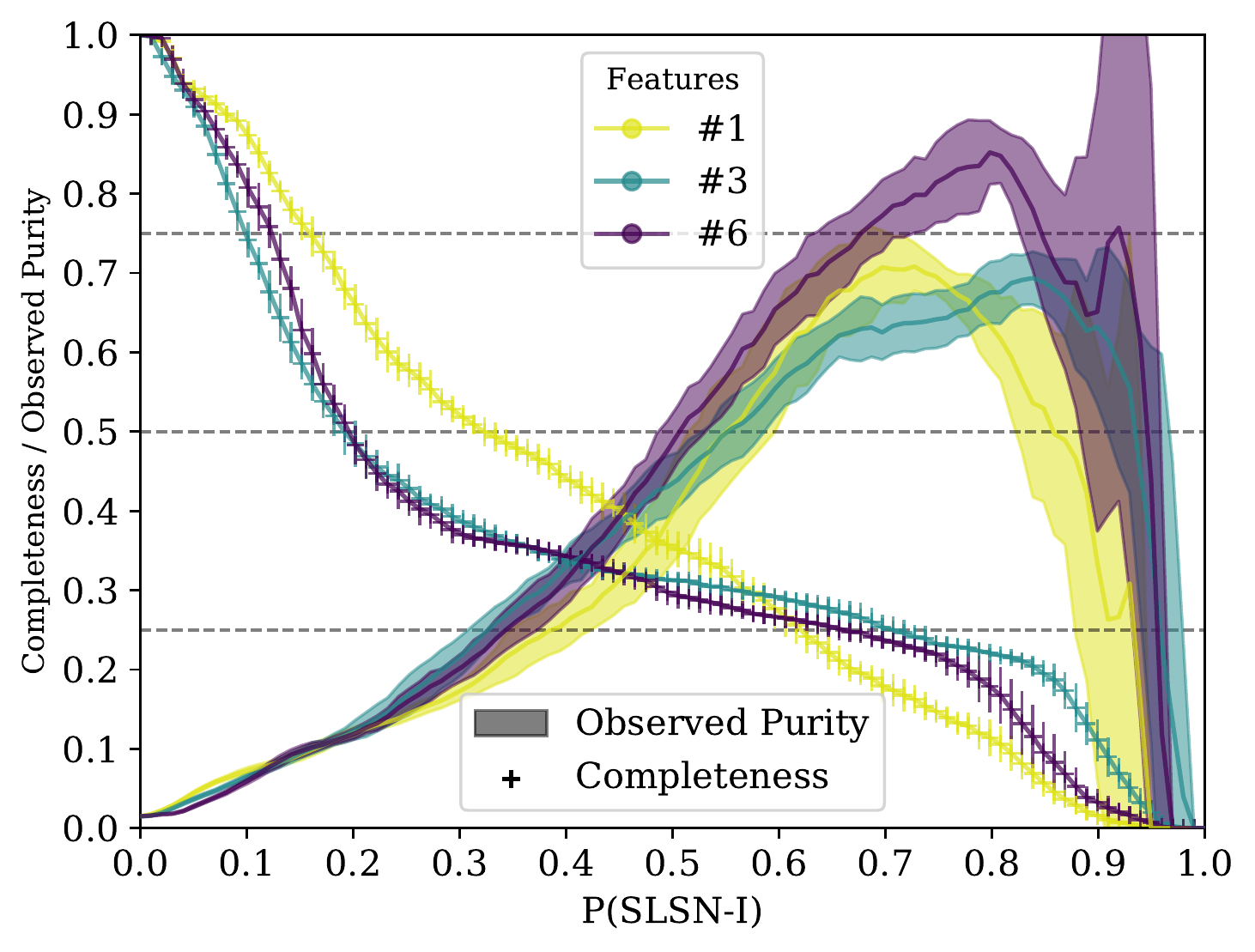}}
    \caption{The observed purity and completeness for the best performing set of features described in Table~\ref{tab:features}. The purity curve represents the percent of transients that are SLSN-I as a function of the classifier confidence p(SLSN-I). The shaded region for the purity and error-bars for the completeness represent $1\sigma$ uncertainties. \label{fig:features}}
    \end{center}
\end{figure}

\begin{figure}
    \begin{center}
    \centering
    {\includegraphics[width=\columnwidth]{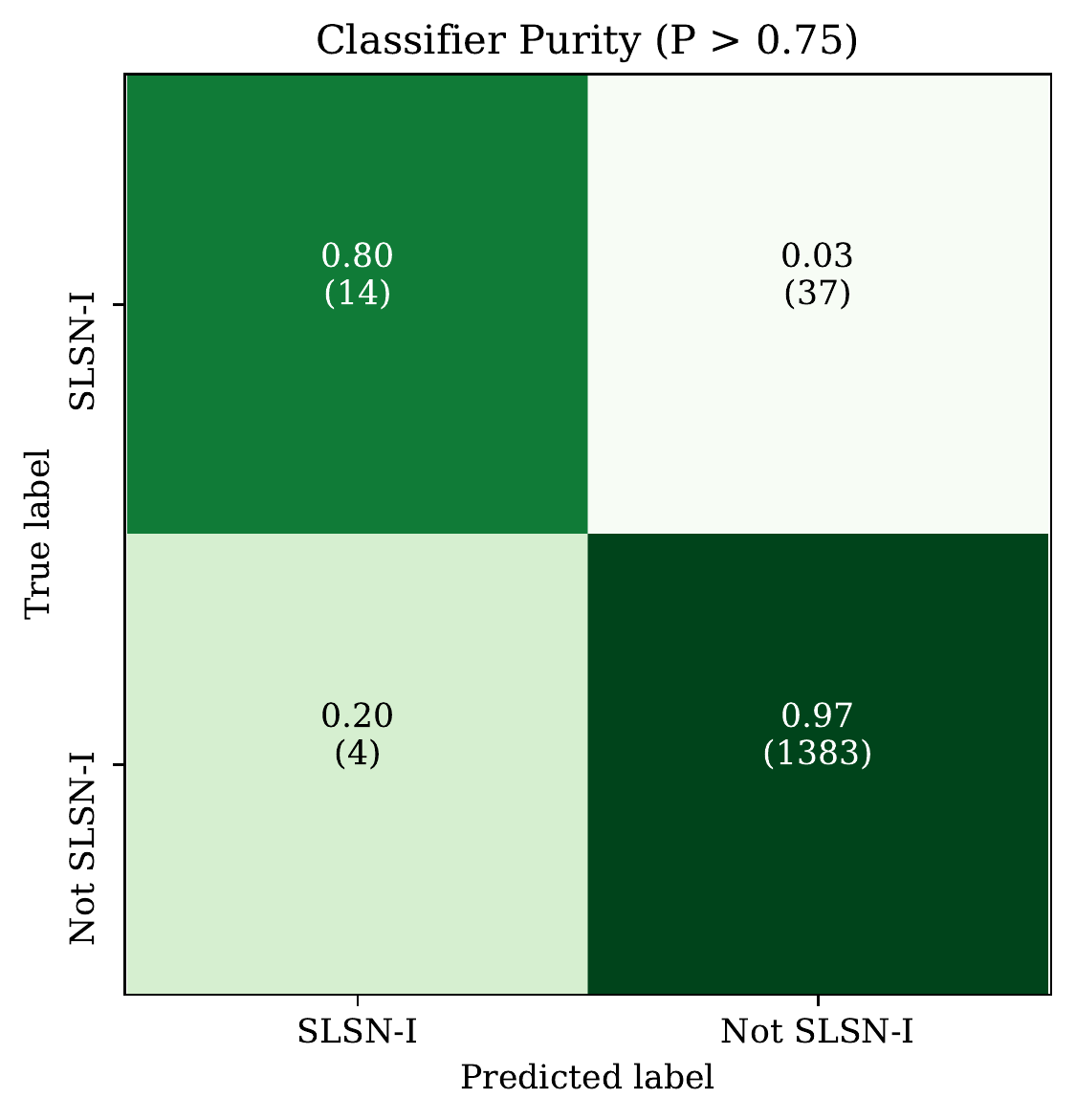}}
    \caption{Confusion matrix that indicates a purity of 80\% for SLSN-I. Based only on objects with a classification probability of P(SLSN-I) $> 0.75$ or P(not-SLSN-I) $> 0.75$, for a total of 1438 transients. \label{fig:confusion}}
    \end{center}
\end{figure}

We use three different methods to evaluate the performance of our classifier: a confusion matrix, a purity/completeness curve, and the fraction of SLSN-I recovered. Unless otherwise stated, in this section the values being reported have been corrected for the observational rates expected in a magnitude-limited survey as described in \S\ref{sec:testing}; i.e., we use the {\it observed purity}. Since we are not concerned with the classification of transients other than SLSN-I, we collapse the individual transient classifications into a binary SLSN-I versus non-SLSN-I classification. To calculate the probability of non-SLSN-I for each transient we sum the probabilities of all other transient classes.

In Figure~\ref{fig:crossy} we show how the rapid classifier performs at classifying SLSN-I and not misclassifying other objects, as a function of classification confidence level. We find that most of the misclassified SLSN-I are at \pslsn$\lesssim 0.6$, with only 4 misclassified SLSN-I at higher values of \pslsn. The few objects that are true SLSN-I but were missclassified as something else with high confidence are usually SLSN-I with relatively bright host galaxies that got missclassified as Type-II SNe, which have light curves that might also appear broad due to their late-time plateau.

The completeness and purity of the rapid classifier for the three top performing feature sets are shown in Figure~\ref{fig:features}. As expected, the purity increases and the completeness declines as we restrict the sample to events with progressively higher values of classification confidence. For \pslsn$>0.5$, the observed purity is $\approx 50\%$, with a completeness of $\approx 30-40\%$. This represents about a factor of 30 times improvement over a random selection of SLSN-I, which would yield $\approx 1.5\%$ success rate in a magnitude-limited survey \citep{Fremling20,Villar19}. The peak observed purity achieved by our classifier is even higher, $\approx 85\%$; however, we note that the completeness achieved at peak observed purity is about $20\%$. This low completeness level may still be acceptable given that current surveys are reporting $\sim 18,000$ transients a year, assuming an observational rate of 1.5\% for SLSN-I (Table~\ref{tab:rates}), a 20\% completeness corresponds to $\sim 50$ SLSN-I a year that could be discovered.

In Figure~\ref{fig:confusion} we show the confusion matrix, namely, the label predicted by our classifier compared to the true label of the transient. We impose a confidence cut of $P>0.75$ for either the SLSN-I or not-SLSN-I classes, corresponding to the peak classifier purity (Figure~\ref{fig:features}); this leads to a sample of 1438 events. We see that 14 out of the 18 transients predicted to be SLSN-I are correctly labeled, indicating a classifier purity of $80\%$.

We run an additional model validation to test for over-fitting. Given the relatively small sample size of our data set we split the entire data set into two independent sets, a training set (with 1209 objects) and a test set (with 604 objects), as opposed to a traditional training/test/validation set. We optimize the combination of transient class grouping, depth or the RF trees, and included features using a leave-one-out cross-validation method on the training set. We find that the best results (in terms of purity and completeness) are consistent with the main classifier presented in this section, with the exception that a depth of 5 is slightly preferred over a depth of 7 for the RF trees. We then test this classifier on the 604 object test set and find it performs as expected with a maximum classifier purity of 75\% and a corresponding completeness of 15\% for objects with p(SLSN-I)$ > 0.75$.

Running FLEET to classify a new transient takes in the order of $10-20$ on a personal computer, and about half the time to re-run on an existing transient once the required catalog data has been downloaded and stored locally. We note that since FLEET is designed to rapidly select the most promising SLSN-I candidates for follow-up, manual vetting of the top candidate events can further increase the sample purity. This is because some candidates might be due to obvious failure modes; for example, an AGN with a highly variable light curve might be classified as a SLSN-I due to its ``broad'' light curve, but manual inspection will reveal a variable nuclear source that is not SN-like. Another potential failure mode that can be mitigated with manual inspection, is when SDSS and/or PS1/$3\pi$ report large galaxies as multiple individual sources, causing the classifier to associate the transient to a small dim source, instead of the main galaxy. 

To summarize, our rapid classifer, using basic light curve and contextual information (and no redshift information) can achieve a factor of $30-60$ times improvement over random selection for SLSN-I, with a completeness of $\sim20\%$

\section{Alternative Classifiers}
\label{sec:alternatives}

The rapid version of the FLEET classifier presented above is tailored to find a pure sample of SLSN-I before or near peak, as to enable real-time follow-up. In this section we explore two alternative classifiers that utilize additional information: (i) using redshift as a feature, based on the expectation that LSST will provide photometric redshifts with $\sim 5\%$ uncertainty for galaxies down to $i\approx 25$ mag \citep{Graham18}; and (ii) using more complete light curve information, including the decline phase, which may hinder spectroscopic classification, but will provide samples of SLSN-I for pure photometric population studies. We optimize these alternative classifiers in terms of feature selection, depth of the classifier's trees, and time span of the light curve used in the same manner as for the main rapid classifier, described in \S\ref{sec:algorithm}.

\begin{figure}
    \begin{center}
    \centering
    {\includegraphics[width=\columnwidth]{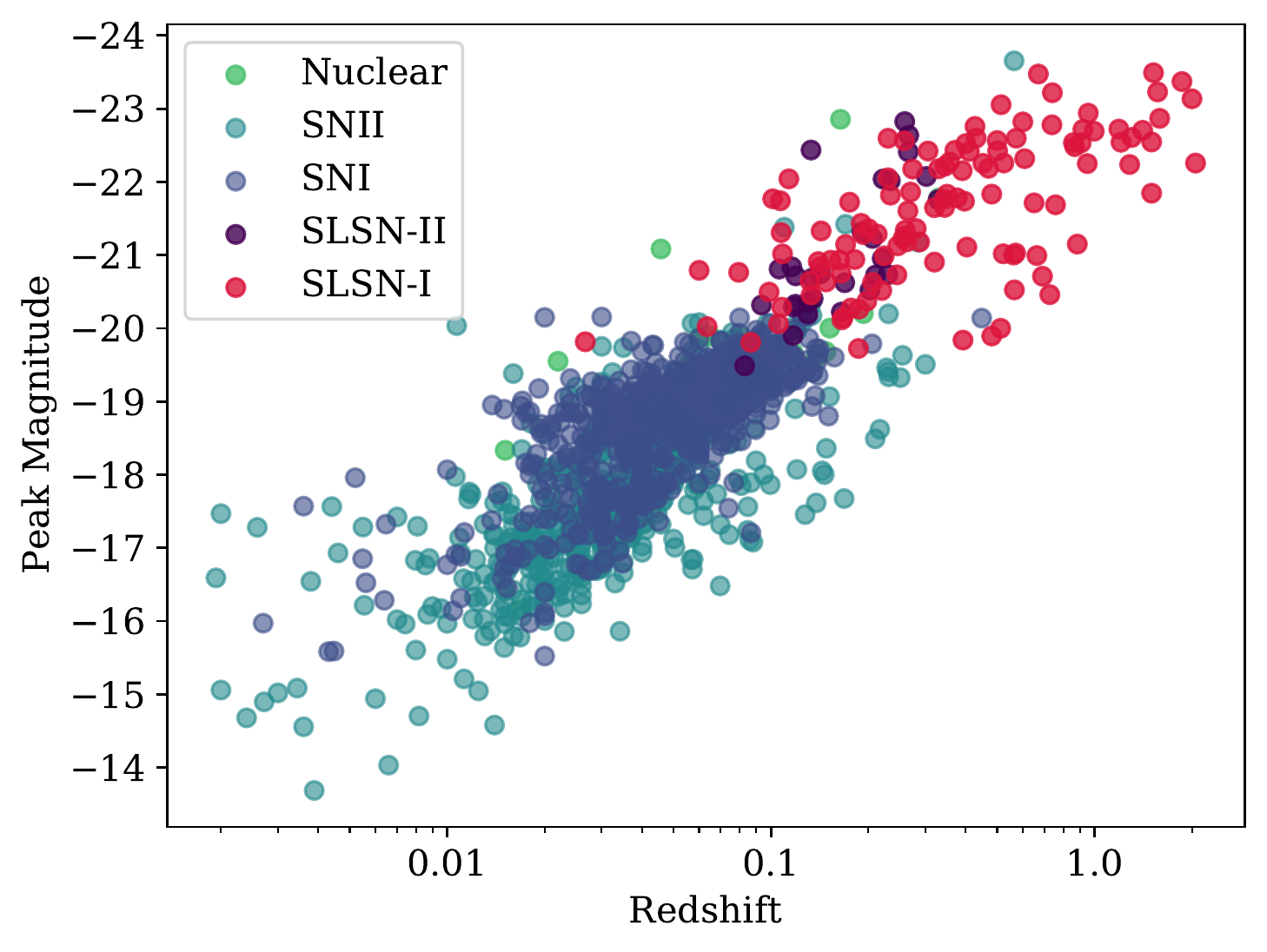}}
    \caption{Peak absolute magnitude in $r$-band versus spectroscopic redshift for all the transients in our sample (excluding stars). As expected, SLSN-I separate well from other types of transients when the redshift is known. \label{fig:redshift_class}}
    \end{center}
\end{figure}

\begin{figure*}
    \begin{center}
    \centering
    {\includegraphics[width=\columnwidth]{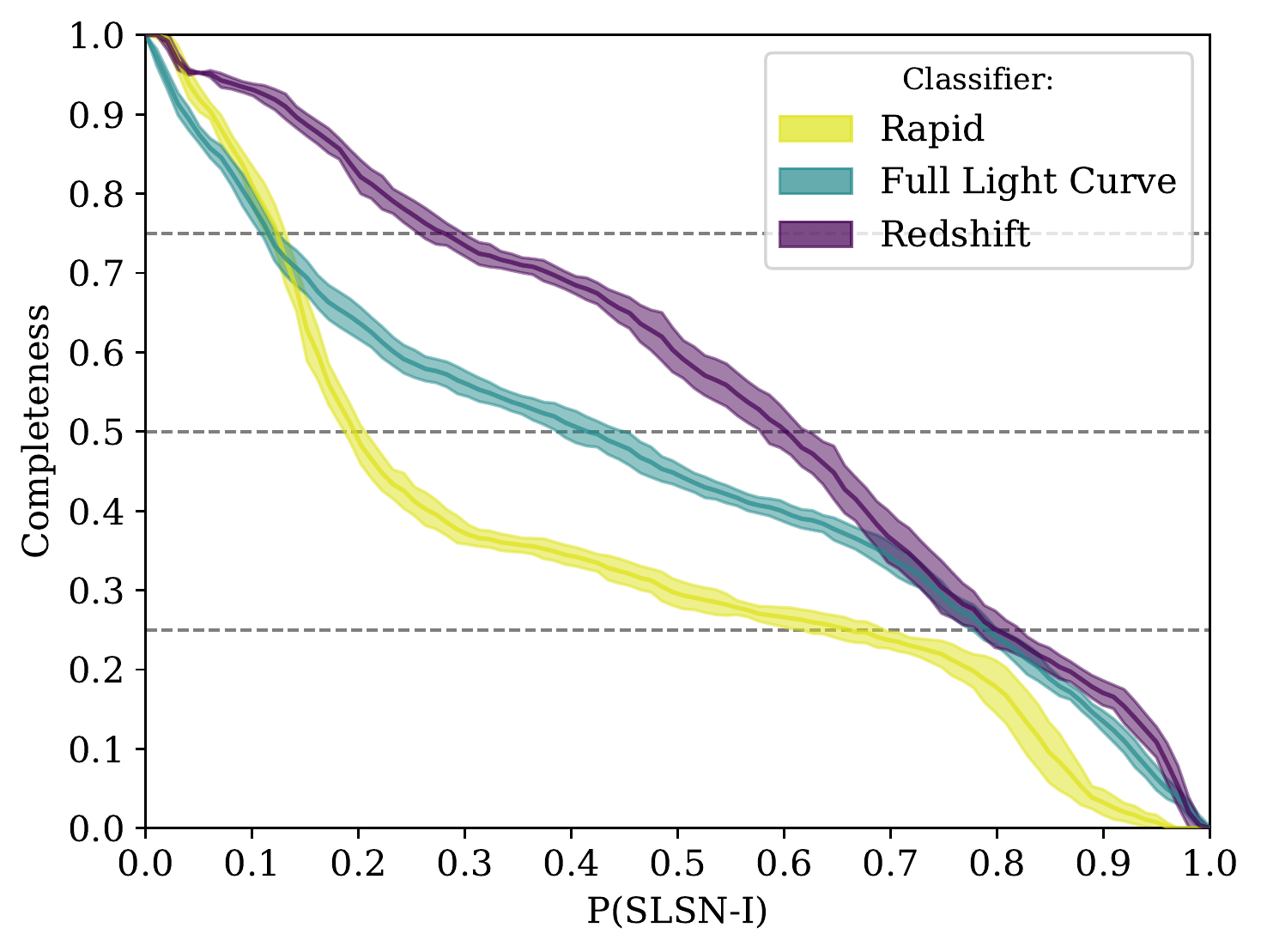}}
    {\includegraphics[width=\columnwidth]{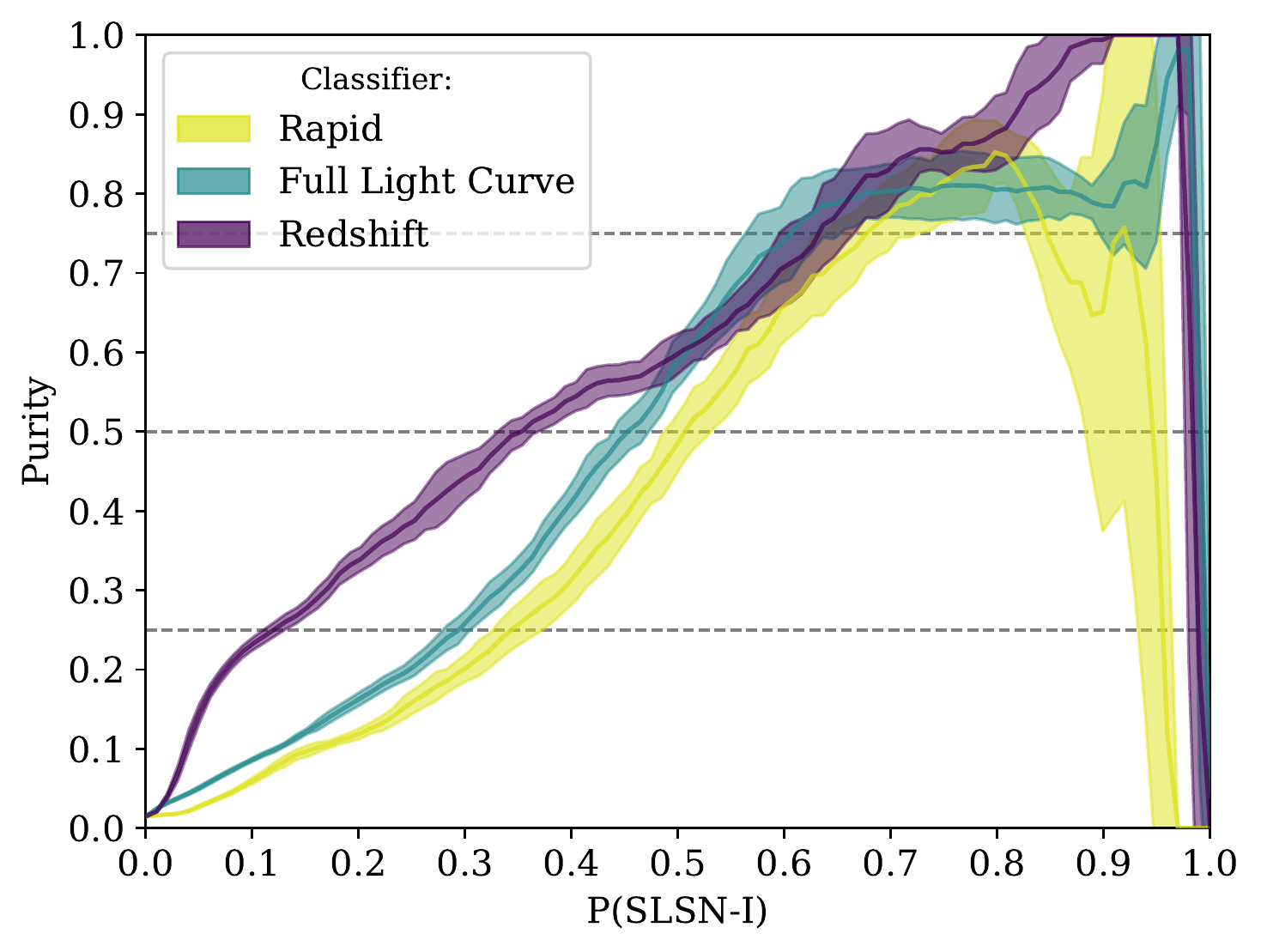}}
    \caption{\textit{Left}: Completeness as a function of confidence for all three classifiers presented here. \textit{Right}: Purity, corrected for observational rates for the same classifiers. The shaded regions represent the $1\sigma$ uncertainties. \label{fig:all_features}}
    \end{center}
\end{figure*}

\begin{figure*}
    \begin{center}
    \centering
    {\includegraphics[width=\columnwidth]{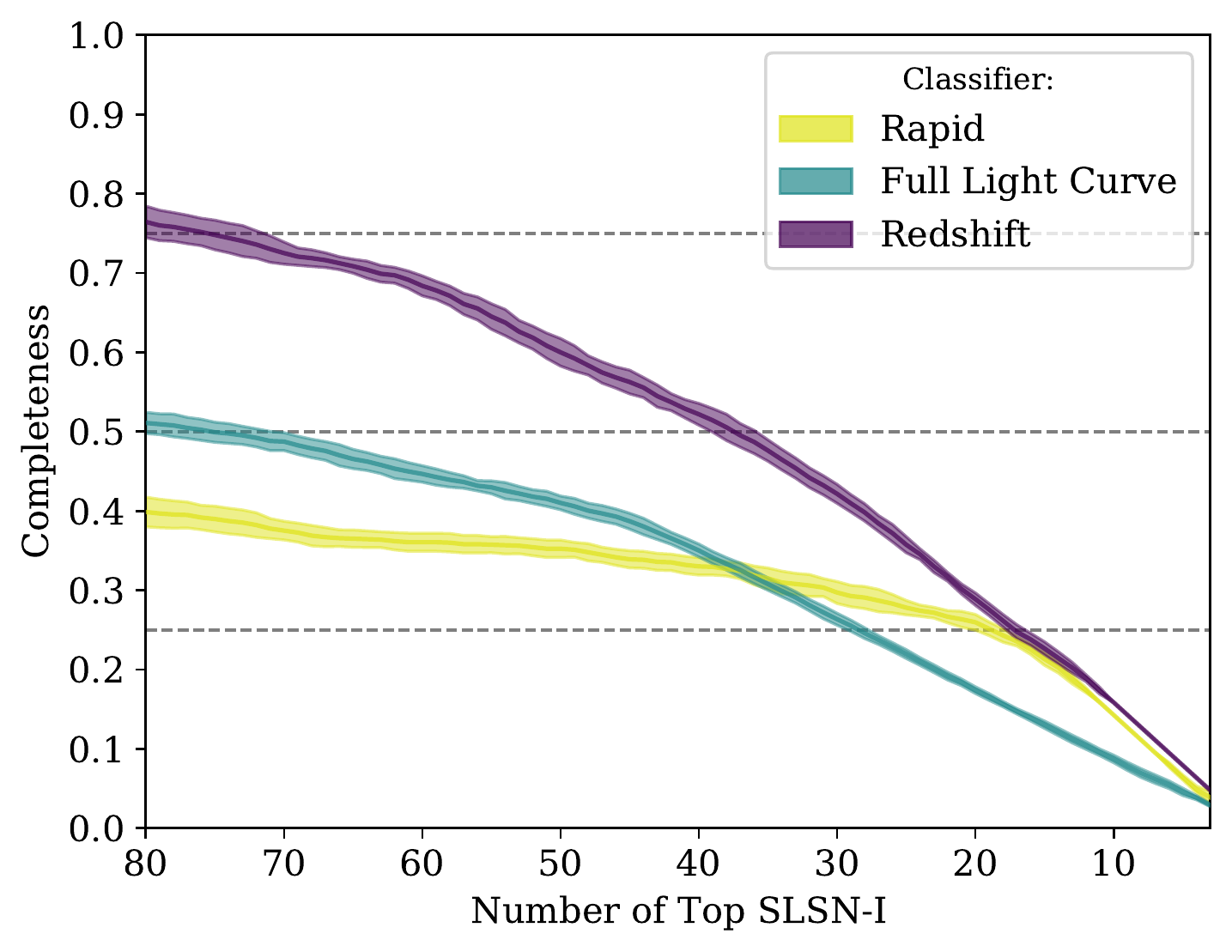}}
    {\includegraphics[width=\columnwidth]{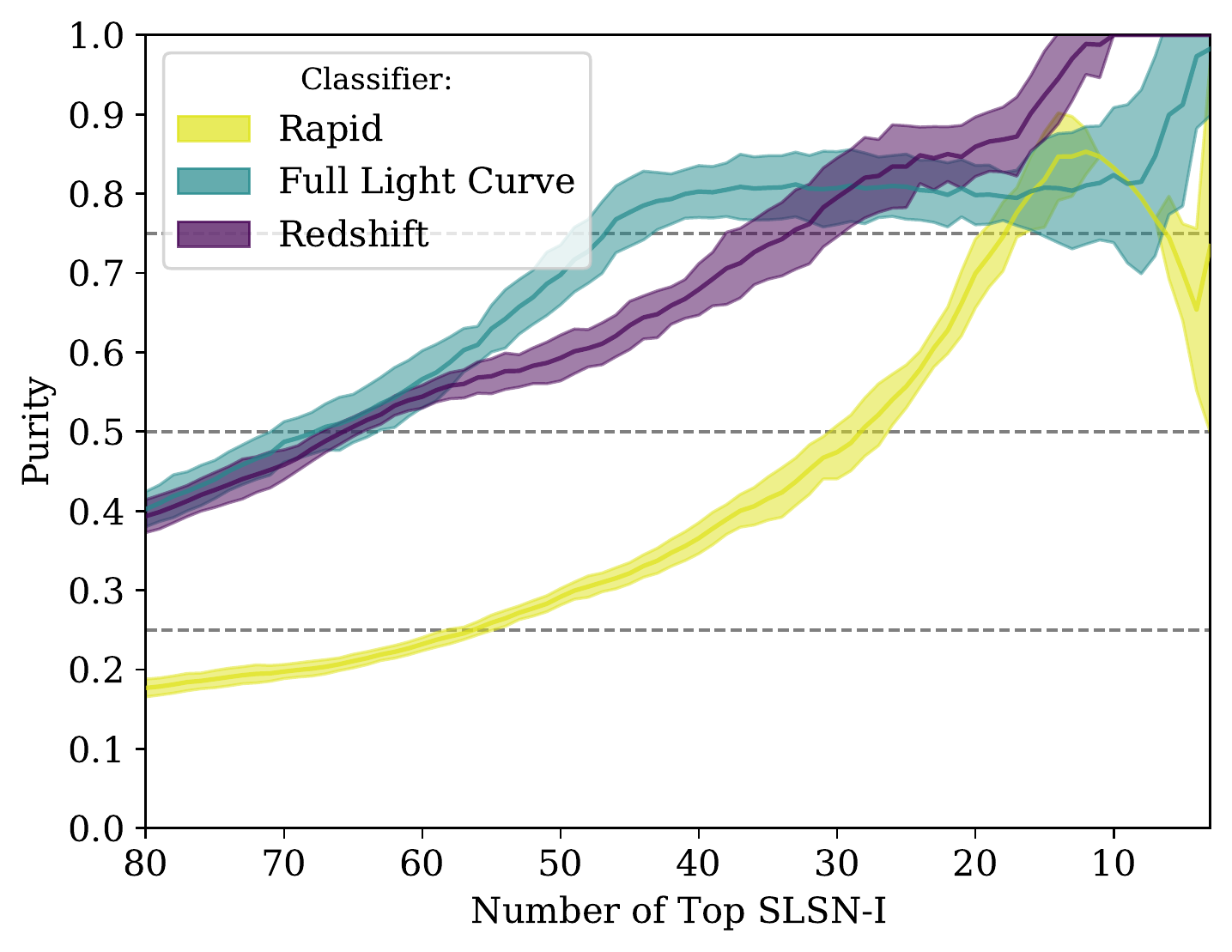}}
    \caption{\textit{Left}: Completeness as a function of the top transients classified as SLSN-I for all three classifiers presented here. \textit{Right}: Purity, corrected for observational rates for the same classifiers. The shaded regions represent the $1\sigma$ uncertainties. \label{fig:all_top}}
    \end{center}
\end{figure*}

\subsection{Redshift Classifier}

A key advantage of our rapid classifier is that it does not rely on redshift information. However, with the advent of LSST it is expected that robust photometric redshifts will be available for galaxies down to $i\approx 25$ mag. Since SLSN-I are generally more luminous than other SN classes, redshift information is certain to aid in the classification confidence. In Figure~\ref{fig:redshift_class} we plot the peak absolute $r$-band magnitude as a function of redshift for all of the extragalactic transients in our training set, indicating how well SLSN-I can be separated when redshift information is available.

To test this effect, we use here the known spectroscopic redshift of each transient in our training set (assigning Galactic transients a redshift of 0). As in the rapid classifier, we only use the first 20 days of data (designed to enable rapid follow-up) and optimize for RF depth and features. We find that feature set \#6 (Table~\ref{tab:features}) performs best, with an optimal depth of 9 for the RF trees. 

In Figure~\ref{fig:all_features} we show the observed purity and completeness of this classifier as a function of classification confidence. We find that the redshift classifier performs better than the main rapid classifier, for essentially all values of \pslsn, with an observed purity of about $60\%$ and completeness of about $60\%$ at \pslsn$>0.5$ (compared to $50\%$ and $30\%$, respectively, for the main rapid classifier). The peak purity is effectively $100\%$ with a corresponding completeness of about $15\%$ at \pslsn$>0.85$.

In Figure~\ref{fig:all_top} we show the classifier's performance in terms of the number of top SLSN-I candidates selected. The redshift classifier achieves a purity of $50$\% for the top $\approx 65$ candidate SLSN-I, significantly higher than the $\approx 27$ candidate SLSN-I at $50\%$ purity for the main classifier. Stated differently, the redshift classifier achieves $80\%$ observed purity for the 27 top candidate SLSN-I, compared to the $50\%$ observed purity for the main classifier. We therefore conclude that when robust redshift information is available it can significantly aid in the purity and completeness of the classifier.

\subsection{Full Light Curve Classifier}\label{sec:full}

The rapid classifier is trained on only the first 20 days of light curve data. Here we investigate the efficacy of using more complete light curves. This may inhibit the success of spectroscopic classification, since SLSN-I are on average about 2 mag fainter on a timescale of 70 days after discovery compared to at 20 days after discovery. But using light curves well beyond peak allows for a more robust classification and can aid in the construction of more complete photometric SLSN-I samples once they fade away. For this full light curve classifier we measure the decline rate by fitting for $A$ in Equation~\ref{eq:lc}.

After optimizing the classifier we find that feature set \#5, which includes $W$ and $A$ (Table~\ref{tab:features}), and a depth of 9 for the RF trees provide the best results. We similarly find that using the first 70 days of light curve data provides optimal results; later time data tend to be of lower quality and are more greatly affected by non-monotonic light curve features that cannot be captured in our simple light curve model. In Figure~\ref{fig:all_features} we show how the full light curve classifier performs in terms of classification probability. We find an overall better performance than for the rapid classifier, achieving a comparable peak observed purity, but at \pslsn$\approx 0.65$ instead of $\approx 0.80$, and hence with a higher completeness of about $40\%$ compared to $20\%$ for the rapid classifier. As shown in Figure~\ref{fig:all_top}, this essentially means that the full light curve classifier can achieve $50\%$ purity for a comparable number of top SLSN-I candidates as the redshift classifier, $\approx 65$ events. Similarly, it can achieve an observed purity comparable to the peak observed purity of the rapid classifier, but for about 45 top SLSN-I candidates as opposed to about 27.

\section{Conclusions}
\label{sec:conclusions}

We have presented a random forest classifier, FLEET, designed specifically to rapidly identify SLSN-I with a high purity, without the need for redshift information. We trained this classifier on a sample of about 1800 classified transients reported to the TNS, including 156 SLSN-I (i.e., 8.6\% of the total sample). The classifier uses both light curve and contextual host galaxy information. We assess the {\it observed purity} achieved by FLEET for the actual rate of SLSN-I in a magnitude-limited survey of $\approx 1.5\%$. Our key findings for the rapid FLEET classifier are as follows:
\begin{itemize}

\item We find that the most important features are the light curve width, $g-r$ color at peak, and the projected angular separation between the transient and host galaxy normalized by host radius.

\item We find an observed purity of about 50\% for events classified as SLSN-I with a probability confidence of \pslsn$>0.5$. This is a factor of 33 times improvement compared to a random selection (i.e., compared to the fraction of 1.5\% of SLSN-I in a magnitude-limited survey). The completeness for this classification confidence threshold is about 30\%.

\item We find a peak observed purity of about $85\%$ for SLSN-I, corresponding to a classification probability threshold of \pslsn$>0.80$ and a total of $\sim 15$ objects. The completeness for this classification confidence threshold is about 20\%. 

\end{itemize}

In addition to the main rapid classifier we also explored two alternative classifiers that use redshift information and full light curves, respectively. As expected, we find that these classifiers achieve better results, with a significant increase in completeness by about a factor of 2, for an observed purity that matches the peak performance of the main rapid classifier. 

Placing our results in context we note that at present, current surveys are reporting $\sim 18,000$ transients a year, out of which $\sim 6000$ transients per year have the requisite photometry (minimum of 2 data points in each of $g$- and $r$-band) and localization (within the footprint of PS1/$3\pi$) to be classified by our algorithm. For an observational SLSN-I fraction of 1.5\%, this sample contains about 90 SLSN-I per year. Our rapid classifier can therefore recover about 30 SLSN-I with a purity of $50\%$, thereby requiring about 60 follow-up spectra per year; or alternatively, about 18 SLSN-I per year with a purity of about 85\%, requiring about 21 follow-up spectra. Looking forward to LSST, which is expected to have $\sim 10^4$ SLSN-I in its data stream \citep{Villar18}, our classifier could discover $\sim 140$ SLSN-I a month, with $\sim 170$ follow-up spectra. This would increase the existing sample by two orders of magnitude over the lifetime of LSST.

\acknowledgements

The Berger Time-Domain Group is supported in part by NSF grant AST-1714498. V.A.V. acknowledges support from a Ford Foundation Dissertation Fellowship. Operation of the Pan-STARRS1 telescope is supported by the National Aeronautics and Space Administration under grant No. NNX12AR65G and grant No. NNX14AM74G issued through the NEO Observation Program. This work has made use of data from the European Space Agency (ESA) mission {\it Gaia} (\url{https://www.cosmos.esa.int/gaia}), processed by the {\it Gaia} Data Processing and Analysis Consortium (DPAC, \url{https://www.cosmos.esa.int/web/gaia/dpac/consortium}). Funding for the DPAC has been provided by national institutions, in particular the institutions participating in the {\it Gaia} Multilateral Agreement. This research has made use of NASA’s Astrophysics Data System. This research has made use of the SIMBAD database, operated at CDS, Strasbourg, France. Based on observations obtained with MegaPrime/MegaCam, a joint project of CFHT and CEA/IRFU, at the Canada-France-Hawaii Telescope (CFHT) which is operated by the National Research Council (NRC) of Canada, the Institut National des Science de l'Univers of the Centre National de la Recherche Scientifique (CNRS) of France, and the University of Hawaii. This work is based in part on data products produced at Terapix available at the Canadian Astronomy Data Centre as part of the Canada-France-Hawaii Telescope Legacy Survey, a collaborative project of NRC and CNRS. This research has made use of the NASA/IPAC Extragalactic Database, which is funded by the National Aeronautics and Space Administration and operated by the California Institute of Technology.

\facilities{ADS, TNS}
\software{Astropy \citep{astropy}, extinction(\citep{Barbary16}), Matplotlib \citep{matplotlib}, emcee\citep{foreman13}, NumPy \citep{numpy}, scikit-learn \citep{Pedregosa12}, SMOTE \citealt{Chawla02}}

\bibliography{references}

\begin{thebibliography}{}
\expandafter\ifx\csname natexlab\endcsname\relax\def\natexlab#1{#1}\fi
\providecommand{\url}[1]{\href{#1}{#1}}
\providecommand{\dodoi}[1]{doi:~\href{http://doi.org/#1}{\nolinkurl{#1}}}
\footnotesize

\bibitem[{{Ahumada} {et~al.}(2019){Ahumada}, {Allende Prieto}, {Almeida},
  {Anders}, {Anderson}, {Andrews}, {Anguiano}, {Arcodia}, {Armengaud},
  {Aubert}, {Avila}, {Avila-Reese}, {Badenes}, {Balland}, {Barger},
  {Barrera-Ballesteros}, {Basu}, {Bautista}, {Beaton}, {Beers}, {Benavides},
  {Bender}, {Bernardi}, {Bershady}, {Beutler}, {Moni Bidin}, {Bird}, {Bizyaev},
  {Blanc}, {Blanton}, {Boquien}, {Borissova}, {Bovy}, {Brandt}, {Brinkmann},
  {Brownstein}, {Bundy}, {Bureau}, {Burgasser}, {Burtin}, {Cano-Diaz},
  {Capasso}, {Cappellari}, {Carrera}, {Chabanier}, {Chaplin}, {Chapman},
  {Cherinka}, {Chiappini}, {Choi}, {Chojnowski}, {Chung}, {Clerc}, {Coffey},
  {Comerford}, {Comparat}, {da Costa}, {Cousinou}, {Covey}, {Crane}, {Cunha},
  {da Silva Ilha}, {Dai}, {Damsted}, {Darling}, {Davidson}, {Davies}, {Dawson},
  {De}, {de la Macorra}, {De Lee}, {Queiroz}, {Deconto Machado}, {de la Torre},
  {Dell'Agli}, {du Mas des Bourboux}, {Diamond-Stanic}, {Dillon}, {Donor},
  {Drory}, {Duckworth}, {Dwelly}, {Ebelke}, {Eftekharzadeh}, {Davis Eigenbrot},
  {Elsworth}, {Eracleous}, {Erfanianfar}, {Escoffier}, {Fan}, {Farr},
  {Fernandez-Trincado}, {Feuillet}, {Finoguenov}, {Fofie}, {Fraser-McKelvie},
  {Frinchaboy}, {Fromenteau}, {Fu}, {Galbany}, {Garcia}, {Garcia-Hernandez},
  {Garma Oehmichen}, {Ge}, {Geimba Maia}, {Geisler}, {Gelfand }, {Goddy}, {Le
  Goff}, {Gonzalez-Perez}, {Grabowski}, {Green}, {Grier}, {Guo}, {Guy},
  {Harding}, {Hasselquist}, {Hawken}, {Hayes}, {Hearty}, {Hekker}, {Hogg},
  {Holtzman}, {Horta}, {Hou}, {Hsieh}, {Huber}, {Hunt}, {Ider Chitham}, {Imig},
  {Jaber}, {Jimenez Angel}, {Johnson}, {Jones}, {Jonsson}, {Jullo}, {Kim},
  {Kinemuchi}, {Kirkpatrick}, {Kite}, {Klaene}, {Kneib}, {Kollmeier}, {Kong},
  {Kounkel}, {Krishnarao}, {Lacerna}, {Lan}, {Lane}, {Law}, {Leung}, {Lewis},
  {Li}, {Lian}, {Lin}, {Long}, {Longa-Pena}, {Lundgren}, {Lyke}, {Mackereth},
  {MacLeod}, {Majewski}, {Manchado}, {Maraston}, {Martini}, {Masseron},
  {Masters}, {Mathur}, {McDermid}, {Merloni}, {Merrifield}, {Meszaros},
  {Miglio}, {Minniti}, {Minsley}, {Miyaji}, {Gohar Mohammad}, {Mosser},
  {Mueller}, {Muna}, {Munoz-Gutierrez}, {Myers}, {Nadathur}, {Nair}, {Nandra},
  {Correa do Nascimento}, {Nevin}, {Newman}, {Nidever}, {Nitschelm},
  {Noterdaeme}, {O'Connell}, {Olmstead}, {Oravetz}, {Oravetz}, {Osorio},
  {Pace}, {Padilla}, {Palanque-Delabrouille}, {Palicio}, {Pan}, {Pan},
  {Parker}, {Paviot}, {Peirani}, {Pena Ramrez}, {Penny}, {Percival},
  {Perez-Fournon}, {Perez-Rafols}, {Petitjean}, {Pieri}, {Pinsonneault},
  {Poovelil}, {Povick}, {Prakash}, {Price-Whelan}, {Raddick}, {Raichoor},
  {Ray}, {Barboza Rembold}, {Rezaie}, {Riffel}, {Riffel}, {Rix}, {Robin},
  {Roman-Lopes}, {Roman-Zuniga}, {Rose}, {Ross}, {Rossi}, {Rowlands}, {Rubin},
  {Salvato}, {Sanchez}, {Sanchez-Menguiano}, {Sanchez-Gallego}, {Sayres},
  {Schaefer}, {Schiavon}, {Schimoia}, {Schlafly}, {Schlegel}, {Schneider},
  {Schultheis}, {Schwope}, {Seo}, {Serenelli}, {Shafieloo}, {Shamsi}, {Shao},
  {Shen}, {Shetrone}, {Shirley}, {Silva Aguirre}, {Simon}, {Skrutskie},
  {Slosar}, {Smethurst}, {Sobeck}, {Cervantes Sodi}, {Souto}, {Stark},
  {Stassun}, {Steinmetz}, {Stello}, {Stermer}, {Storchi-Bergmann},
  {Streblyanska}, {Stringfellow}, {Stutz}, {Suarez}, {Sun}, {Taghizadeh-Popp},
  {Talbot}, {Tayar}, {Thakar}, {Theriault}, {Thomas}, {Thomas}, {Tinker},
  {Tojeiro}, {Hernandez Toledo}, {Tremonti}, {Troup}, {Tuttle}, {Unda-Sanzana},
  {Valentini}, {Vargas-Gonzalez}, {Vargas-Magana}, {Vazquez-Mata}, {Vivek},
  {Wake}, {Wang}, {Weaver}, {Weijmans}, {Wild}, {Wilson}, {Wilson}, {Wolthuis},
  {Wood-Vasey}, {Yan}, {Yang}, {Yeche}, {Zamora}, {Zarrouk}, {Zasowski},
  {Zhang}, {Zhao}, {Zhao}, {Zheng}, {Zheng}, {Zhu}, \& {Zou}}]{Ahumada19}
{Ahumada}, R., {Allende Prieto}, C., {Almeida}, A., {et~al.} 2019, arXiv
  e-prints,
  \hypersetup{urlcolor=magenta}\href{https://arxiv.org/abs/1912.02905}{arXiv}{:}\hypersetup{urlcolor=blue}\href{https://ui.adsabs.harvard.edu/abs/2019arXiv191202905A}{1912.02905}

\bibitem[{{Alam} {et~al.}(2015){Alam}, {Albareti}, {Allende Prieto}, {Anders},
  {Anderson}, {Anderton}, {Andrews}, {Armengaud}, {Aubourg}, {Bailey}, {Basu},
  {Bautista}, {Beaton}, {Beers}, {Bender}, {Berlind}, {Beutler}, {Bhardwaj},
  {Bird}, {Bizyaev}, {Blake}, {Blanton}, {Blomqvist}, {Bochanski}, {Bolton},
  {Bovy}, {Shelden Bradley}, {Brandt}, {Brauer}, {Brinkmann}, {Brown},
  {Brownstein}, {Burden}, {Burtin}, {Busca}, {Cai}, {Capozzi}, {Carnero
  Rosell}, {Carr}, {Carrera}, {Chambers}, {Chaplin}, {Chen}, {Chiappini},
  {Chojnowski}, {Chuang}, {Clerc}, {Comparat}, {Covey}, {Croft}, {Cuesta},
  {Cunha}, {da Costa}, {Da Rio}, {Davenport}, {Dawson}, {De Lee}, {Delubac},
  {Deshpande}, {Dhital}, {Dutra-Ferreira}, {Dwelly}, {Ealet}, {Ebelke},
  {Edmondson}, {Eisenstein}, {Ellsworth}, {Elsworth}, {Epstein}, {Eracleous},
  {Escoffier}, {Esposito}, {Evans}, {Fan}, {Fern{\'a}ndez-Alvar}, {Feuillet},
  {Filiz Ak}, {Finley}, {Finoguenov}, {Flaherty}, {Fleming}, {Font-Ribera},
  {Foster}, {Frinchaboy}, {Galbraith-Frew}, {Garc{\'\i}a},
  {Garc{\'\i}a-Hern{\'a}ndez}, {Garc{\'\i}a P{\'e}rez}, {Gaulme}, {Ge},
  {G{\'e}nova-Santos}, {Georgakakis}, {Ghezzi}, {Gillespie}, {Girardi},
  {Goddard}, {Gontcho}, {Gonz{\'a}lez Hern{\'a}ndez}, {Grebel}, {Green},
  {Grieb}, {Grieves}, {Gunn}, {Guo}, {Harding}, {Hasselquist}, {Hawley},
  {Hayden}, {Hearty}, {Hekker}, {Ho}, {Hogg}, {Holley-Bockelmann}, {Holtzman},
  {Honscheid}, {Huber}, {Huehnerhoff}, {Ivans}, {Jiang}, {Johnson},
  {Kinemuchi}, {Kirkby}, {Kitaura}, {Klaene}, {Knapp}, {Kneib}, {Koenig},
  {Lam}, {Lan}, {Lang}, {Laurent}, {Le Goff}, {Leauthaud}, {Lee}, {Lee},
  {Licquia}, {Liu}, {Long}, {L{\'o}pez-Corredoira}, {Lorenzo-Oliveira},
  {Lucatello}, {Lundgren}, {Lupton}, {Mack}, {Mahadevan}, {Maia}, {Majewski},
  {Malanushenko}, {Malanushenko}, {Manchado}, {Manera}, {Mao}, {Maraston},
  {Marchwinski}, {Margala}, {Martell}, {Martig}, {Masters}, {Mathur},
  {McBride}, {McGehee}, {McGreer}, {McMahon}, {M{\'e}nard}, {Menzel},
  {Merloni}, {M{\'e}sz{\'a}ros}, {Miller}, {Miralda-Escud{\'e}}, {Miyatake},
  {Montero-Dorta}, {More}, {Morganson}, {Morice-Atkinson}, {Morrison},
  {Mosser}, {Muna}, {Myers}, {Nand ra}, {Newman}, {Neyrinck}, {Nguyen},
  {Nichol}, {Nidever}, {Noterdaeme}, {Nuza}, {O'Connell}, {O'Connell},
  {O'Connell}, {Ogando}, {Olmstead}, {Oravetz}, {Oravetz}, {Osumi}, {Owen},
  {Padgett}, {Padmanabhan}, {Paegert}, {Palanque-Delabrouille}, {Pan},
  {Parejko}, {P{\^a}ris}, {Park}, {Pattarakijwanich}, {Pellejero-Ibanez},
  {Pepper}, {Percival}, {P{\'e}rez-Fournon}, {Ṕrez-Ra`fols}, {Petitjean},
  {Pieri}, {Pinsonneault}, {Porto de Mello}, {Prada}, {Prakash},
  {Price-Whelan}, {Protopapas}, {Raddick}, {Rahman}, {Reid}, {Rich}, {Rix},
  {Robin}, {Rockosi}, {Rodrigues}, {Rodr{\'\i}guez-Torres}, {Roe}, {Ross},
  {Ross}, {Rossi}, {Ruan}, {Rubi{\~n}o-Mart{\'\i}n}, {Rykoff},
  {Salazar-Albornoz}, {Salvato}, {Samushia}, {S{\'a}nchez}, {Santiago},
  {Sayres}, {Schiavon}, {Schlegel}, {Schmidt}, {Schneider}, {Schultheis},
  {Schwope}, {Sc{\'o}ccola}, {Scott}, {Sellgren}, {Seo}, {Serenelli}, {Shane},
  {Shen}, {Shetrone}, {Shu}, {Silva Aguirre}, {Sivarani}, {Skrutskie},
  {Slosar}, {Smith}, {Sobreira}, {Souto}, {Stassun}, {Steinmetz}, {Stello},
  {Strauss}, {Streblyanska}, {Suzuki}, {Swanson}, {Tan}, {Tayar}, {Terrien},
  {Thakar}, {Thomas}, {Thomas}, {Thompson}, {Tinker}, {Tojeiro}, {Troup},
  {Vargas-Maga{\~n}a}, {Vazquez}, {Verde}, {Viel}, {Vogt}, {Wake}, {Wang},
  {Weaver}, {Weinberg}, {Weiner}, {White}, {Wilson}, {Wisniewski},
  {Wood-Vasey}, {Ye`che}, {York}, {Zakamska}, {Zamora}, {Zasowski}, {Zehavi},
  {Zhao}, {Zheng}, {Zhou}, {Zhou}, {Zou}, \& {Zhu}}]{Alam15}
{Alam}, S., {Albareti}, F.~D., {Allende Prieto}, C., {et~al.} 2015,
  \hypersetup{urlcolor=magenta}\href{https://dx.doi.org/10.1088/0067-0049/219/1/12}{\apjs},
  \hypersetup{urlcolor=blue}\href{https://ui.adsabs.harvard.edu/abs/2015ApJS..219...12A}{219,
  12}

\bibitem[{{Angus} {et~al.}(2019){Angus}, {Smith}, {Sullivan}, {Inserra},
  {Wiseman}, {D'Andrea}, {Thomas}, {Nichol}, {Galbany}, {Childress}, {Asorey},
  {Brown}, {Casas}, {Castander}, {Curtin}, {Frohmaier}, {Glazebrook}, {Gruen},
  {Gutierrez}, {Kessler}, {Kim}, {Lidman}, {Macaulay}, {Nugent}, {Pursiainen},
  {Sako}, {Soares-Santos}, {Thomas}, {Abbott}, {Avila}, {Bertin}, {Brooks},
  {Buckley-Geer}, {Burke}, {Carnero Rosell}, {Carretero}, {da Costa}, {De
  Vicente}, {Desai}, {Diehl}, {Doel}, {Eifler}, {Flaugher}, {Fosalba},
  {Frieman}, {Garc{\'\i}a-Bellido}, {Gruendl}, {Gschwend}, {Hartley},
  {Hollowood}, {Honscheid}, {Hoyle}, {James}, {Kuehn}, {Kuropatkin}, {Lahav},
  {Lima}, {Maia}, {March}, {Marshall}, {Menanteau}, {Miller}, {Miquel}, {Ogand
  o}, {Plazas}, {Romer}, {Sanchez}, {Schindler}, {Schubnell}, {Sobreira},
  {Suchyta}, {Swanson}, {Tarle}, {Thomas}, {Tucker}, \& {DES
  Collaboration}}]{Angus19}
{Angus}, C.~R., {Smith}, M., {Sullivan}, M., {et~al.} 2019,
  \hypersetup{urlcolor=magenta}\href{https://dx.doi.org/10.1093/mnras/stz1321}{\mnras},
  \hypersetup{urlcolor=blue}\href{https://ui.adsabs.harvard.edu/abs/2019MNRAS.487.2215A}{487,
  2215}

\bibitem[{{Astropy Collaboration}(2018)}]{astropy}
{Astropy Collaboration}. 2018,
  \hypersetup{urlcolor=magenta}\href{https://dx.doi.org/10.3847/1538-3881/aabc4f}{\aj},
  \hypersetup{urlcolor=blue}\href{https://ui.adsabs.harvard.edu/abs/2018AJ....156..123A}{156,
  123}

\bibitem[{Barbary(2016)}]{Barbary16}
Barbary, K. 2016, extinction, v0.3.0,  Zenodo,
  \hypersetup{urlcolor=magenta}doi:\href{https://dx.doi.org/10.5281/zenodo.804967}{10.5281/zenodo.804967}

\bibitem[{{Bellm} {et~al.}(2019){Bellm}, {Kulkarni}, {Graham}, {Dekany},
  {Smith}, {Riddle}, {Masci}, {Helou}, {Prince}, {Adams}, {Barbarino},
  {Barlow}, {Bauer}, {Beck}, {Belicki}, {Biswas}, {Blagorodnova}, {Bodewits},
  {Bolin}, {Brinnel}, {Brooke}, {Bue}, {Bulla}, {Burruss}, {Cenko}, {Chang},
  {Connolly}, {Coughlin}, {Cromer}, {Cunningham}, {De}, {Delacroix}, {Desai},
  {Duev}, {Eadie}, {Farnham}, {Feeney}, {Feindt}, {Flynn}, {Franckowiak},
  {Frederick}, {Fremling}, {Gal-Yam}, {Gezari}, {Giomi}, {Goldstein},
  {Golkhou}, {Goobar}, {Groom}, {Hacopians}, {Hale}, {Henning}, {Ho}, {Hover},
  {Howell}, {Hung}, {Huppenkothen}, {Imel}, {Ip}, {Ivezi{\'c}}, {Jackson},
  {Jones}, {Juric}, {Kasliwal}, {Kaspi}, {Kaye}, {Kelley}, {Kowalski},
  {Kramer}, {Kupfer}, {Landry}, {Laher}, {Lee}, {Lin}, {Lin}, {Lunnan},
  {Giomi}, {Mahabal}, {Mao}, {Miller}, {Monkewitz}, {Murphy}, {Ngeow},
  {Nordin}, {Nugent}, {Ofek}, {Patterson}, {Penprase}, {Porter}, {Rauch},
  {Rebbapragada}, {Reiley}, {Rigault}, {Rodriguez}, {van Roestel}, {Rusholme},
  {van Santen}, {Schulze}, {Shupe}, {Singer}, {Soumagnac}, {Stein}, {Surace},
  {Sollerman}, {Szkody}, {Taddia}, {Terek}, {Van Sistine}, {van Velzen},
  {Vestrand}, {Walters}, {Ward}, {Ye}, {Yu}, {Yan}, \& {Zolkower}}]{Bellm19}
{Bellm}, E.~C., {Kulkarni}, S.~R., {Graham}, M.~J., {et~al.} 2019,
  \hypersetup{urlcolor=magenta}\href{https://dx.doi.org/10.1088/1538-3873/aaecbe}{\pasp},
  \hypersetup{urlcolor=blue}\href{https://ui.adsabs.harvard.edu/abs/2019PASP..131a8002B}{131,
  018002}

\bibitem[{{Berger}(2010)}]{Berger10}
{Berger}, E. 2010,
  \hypersetup{urlcolor=magenta}\href{https://dx.doi.org/10.1088/0004-637X/722/2/1946}{\apj},
  \hypersetup{urlcolor=blue}\href{https://ui.adsabs.harvard.edu/abs/2010ApJ...722.1946B}{722,
  1946}

\bibitem[{{Bertin} \& {Arnouts}(1996)}]{Bertin96}
{Bertin}, E., \& {Arnouts}, S. 1996,
  \hypersetup{urlcolor=magenta}\href{https://dx.doi.org/10.1051/aas:1996164}{\aaps},
  \hypersetup{urlcolor=blue}\href{https://ui.adsabs.harvard.edu/abs/1996A&AS..117..393B}{117,
  393}

\bibitem[{{Blanchard}(2019)}]{Blanchard_thesis}
{Blanchard}, P.~K. 2019, PhD thesis, Harvard University,
  \hypersetup{urlcolor=magenta}\url{http://nrs.harvard.edu/urn-3:HUL.InstRepos:42029690}

\bibitem[{{Blanchard} {et~al.}(2020){Blanchard}, {Berger}, {Nicholl}, \&
  {Villar}}]{Blanchard20}
{Blanchard}, P.~K., {Berger}, E., {Nicholl}, M., \& {Villar}, V.~A. 2020,
  \hypersetup{urlcolor=magenta}\href{https://dx.doi.org/10.3847/1538-4357/ab9638}{\apj},
  \hypersetup{urlcolor=blue}\href{https://ui.adsabs.harvard.edu/abs/2020ApJ...897..114B}{897,
  114}

\bibitem[{{Blanchard} {et~al.}(2019){Blanchard}, {Nicholl}, {Berger},
  {Chornock}, {Milisavljevic}, {Margutti}, \& {Gomez}}]{Blanchard19}
{Blanchard}, P.~K., {Nicholl}, M., {Berger}, E., {et~al.} 2019,
  \hypersetup{urlcolor=magenta}\href{https://dx.doi.org/10.3847/1538-4357/aafa13}{\apj},
  \hypersetup{urlcolor=blue}\href{https://ui.adsabs.harvard.edu/abs/2019ApJ...872...90B}{872,
  90}

\bibitem[{{Blanchard} {et~al.}(2017){Blanchard}, {Nicholl}, {Berger},
  {Guillochon}, {Margutti}, {Chornock}, {Alexand er}, {Leja}, \&
  {Drout}}]{Blanchard17}
{Blanchard}, P.~K., {Nicholl}, M., {Berger}, E., {et~al.} 2017,
  \hypersetup{urlcolor=magenta}\href{https://dx.doi.org/10.3847/1538-4357/aa77f7}{\apj},
  \hypersetup{urlcolor=blue}\href{https://ui.adsabs.harvard.edu/abs/2017ApJ...843..106B}{843,
  106}

\bibitem[{{Blanchard} {et~al.}(2018){Blanchard}, {Nicholl}, {Berger},
  {Chornock}, {Margutti}, {Milisavljevic}, {Fong}, {MacLeod}, \&
  {Bhirombhakdi}}]{Blanchard18}
{Blanchard}, P.~K., {Nicholl}, M., {Berger}, E., {et~al.} 2018,
  \hypersetup{urlcolor=magenta}\href{https://dx.doi.org/10.3847/1538-4357/aad8b9}{\apj},
  \hypersetup{urlcolor=blue}\href{https://ui.adsabs.harvard.edu/abs/2018ApJ...865....9B}{865,
  9}

\bibitem[{{Bloom} {et~al.}(2002){Bloom}, {Kulkarni}, \& {Djorgovski}}]{Bloom02}
{Bloom}, J.~S., {Kulkarni}, S.~R., \& {Djorgovski}, S.~G. 2002,
  \hypersetup{urlcolor=magenta}\href{https://dx.doi.org/10.1086/338893}{\aj},
  \hypersetup{urlcolor=blue}\href{https://ui.adsabs.harvard.edu/abs/2002AJ....123.1111B}{123,
  1111}

\bibitem[{{Boone}(2019)}]{Boone19}
{Boone}, K. 2019,
  \hypersetup{urlcolor=magenta}\href{https://dx.doi.org/10.3847/1538-3881/ab5182}{\aj},
  \hypersetup{urlcolor=blue}\href{https://ui.adsabs.harvard.edu/abs/2019AJ....158..257B}{158,
  257}

\bibitem[{{Breiman}(2001)}]{Breiman01}
{Breiman}, L. 2001,
  \hypersetup{urlcolor=magenta}\href{https://dx.doi.org/10.1023/A:1010933404324}{Machine
  Learning},
  \hypersetup{urlcolor=blue}\href{https://dx.doi.org/10.1023/A:1010933404324}{45,
  5}

\bibitem[{{Chambers} \& {Pan-STARRS Team}(2018)}]{Chambers18}
{Chambers}, K., \& {Pan-STARRS Team}. 2018, in American Astronomical Society
  Meeting Abstracts, Vol. 231, American Astronomical Society Meeting Abstracts
  \#231, 102.01

\bibitem[{{Chawla} {et~al.}(2002){Chawla}, {Bowyer}, {Hall}, \&
  {Kegelmeyer}}]{Chawla02}
{Chawla}, N.~V., {Bowyer}, K.~W., {Hall}, L.~O., \& {Kegelmeyer}, W.~P. 2002,
  Journal of artificial intelligence research,
  \hypersetup{urlcolor=blue}\href{https://www.jair.org/index.php/jair/article/view/10302}{16,
  321}

\bibitem[{{Chen}(2019)}]{Chen19}
{Chen}, T. 2019, Transient Name Server Classification Report,
  \hypersetup{urlcolor=blue}\href{https://ui.adsabs.harvard.edu/abs/2019TNSCR.938....1C}{2019-938,
  1}

\bibitem[{{Chen} {et~al.}(2017){Chen}, {Nicholl}, {Smartt}, {Mazzali}, {Yates},
  {Moriya}, {Inserra}, {Langer}, {Kr{\"u}hler}, {Pan}, {Kotak}, {Galbany},
  {Schady}, {Wiseman}, {Greiner}, {Schulze}, {Man}, {Jerkstrand }, {Smith},
  {Dennefeld}, {Baltay}, {Bolmer}, {Kankare}, {Knust}, {Maguire}, {Rabinowitz},
  {Rostami}, {Sullivan}, \& {Young}}]{Chen17}
{Chen}, T.~W., {Nicholl}, M., {Smartt}, S.~J., {et~al.} 2017,
  \hypersetup{urlcolor=magenta}\href{https://dx.doi.org/10.1051/0004-6361/201630163}{\aap},
  \hypersetup{urlcolor=blue}\href{https://ui.adsabs.harvard.edu/abs/2017A&A...602A...9C}{602,
  A9}

\bibitem[{{Chen} {et~al.}(2018){Chen}, {Inserra}, {Fraser}, {Moriya}, {Schady},
  {Schweyer}, {Filippenko}, {Perley}, {Ruiter}, {Seitenzahl}, {Sollerman},
  {Taddia}, {Anderson}, {Foley}, {Jerkstrand}, {Ngeow}, {Pan}, {Pastorello},
  {Points}, {Smartt}, {Smith}, {Taubenberger}, {Wiseman}, {Young}, {Benetti},
  {Berton}, {Bufano}, {Clark}, {Della Valle}, {Galbany}, {Gal-Yam},
  {Gromadzki}, {Guti{\'e}rrez}, {Heinze}, {Kankare}, {Kilpatrick},
  {Kuncarayakti}, {Leloudas}, {Lin}, {Maguire}, {Mazzali}, {McBrien},
  {Prentice}, {Rau}, {Rest}, {Siebert}, {Stalder}, {Tonry}, \& {Yu}}]{Chen18}
{Chen}, T.~W., {Inserra}, C., {Fraser}, M., {et~al.} 2018,
  \hypersetup{urlcolor=magenta}\href{https://dx.doi.org/10.3847/2041-8213/aaeb2e}{\apjl},
  \hypersetup{urlcolor=blue}\href{https://ui.adsabs.harvard.edu/abs/2018ApJ...867L..31C}{867,
  L31}

\bibitem[{{Chomiuk} {et~al.}(2011){Chomiuk}, {Chornock}, {Soderberg}, {Berger},
  {Chevalier}, {Foley}, {Huber}, {Narayan}, {Rest}, {Gezari}, {Kirshner},
  {Riess}, {Rodney}, {Smartt}, {Stubbs}, {Tonry}, {Wood-Vasey}, {Burgett},
  {Chambers}, {Czekala}, {Flewelling}, {Forster}, {Kaiser}, {Kudritzki},
  {Magnier}, {Martin}, {Morgan}, {Neill}, {Price}, {Roth}, {Sanders}, \&
  {Wainscoat}}]{Chomiuk11}
{Chomiuk}, L., {Chornock}, R., {Soderberg}, A.~M., {et~al.} 2011,
  \hypersetup{urlcolor=magenta}\href{https://dx.doi.org/10.1088/0004-637X/743/2/114}{\apj},
  \hypersetup{urlcolor=blue}\href{https://ui.adsabs.harvard.edu/abs/2011ApJ...743..114C}{743,
  114}

\bibitem[{{Cooke} {et~al.}(2012){Cooke}, {Sullivan}, {Gal-Yam}, {Barton},
  {Carlberg}, {Ryan-Weber}, {Horst}, {Omori}, \& {D{\'\i}az}}]{Cooke12}
{Cooke}, J., {Sullivan}, M., {Gal-Yam}, A., {et~al.} 2012,
  \hypersetup{urlcolor=magenta}\href{https://dx.doi.org/10.1038/nature11521}{\nat},
  \hypersetup{urlcolor=blue}\href{https://ui.adsabs.harvard.edu/abs/2012Natur.491..228C}{491,
  228}

\bibitem[{{Dahiwale} \& {Fremling}(2020)}]{Dahiwale20}
{Dahiwale}, A., \& {Fremling}, C. 2020, Transient Name Server Classification
  Report,
  \hypersetup{urlcolor=blue}\href{https://ui.adsabs.harvard.edu/abs/2020TNSCR1756....1D}{2020-1756,
  1}

\bibitem[{{De Cia} {et~al.}(2018){De Cia}, {Gal-Yam}, {Rubin}, {Leloudas},
  {Vreeswijk}, {Perley}, {Quimby}, {Yan}, {Sullivan}, {Fl{\"o}rs}, {Sollerman},
  {Bersier}, {Cenko}, {Gal-Yam}, {Maguire}, {Ofek}, {Prentice}, {Schulze},
  {Spyromilio}, {Valenti}, {Arcavi}, {Corsi}, {Howell}, {Mazzali}, {Kasliwal},
  {Taddia}, \& {Yaron}}]{Cia18}
{De Cia}, A., {Gal-Yam}, A., {Rubin}, A., {et~al.} 2018,
  \hypersetup{urlcolor=magenta}\href{https://dx.doi.org/10.3847/1538-4357/aab9b6}{\apj},
  \hypersetup{urlcolor=blue}\href{https://ui.adsabs.harvard.edu/abs/2018ApJ...860..100D}{860,
  100}

\bibitem[{{Dessart} {et~al.}(2012){Dessart}, {Hillier}, {Waldman}, {Livne}, \&
  {Blondin}}]{Dessart12}
{Dessart}, L., {Hillier}, D.~J., {Waldman}, R., {Livne}, E., \& {Blondin}, S.
  2012,
  \hypersetup{urlcolor=magenta}\href{https://dx.doi.org/10.1111/j.1745-3933.2012.01329.x}{\mnras},
  \hypersetup{urlcolor=blue}\href{https://ui.adsabs.harvard.edu/abs/2012MNRAS.426L..76D}{426,
  L76}

\bibitem[{{Drake} {et~al.}(2009){Drake}, {Djorgovski}, {Mahabal}, {Beshore},
  {Larson}, {Graham}, {Williams}, {Christensen}, {Catelan}, {Boattini},
  {Gibbs}, {Hill}, \& {Kowalski}}]{Drake09}
{Drake}, A.~J., {Djorgovski}, S.~G., {Mahabal}, A., {et~al.} 2009,
  \hypersetup{urlcolor=magenta}\href{https://dx.doi.org/10.1088/0004-637X/696/1/870}{\apj},
  \hypersetup{urlcolor=blue}\href{https://ui.adsabs.harvard.edu/abs/2009ApJ...696..870D}{696,
  870}

\bibitem[{{Foreman-Mackey} {et~al.}(2013){Foreman-Mackey}, {Hogg}, {Lang}, \&
  {Goodman}}]{foreman13}
{Foreman-Mackey}, D., {Hogg}, D.~W., {Lang}, D., \& {Goodman}, J. 2013,
  \hypersetup{urlcolor=magenta}\href{https://dx.doi.org/10.1086/670067}{\pasp},
  \hypersetup{urlcolor=blue}\href{https://ui.adsabs.harvard.edu/abs/2013PASP..125..306F}{125,
  306}

\bibitem[{{Fraser} {et~al.}(2016){Fraser}, {Reynolds}, {Mattila}, \&
  {Yaron}}]{Fraser16}
{Fraser}, M., {Reynolds}, T., {Mattila}, S., \& {Yaron}, O. 2016, Transient
  Name Server Classification Report,
  \hypersetup{urlcolor=blue}\href{https://ui.adsabs.harvard.edu/abs/2016TNSCR.521....1F}{2016-521,
  1}

\bibitem[{{Fremling} \& {Dahiwale}(2019)}]{Fremling19_2019lsq}
{Fremling}, C., \& {Dahiwale}, A. 2019, Transient Name Server Classification
  Report,
  \hypersetup{urlcolor=blue}\href{https://ui.adsabs.harvard.edu/abs/2019TNSCR1774....1F}{2019-1774,
  1}

\bibitem[{{Fremling} {et~al.}(2018{\natexlab{\hspace{0pt}a}}){Fremling},
  {Dugas}, \& {Sharma}}]{Fremling18_2018gft}
{Fremling}, C., {Dugas}, A., \& {Sharma}, Y. 2018{\natexlab{\hspace{0pt}a}},
  Transient Name Server Classification Report,
  \hypersetup{urlcolor=blue}\href{https://ui.adsabs.harvard.edu/abs/2018TNSCR1411....1F}{2018-1411,
  1}

\bibitem[{{Fremling} {et~al.}(2018{\natexlab{\hspace{0pt}b}}){Fremling},
  {Dugas}, \& {Sharma}}]{Fremling18}
{Fremling}, C., {Dugas}, A., \& {Sharma}, Y. 2018{\natexlab{\hspace{0pt}b}},
  Transient Name Server Classification Report,
  \hypersetup{urlcolor=blue}\href{https://ui.adsabs.harvard.edu/abs/2018TNSCR1416....1F}{2018-1416,
  1}

\bibitem[{{Fremling} {et~al.}(2018{\natexlab{\hspace{0pt}c}}){Fremling},
  {Dugas}, \& {Sharma}}]{Fremling18_2018ibb}
{Fremling}, C., {Dugas}, A., \& {Sharma}, Y. 2018{\natexlab{\hspace{0pt}c}},
  Transient Name Server Classification Report,
  \hypersetup{urlcolor=blue}\href{https://ui.adsabs.harvard.edu/abs/2018TNSCR1877....1F}{2018-1877,
  1}

\bibitem[{{Fremling} {et~al.}(2019{\natexlab{\hspace{0pt}a}}){Fremling},
  {Dugas}, \& {Sharma}}]{Fremling19_2018lfd}
{Fremling}, C., {Dugas}, A., \& {Sharma}, Y. 2019{\natexlab{\hspace{0pt}a}},
  Transient Name Server Classification Report,
  \hypersetup{urlcolor=blue}\href{https://ui.adsabs.harvard.edu/abs/2019TNSCR..32....1F}{2019-32,
  1}

\bibitem[{{Fremling} {et~al.}(2019{\natexlab{\hspace{0pt}b}}){Fremling},
  {Dugas}, \& {Sharma}}]{Fremling19_2018kyt}
{Fremling}, C., {Dugas}, A., \& {Sharma}, Y. 2019{\natexlab{\hspace{0pt}b}},
  Transient Name Server Classification Report,
  \hypersetup{urlcolor=blue}\href{https://ui.adsabs.harvard.edu/abs/2019TNSCR.188....1F}{2019-188,
  1}

\bibitem[{{Fremling} {et~al.}(2019{\natexlab{\hspace{0pt}c}}){Fremling},
  {Dugas}, \& {Sharma}}]{Fremling19_2019bgu}
{Fremling}, C., {Dugas}, A., \& {Sharma}, Y. 2019{\natexlab{\hspace{0pt}c}},
  Transient Name Server Classification Report,
  \hypersetup{urlcolor=blue}\href{https://ui.adsabs.harvard.edu/abs/2019TNSCR.598....1F}{2019-598,
  1}

\bibitem[{{Fremling} {et~al.}(2019{\natexlab{\hspace{0pt}d}}){Fremling},
  {Dugas}, \& {Sharma}}]{Fremling19_2019cdt}
{Fremling}, C., {Dugas}, A., \& {Sharma}, Y. 2019{\natexlab{\hspace{0pt}d}},
  Transient Name Server Classification Report,
  \hypersetup{urlcolor=blue}\href{https://ui.adsabs.harvard.edu/abs/2019TNSCR.636....1F}{2019-636,
  1}

\bibitem[{{Fremling} {et~al.}(2019{\natexlab{\hspace{0pt}e}}){Fremling},
  {Dugas}, \& {Sharma}}]{Fremling19_2019eot}
{Fremling}, C., {Dugas}, A., \& {Sharma}, Y. 2019{\natexlab{\hspace{0pt}e}},
  Transient Name Server Classification Report,
  \hypersetup{urlcolor=blue}\href{https://ui.adsabs.harvard.edu/abs/2019TNSCR.952....1F}{2019-952,
  1}

\bibitem[{{Fremling} {et~al.}(2020){Fremling}, {Miller}, {Sharma}, {Dugas},
  {Perley}, {Taggart}, {Sollerman}, {Goobar}, {Graham}, {Neill}, {Nordin},
  {Rigault}, {Walters}, {Andreoni}, {Bagdasaryan}, {Belicki}, {Cannella},
  {Bellm}, {Cenko}, {De}, {Dekany}, {Frederick}, {Golkhou}, {Graham}, {Helou},
  {Ho}, {Kasliwal}, {Kupfer}, {Laher}, {Mahabal}, {Masci}, {Riddle},
  {Rusholme}, {Schulze}, {Shupe}, {Smith}, {Velzen}, {Yan}, {Yao}, {Zhuang}, \&
  {Kulkarni}}]{Fremling20}
{Fremling}, C., {Miller}, A.~A., {Sharma}, Y., {et~al.} 2020,
  \hypersetup{urlcolor=magenta}\href{https://dx.doi.org/10.3847/1538-4357/ab8943}{\apj},
  \hypersetup{urlcolor=blue}\href{https://ui.adsabs.harvard.edu/abs/2020ApJ...895...32F}{895,
  32}

\bibitem[{{Gagliano} {et~al.}(2020){Gagliano}, {Narayan}, {Engel}, \& {Carrasco
  Kind}}]{Gagliano20}
{Gagliano}, A., {Narayan}, G., {Engel}, A., \& {Carrasco Kind}, M. 2020, arXiv
  e-prints,
  \hypersetup{urlcolor=magenta}\href{https://arxiv.org/abs/2008.09630}{arXiv}{:}\hypersetup{urlcolor=blue}\href{https://ui.adsabs.harvard.edu/abs/2020arXiv200809630G}{2008.09630}

\bibitem[{Gomez {et~al.}(2020)Gomez, Berger, Blanchard, Hosseinzadeh, Nicholl,
  Villar, \& Yin}]{Gomez20}
Gomez, S., Berger, E., Blanchard, P.~K., {et~al.} 2020, {FLEET Finding Luminous
  and Exotic Extragalactic Transients}, 1.0.0,  Zenodo,
  \hypersetup{urlcolor=magenta}doi:\href{https://dx.doi.org/10.5281/zenodo.4013965}{10.5281/zenodo.4013965}

\bibitem[{{Gomez} {et~al.}(2019){Gomez}, {Berger}, {Nicholl}, {Blanchard},
  {Villar}, {Patton}, {Chornock}, {Leja}, {Hosseinzadeh}, \&
  {Cowperthwaite}}]{Gomez19}
{Gomez}, S., {Berger}, E., {Nicholl}, M., {et~al.} 2019,
  \hypersetup{urlcolor=magenta}\href{https://dx.doi.org/10.3847/1538-4357/ab2f92}{\apj},
  \hypersetup{urlcolor=blue}\href{https://ui.adsabs.harvard.edu/abs/2019ApJ...881...87G}{881,
  87}

\bibitem[{{Goodman} \& {Weare}(2010)}]{Goodman10}
{Goodman}, J., \& {Weare}, J. 2010,
  \hypersetup{urlcolor=magenta}\href{https://dx.doi.org/10.2140/camcos.2010.5.65}{Communications
  in Applied Mathematics and Computational Science},
  \hypersetup{urlcolor=blue}\href{https://ui.adsabs.harvard.edu/abs/2010CAMCS...5...65G}{5,
  65}

\bibitem[{{Graham} {et~al.}(2018){Graham}, {Connolly}, {Ivezi{\'c}}, {Schmidt},
  {Jones}, {Juri{\'c}}, {Daniel}, \& {Yoachim}}]{Graham18}
{Graham}, M.~L., {Connolly}, A.~J., {Ivezi{\'c}}, {\v{Z}}., {et~al.} 2018,
  \hypersetup{urlcolor=magenta}\href{https://dx.doi.org/10.3847/1538-3881/aa99d4}{\aj},
  \hypersetup{urlcolor=blue}\href{https://ui.adsabs.harvard.edu/abs/2018AJ....155....1G}{155,
  1}

\bibitem[{{Guillochon} {et~al.}(2017){Guillochon}, {Parrent}, {Kelley}, \&
  {Margutti}}]{guillochon17}
{Guillochon}, J., {Parrent}, J., {Kelley}, L.~Z., \& {Margutti}, R. 2017,
  \hypersetup{urlcolor=magenta}\href{https://dx.doi.org/10.3847/1538-4357/835/1/64}{\apj},
  \hypersetup{urlcolor=blue}\href{https://ui.adsabs.harvard.edu/abs/2017ApJ...835...64G}{835,
  64}

\bibitem[{{Hosseinzadeh} {et~al.}(2020){Hosseinzadeh}, {Dauphin}, {Villar},
  {Berger}, {Jones}, {Challis}, {Chornock}, {Drout}, {Foley}, {Kirshner},
  {Lunnan}, {Margutti}, {Milisavljevic}, {Pan}, {Rest}, {Scolnic}, {Magnier},
  {Metcalfe}, {Wainscoat}, \& {Waters}}]{Hosseinzadeh20}
{Hosseinzadeh}, G., {Dauphin}, F., {Villar}, V.~A., {et~al.} 2020, arXiv
  e-prints,
  \hypersetup{urlcolor=magenta}\href{https://arxiv.org/abs/2008.04912}{arXiv}{:}\hypersetup{urlcolor=blue}\href{https://ui.adsabs.harvard.edu/abs/2020arXiv200804912H}{2008.04912}

\bibitem[{{Howell} {et~al.}(2013){Howell}, {Kasen}, {Lidman}, {Sullivan},
  {Conley}, {Astier}, {Balland}, {Carlberg}, {Fouchez}, {Guy}, {Hardin},
  {Pain}, {Palanque-Delabrouille}, {Perrett}, {Pritchet}, {Regnault}, {Rich},
  \& {Ruhlmann-Kleider}}]{Howell13}
{Howell}, D.~A., {Kasen}, D., {Lidman}, C., {et~al.} 2013,
  \hypersetup{urlcolor=magenta}\href{https://dx.doi.org/10.1088/0004-637X/779/2/98}{\apj},
  \hypersetup{urlcolor=blue}\href{https://ui.adsabs.harvard.edu/abs/2013ApJ...779...98H}{779,
  98}

\bibitem[{{Hudelot} {et~al.}(2012){Hudelot}, {Cuillandre}, {Withington},
  {Goranova}, {McCracken}, {Magnard}, {Mellier}, {Regnault}, {Betoule},
  {Aussel}, {Kavelaars}, {Fernique}, {Bonnarel}, {Ochsenbein}, \&
  {Ilbert}}]{Hudelot12}
{Hudelot}, P., {Cuillandre}, J.~C., {Withington}, K., {et~al.} 2012, VizieR
  Online Data Catalog

\bibitem[{Hunter(2007)}]{matplotlib}
Hunter, J.~D. 2007,
  \hypersetup{urlcolor=magenta}\href{https://dx.doi.org/10.1109/MCSE.2007.55}{CSE},
  \hypersetup{urlcolor=blue}\href{https://ui.adsabs.harvard.edu/#abs/2007CSE.....9...90H}{9,
  90}

\bibitem[{{Inserra} {et~al.}(2013){Inserra}, {Smartt}, {Jerkstrand}, {Valenti},
  {Fraser}, {Wright}, {Smith}, {Chen}, {Kotak}, {Pastorello}, {Nicholl},
  {Bresolin}, {Kudritzki}, {Benetti}, {Botticella}, {Burgett}, {Chambers},
  {Ergon}, {Flewelling}, {Fynbo}, {Geier}, {Hodapp}, {Howell}, {Huber},
  {Kaiser}, {Leloudas}, {Magill}, {Magnier}, {McCrum}, {Metcalfe}, {Price},
  {Rest}, {Sollerman}, {Sweeney}, {Taddia}, {Taubenberger}, {Tonry},
  {Wainscoat}, {Waters}, \& {Young}}]{Inserra13}
{Inserra}, C., {Smartt}, S.~J., {Jerkstrand}, A., {et~al.} 2013,
  \hypersetup{urlcolor=magenta}\href{https://dx.doi.org/10.1088/0004-637X/770/2/128}{\apj},
  \hypersetup{urlcolor=blue}\href{https://ui.adsabs.harvard.edu/abs/2013ApJ...770..128I}{770,
  128}

\bibitem[{{Inserra} {et~al.}(2017){Inserra}, {Nicholl}, {Chen}, {Jerkstrand},
  {Smartt}, {Kr{\"u}hler}, {Anderson}, {Baltay}, {Della Valle}, {Fraser},
  {Gal-Yam}, {Galbany}, {Kankare}, {Maguire}, {Rabinowitz}, {Smith}, {Valenti},
  \& {Young}}]{Inserra17}
{Inserra}, C., {Nicholl}, M., {Chen}, T.~W., {et~al.} 2017,
  \hypersetup{urlcolor=magenta}\href{https://dx.doi.org/10.1093/mnras/stx834}{\mnras},
  \hypersetup{urlcolor=blue}\href{https://ui.adsabs.harvard.edu/abs/2017MNRAS.468.4642I}{468,
  4642}

\bibitem[{{Kasen} \& {Bildsten}(2010)}]{Kasen10}
{Kasen}, D., \& {Bildsten}, L. 2010,
  \hypersetup{urlcolor=magenta}\href{https://dx.doi.org/10.1088/0004-637X/717/1/245}{\apj},
  \hypersetup{urlcolor=blue}\href{https://ui.adsabs.harvard.edu/abs/2010ApJ...717..245K}{717,
  245}

\bibitem[{{Kasliwal} \& {Cao}(2019)}]{Kasliwal19}
{Kasliwal}, M., \& {Cao}, Y. 2019, Transient Name Server Discovery Report,
  \hypersetup{urlcolor=blue}\href{https://ui.adsabs.harvard.edu/abs/2019TNSTR.259....1K}{2019-259,
  1}

\bibitem[{{Kessler} {et~al.}(2019){Kessler}, {Narayan}, {Avelino}, {Bachelet},
  {Biswas}, {Brown}, {Chernoff}, {Connolly}, {Dai}, {Daniel}, {Di Stefano},
  {Drout}, {Galbany}, {Gonz{\'a}lez-Gait{\'a}n}, {Graham}, {Hlo{\v{z}}ek},
  {Ishida}, {Guillochon}, {Jha}, {Jones}, {Mand el}, {Muthukrishna}, {O'Grady},
  {Peters}, {Pierel}, {Ponder}, {Pr{\v{s}}a}, {Rodney}, {Villar}, {LSST Dark
  Energy Science Collaboration}, \& {Transient and Variable Stars Science
  Collaboration}}]{Kessler19}
{Kessler}, R., {Narayan}, G., {Avelino}, A., {et~al.} 2019,
  \hypersetup{urlcolor=magenta}\href{https://dx.doi.org/10.1088/1538-3873/ab26f1}{\pasp},
  \hypersetup{urlcolor=blue}\href{https://ui.adsabs.harvard.edu/abs/2019PASP..131i4501K}{131,
  094501}

\bibitem[{{Leloudas} {et~al.}(2012){Leloudas}, {Chatzopoulos}, {Dilday},
  {Gorosabel}, {Vinko}, {Gallazzi}, {Wheeler}, {Bassett}, {Fischer}, {Frieman},
  {Fynbo}, {Goobar}, {Jel{\'\i}nek}, {Malesani}, {Nichol}, {Nordin},
  {{\"O}stman}, {Sako}, {Schneider}, {Smith}, {Sollerman}, {Stritzinger},
  {Th{\"o}ne}, \& {de Ugarte Postigo}}]{Leloudas12}
{Leloudas}, G., {Chatzopoulos}, E., {Dilday}, B., {et~al.} 2012,
  \hypersetup{urlcolor=magenta}\href{https://dx.doi.org/10.1051/0004-6361/201118498}{\aap},
  \hypersetup{urlcolor=blue}\href{https://ui.adsabs.harvard.edu/abs/2012A&A...541A.129L}{541,
  A129}

\bibitem[{{Lin} {et~al.}(2020){Lin}, {Wang}, {Li}, {Zhang}, {Mo}, {Sai},
  {Zhang}, {Filippenko}, {Zheng}, {Brink}, {Baron}, {DerKacy}, {Ehgamberdiev},
  {Mirzaqulov}, {Li}, {Zhang}, {Yan}, {Xi}, {Hsiao}, {Zhang}, {Wang}, {Liu},
  {Xiang}, {Wu}, {Rui}, \& {Chen}}]{Lin20}
{Lin}, W.~L., {Wang}, X.~F., {Li}, W.~X., {et~al.} 2020, arXiv e-prints,
  \hypersetup{urlcolor=magenta}\href{https://arxiv.org/abs/2006.16443}{arXiv}{:}\hypersetup{urlcolor=blue}\href{https://ui.adsabs.harvard.edu/abs/2020arXiv200616443L}{2006.16443}

\bibitem[{{Liu} {et~al.}(2018){Liu}, {Wang}, {Wang}, \& {Dai}}]{Liu18}
{Liu}, L.-D., {Wang}, L.-J., {Wang}, S.-Q., \& {Dai}, Z.-G. 2018,
  \hypersetup{urlcolor=magenta}\href{https://dx.doi.org/10.3847/1538-4357/aab157}{\apj},
  \hypersetup{urlcolor=blue}\href{https://ui.adsabs.harvard.edu/abs/2018ApJ...856...59L}{856,
  59}

\bibitem[{{Lunnan} {et~al.}(2013){Lunnan}, {Chornock}, {Berger},
  {Milisavljevic}, {Drout}, {Sanders}, {Challis}, {Czekala}, {Foley}, {Fong},
  {Huber}, {Kirshner}, {Leibler}, {Marion}, {McCrum}, {Narayan}, {Rest},
  {Roth}, {Scolnic}, {Smartt}, {Smith}, {Soderberg}, {Stubbs}, {Tonry},
  {Burgett}, {Chambers}, {Kudritzki}, {Magnier}, \& {Price}}]{Lunnan13}
{Lunnan}, R., {Chornock}, R., {Berger}, E., {et~al.} 2013,
  \hypersetup{urlcolor=magenta}\href{https://dx.doi.org/10.1088/0004-637X/771/2/97}{\apj},
  \hypersetup{urlcolor=blue}\href{https://ui.adsabs.harvard.edu/abs/2013ApJ...771...97L}{771,
  97}

\bibitem[{{Lunnan} {et~al.}(2014){Lunnan}, {Chornock}, {Berger}, {Laskar},
  {Fong}, {Rest}, {Sanders}, {Challis}, {Drout}, {Foley}, {Huber}, {Kirshner},
  {Leibler}, {Marion}, {McCrum}, {Milisavljevic}, {Narayan}, {Scolnic},
  {Smartt}, {Smith}, {Soderberg}, {Tonry}, {Burgett}, {Chambers}, {Flewelling},
  {Hodapp}, {Kaiser}, {Magnier}, {Price}, \& {Wainscoat}}]{Lunnan14}
{Lunnan}, R., {Chornock}, R., {Berger}, E., {et~al.} 2014,
  \hypersetup{urlcolor=magenta}\href{https://dx.doi.org/10.1088/0004-637X/787/2/138}{\apj},
  \hypersetup{urlcolor=blue}\href{https://ui.adsabs.harvard.edu/abs/2014ApJ...787..138L}{787,
  138}

\bibitem[{{Lunnan} {et~al.}(2018{\natexlab{\hspace{0pt}a}}){Lunnan},
  {Fransson}, {Vreeswijk}, {Woosley}, {Leloudas}, {Perley}, {Quimby}, {Yan},
  {Blagorodnova}, {Bue}, {Cenko}, {De Cia}, {Cook}, {Fremling}, {Gatkine},
  {Gal-Yam}, {Kasliwal}, {Kulkarni}, {Masci}, {Nugent}, {Nyholm}, {Rubin},
  {Suzuki}, \& {Wozniak}}]{Lunnan18_iPTF16eh}
{Lunnan}, R., {Fransson}, C., {Vreeswijk}, P.~M., {et~al.}
  2018{\natexlab{\hspace{0pt}a}},
  \hypersetup{urlcolor=magenta}\href{https://dx.doi.org/10.1038/s41550-018-0568-z}{Nature
  Astronomy},
  \hypersetup{urlcolor=blue}\href{https://ui.adsabs.harvard.edu/abs/2018NatAs...2..887L}{2,
  887}

\bibitem[{{Lunnan} {et~al.}(2018{\natexlab{\hspace{0pt}b}}){Lunnan},
  {Chornock}, {Berger}, {Jones}, {Rest}, {Czekala}, {Dittmann}, {Drout},
  {Foley}, {Fong}, {Kirshner}, {Laskar}, {Leibler}, {Margutti},
  {Milisavljevic}, {Narayan}, {Pan}, {Riess}, {Roth}, {Sanders}, {Scolnic},
  {Smartt}, {Smith}, {Chambers}, {Draper}, {Flewelling}, {Huber}, {Kaiser},
  {Kudritzki}, {Magnier}, {Metcalfe}, {Wainscoat}, {Waters}, \&
  {Willman}}]{Lunnan18}
{Lunnan}, R., {Chornock}, R., {Berger}, E., {et~al.}
  2018{\natexlab{\hspace{0pt}b}},
  \hypersetup{urlcolor=magenta}\href{https://dx.doi.org/10.3847/1538-4357/aa9f1a}{\apj},
  \hypersetup{urlcolor=blue}\href{https://ui.adsabs.harvard.edu/abs/2018ApJ...852...81L}{852,
  81}

\bibitem[{{Lunnan} {et~al.}(2019){Lunnan}, {Yan}, {Perley}, {Schulze},
  {Taggart}, {Gal-Yam}, {Fremling}, {Soumagnac}, {Ofek}, {Adams}, {Barbarino},
  {Bellm}, {De}, {Fransson}, {Frederick}, {Golkhou}, {Graham}, {Hallakoun},
  {Ho}, {Kasliwal}, {Kaspi}, {Kulkarni}, {Laher}, {Masci}, {Pozo Nunez},
  {Rusholme}, {Quimby}, {Shupe}, {Sollerman}, {Taddia}, {van Roestel}, {Yang},
  \& {Yao}}]{Lunnan19_four}
{Lunnan}, R., {Yan}, L., {Perley}, D.~A., {et~al.} 2019, arXiv e-prints,
  \hypersetup{urlcolor=magenta}\href{https://arxiv.org/abs/1910.02968}{arXiv}{:}\hypersetup{urlcolor=blue}\href{https://ui.adsabs.harvard.edu/abs/2019arXiv191002968L}{1910.02968}

\bibitem[{{Lyman} {et~al.}(2017){Lyman}, {Homan}, {Magee}, \&
  {Yaron}}]{Lyman17}
{Lyman}, J., {Homan}, D., {Magee}, M., \& {Yaron}, O. 2017, Transient Name
  Server Classification Report,
  \hypersetup{urlcolor=blue}\href{https://ui.adsabs.harvard.edu/abs/2017TNSCR.881....1L}{2017-881,
  1}

\bibitem[{{Mazzali} {et~al.}(2016){Mazzali}, {Sullivan}, {Pian}, {Greiner}, \&
  {Kann}}]{Mazzali16}
{Mazzali}, P.~A., {Sullivan}, M., {Pian}, E., {Greiner}, J., \& {Kann}, D.~A.
  2016,
  \hypersetup{urlcolor=magenta}\href{https://dx.doi.org/10.1093/mnras/stw512}{\mnras},
  \hypersetup{urlcolor=blue}\href{https://ui.adsabs.harvard.edu/abs/2016MNRAS.458.3455M}{458,
  3455}

\bibitem[{{McCrum} {et~al.}(2015){McCrum}, {Smartt}, {Rest}, {Smith}, {Kotak},
  {Rodney}, {Young}, {Chornock}, {Berger}, {Foley}, {Fraser}, {Wright},
  {Scolnic}, {Tonry}, {Urata}, {Huang}, {Pastorello}, {Botticella}, {Valenti},
  {Mattila}, {Kankare}, {Farrow}, {Huber}, {Stubbs}, {Kirshner}, {Bresolin},
  {Burgett}, {Chambers}, {Draper}, {Flewelling}, {Jedicke}, {Kaiser},
  {Magnier}, {Metcalfe}, {Morgan}, {Price}, {Sweeney}, {Wainscoat}, \&
  {Waters}}]{McCrum15}
{McCrum}, M., {Smartt}, S.~J., {Rest}, A., {et~al.} 2015,
  \hypersetup{urlcolor=magenta}\href{https://dx.doi.org/10.1093/mnras/stv034}{\mnras},
  \hypersetup{urlcolor=blue}\href{https://ui.adsabs.harvard.edu/abs/2015MNRAS.448.1206M}{448,
  1206}

\bibitem[{{Metzger} {et~al.}(2015){Metzger}, {Margalit}, {Kasen}, \&
  {Quataert}}]{Metzger15}
{Metzger}, B.~D., {Margalit}, B., {Kasen}, D., \& {Quataert}, E. 2015,
  \hypersetup{urlcolor=magenta}\href{https://dx.doi.org/10.1093/mnras/stv2224}{\mnras},
  \hypersetup{urlcolor=blue}\href{https://ui.adsabs.harvard.edu/abs/2015MNRAS.454.3311M}{454,
  3311}

\bibitem[{{Muthukrishna} {et~al.}(2019){Muthukrishna}, {Narayan}, {Mandel},
  {Biswas}, \& {Hlo{\v{z}}ek}}]{Muthukrishna19}
{Muthukrishna}, D., {Narayan}, G., {Mandel}, K.~S., {Biswas}, R., \&
  {Hlo{\v{z}}ek}, R. 2019,
  \hypersetup{urlcolor=magenta}\href{https://dx.doi.org/10.1088/1538-3873/ab1609}{\pasp},
  \hypersetup{urlcolor=blue}\href{https://ui.adsabs.harvard.edu/abs/2019PASP..131k8002M}{131,
  118002}

\bibitem[{{Nicholl} {et~al.}(2019){Nicholl}, {Berger}, {Blanchard}, {Gomez}, \&
  {Chornock}}]{Nicholl19_nebular}
{Nicholl}, M., {Berger}, E., {Blanchard}, P.~K., {Gomez}, S., \& {Chornock}, R.
  2019,
  \hypersetup{urlcolor=magenta}\href{https://dx.doi.org/10.3847/1538-4357/aaf470}{\apj},
  \hypersetup{urlcolor=blue}\href{https://ui.adsabs.harvard.edu/abs/2019ApJ...871..102N}{871,
  102}

\bibitem[{{Nicholl} {et~al.}(2017{\natexlab{\hspace{0pt}a}}){Nicholl},
  {Berger}, {Margutti}, {Blanchard}, {Guillochon}, {Leja}, \&
  {Chornock}}]{Nicholl17_17egm}
{Nicholl}, M., {Berger}, E., {Margutti}, R., {et~al.}
  2017{\natexlab{\hspace{0pt}a}},
  \hypersetup{urlcolor=magenta}\href{https://dx.doi.org/10.3847/2041-8213/aa82b1}{\apjl},
  \hypersetup{urlcolor=blue}\href{https://ui.adsabs.harvard.edu/abs/2017ApJ...845L...8N}{845,
  L8}

\bibitem[{{Nicholl} {et~al.}(2017{\natexlab{\hspace{0pt}b}}){Nicholl},
  {Berger}, {Margutti}, {Blanchard}, {Milisavljevic}, {Challis}, {Metzger}, \&
  {Chornock}}]{Nicholl17_16apd}
{Nicholl}, M., {Berger}, E., {Margutti}, R., {et~al.}
  2017{\natexlab{\hspace{0pt}b}},
  \hypersetup{urlcolor=magenta}\href{https://dx.doi.org/10.3847/2041-8213/aa56c5}{\apjl},
  \hypersetup{urlcolor=blue}\href{https://ui.adsabs.harvard.edu/abs/2017ApJ...835L...8N}{835,
  L8}

\bibitem[{{Nicholl} {et~al.}(2017{\natexlab{\hspace{0pt}c}}){Nicholl},
  {Guillochon}, \& {Berger}}]{Nicholl17_mosfit}
{Nicholl}, M., {Guillochon}, J., \& {Berger}, E.
  2017{\natexlab{\hspace{0pt}c}},
  \hypersetup{urlcolor=magenta}\href{https://dx.doi.org/10.3847/1538-4357/aa9334}{\apj},
  \hypersetup{urlcolor=blue}\href{https://ui.adsabs.harvard.edu/abs/2017ApJ...850...55N}{850,
  55}

\bibitem[{{Nicholl} {et~al.}(2013){Nicholl}, {Smartt}, {Jerkstrand}, {Inserra},
  {McCrum}, {Kotak}, {Fraser}, {Wright}, {Chen}, {Smith}, {Young}, {Sim},
  {Valenti}, {Howell}, {Bresolin}, {Kudritzki}, {Tonry}, {Huber}, {Rest},
  {Pastorello}, {Tomasella}, {Cappellaro}, {Benetti}, {Mattila}, {Kankare},
  {Kangas}, {Leloudas}, {Sollerman}, {Taddia}, {Berger}, {Chornock}, {Narayan},
  {Stubbs}, {Foley}, {Lunnan}, {Soderberg}, {Sanders}, {Milisavljevic},
  {Margutti}, {Kirshner}, {Elias-Rosa}, {Morales-Garoffolo}, {Taubenberger},
  {Botticella}, {Gezari}, {Urata}, {Rodney}, {Riess}, {Scolnic}, {Wood-Vasey},
  {Burgett}, {Chambers}, {Flewelling}, {Magnier}, {Kaiser}, {Metcalfe},
  {Morgan}, {Price}, {Sweeney}, \& {Waters}}]{Nicholl13}
{Nicholl}, M., {Smartt}, S.~J., {Jerkstrand}, A., {et~al.} 2013,
  \hypersetup{urlcolor=magenta}\href{https://dx.doi.org/10.1038/nature12569}{\nat},
  \hypersetup{urlcolor=blue}\href{https://ui.adsabs.harvard.edu/abs/2013Natur.502..346N}{502,
  346}

\bibitem[{{Nicholl} {et~al.}(2014){Nicholl}, {Smartt}, {Jerkstrand}, {Inserra},
  {Anderson}, {Baltay}, {Benetti}, {Chen}, {Elias-Rosa}, {Feindt}, {Fraser},
  {Gal-Yam}, {Hadjiyska}, {Howell}, {Kotak}, {Lawrence}, {Leloudas},
  {Margheim}, {Mattila}, {McCrum}, {McKinnon}, {Mead}, {Nugent}, {Rabinowitz},
  {Rest}, {Smith}, {Sollerman}, {Sullivan}, {Taddia}, {Valenti}, {Walker}, \&
  {Young}}]{Nicholl14}
{Nicholl}, M., {Smartt}, S.~J., {Jerkstrand}, A., {et~al.} 2014,
  \hypersetup{urlcolor=magenta}\href{https://dx.doi.org/10.1093/mnras/stu1579}{\mnras},
  \hypersetup{urlcolor=blue}\href{https://ui.adsabs.harvard.edu/abs/2014MNRAS.444.2096N}{444,
  2096}

\bibitem[{{Nicholl} {et~al.}(2015){Nicholl}, {Smartt}, {Jerkstrand}, {Sim},
  {Inserra}, {Anderson}, {Baltay}, {Benetti}, {Chambers}, {Chen}, {Elias-Rosa},
  {Feindt}, {Flewelling}, {Fraser}, {Gal-Yam}, {Galbany}, {Huber}, {Kangas},
  {Kankare}, {Kotak}, {Kr{\"u}hler}, {Maguire}, {McKinnon}, {Rabinowitz},
  {Rostami}, {Schulze}, {Smith}, {Sullivan}, {Tonry}, {Valenti}, \&
  {Young}}]{Nicholl15_LSQ14bdq}
{Nicholl}, M., {Smartt}, S.~J., {Jerkstrand}, A., {et~al.} 2015,
  \hypersetup{urlcolor=magenta}\href{https://dx.doi.org/10.1088/2041-8205/807/1/L18}{\apjl},
  \hypersetup{urlcolor=blue}\href{https://ui.adsabs.harvard.edu/abs/2015ApJ...807L..18N}{807,
  L18}

\bibitem[{{Nicholl} {et~al.}(2016){Nicholl}, {Berger}, {Margutti}, {Chornock},
  {Blanchard}, {Jerkstrand}, {Smartt}, {Arcavi}, {Challis}, {Chambers}, {Chen},
  {Cowperthwaite}, {Gal-Yam}, {Hosseinzadeh}, {Howell}, {Inserra}, {Kankare},
  {Magnier}, {Maguire}, {Mazzali}, {McCully}, {Milisavljevic}, {Smith},
  {Taubenberger}, {Valenti}, {Wainscoat}, {Yaron}, \& {Young}}]{Nicholl16_15bn}
{Nicholl}, M., {Berger}, E., {Margutti}, R., {et~al.} 2016,
  \hypersetup{urlcolor=magenta}\href{https://dx.doi.org/10.3847/2041-8205/828/2/L18}{\apjl},
  \hypersetup{urlcolor=blue}\href{https://ui.adsabs.harvard.edu/abs/2016ApJ...828L..18N}{828,
  L18}

\bibitem[{{Nicholl} {et~al.}(2018){Nicholl}, {Blanchard}, {Berger}, {Alexand
  er}, {Metzger}, {Bhirombhakdi}, {Chornock}, {Coppejans}, {Gomez}, {Margalit},
  {Margutti}, \& {Terreran}}]{Nicholl18_1000days}
{Nicholl}, M., {Blanchard}, P.~K., {Berger}, E., {et~al.} 2018,
  \hypersetup{urlcolor=magenta}\href{https://dx.doi.org/10.3847/2041-8213/aae70d}{\apjl},
  \hypersetup{urlcolor=blue}\href{https://ui.adsabs.harvard.edu/abs/2018ApJ...866L..24N}{866,
  L24}

\bibitem[{{Nicholl} {et~al.}(2020){Nicholl}, {Blanchard}, {Berger}, {Chornock},
  {Margutti}, {Gomez}, {Lunnan}, {Miller}, {Fong}, {Terreran},
  {Vigna-G{\'o}mez}, {Bhirombhakdi}, {Bieryla}, {Challis}, {Laher}, {Masci}, \&
  {Paterson}}]{Nicholl20_nature}
{Nicholl}, M., {Blanchard}, P.~K., {Berger}, E., {et~al.} 2020,
  \hypersetup{urlcolor=magenta}\href{https://dx.doi.org/10.1038/s41550-020-1066-7}{Nature
  Astronomy},
  \hypersetup{urlcolor=magenta}\href{https://arxiv.org/abs/2004.05840}{arXiv}{:}\hypersetup{urlcolor=blue}\href{https://ui.adsabs.harvard.edu/abs/2020NatAs.tmp...78N}{2004.05840}

\bibitem[{{Papadopoulos} {et~al.}(2015){Papadopoulos}, {D'Andrea}, {Sullivan},
  {Nichol}, {Barbary}, {Biswas}, {Brown}, {Covarrubias}, {Finley}, {Fischer},
  {Foley}, {Goldstein}, {Gupta}, {Kessler}, {Kovacs}, {Kuhlmann}, {Lidman},
  {March}, {Nugent}, {Sako}, {Smith}, {Spinka}, {Wester}, {Abbott}, {Abdalla},
  {Allam}, {Banerji}, {Bernstein}, {Bernstein}, {Carnero}, {da Costa}, {DePoy},
  {Desai}, {Diehl}, {Eifler}, {Evrard}, {Flaugher}, {Frieman}, {Gerdes},
  {Gruen}, {Honscheid}, {James}, {Kuehn}, {Kuropatkin}, {Lahav}, {Maia},
  {Makler}, {Marshall}, {Merritt}, {Miller}, {Miquel}, {Ogando}, {Plazas},
  {Roe}, {Romer}, {Rykoff}, {Sanchez}, {Santiago}, {Scarpine}, {Schubnell},
  {Sevilla}, {Soares-Santos}, {Suchyta}, {Swanson}, {Tarle}, {Thaler},
  {Tucker}, {Wechsler}, \& {Zuntz}}]{Papadopoulos15}
{Papadopoulos}, A., {D'Andrea}, C.~B., {Sullivan}, M., {et~al.} 2015,
  \hypersetup{urlcolor=magenta}\href{https://dx.doi.org/10.1093/mnras/stv174}{\mnras},
  \hypersetup{urlcolor=blue}\href{https://ui.adsabs.harvard.edu/abs/2015MNRAS.449.1215P}{449,
  1215}

\bibitem[{{Pedregosa} {et~al.}(2012){Pedregosa}, {Varoquaux}, {Gramfort},
  {Michel}, {Thirion}, {Grisel}, {Blondel}, {M{\"u}ller}, {Nothman}, {Louppe},
  {Prettenhofer}, {Weiss}, {Dubourg}, {Vanderplas}, {Passos}, {Cournapeau},
  {Brucher}, {Perrot}, \& {Duchesnay}}]{Pedregosa12}
{Pedregosa}, F., {Varoquaux}, G., {Gramfort}, A., {et~al.} 2012, arXiv
  e-prints,
  \hypersetup{urlcolor=magenta}\href{https://arxiv.org/abs/1201.0490}{arXiv}{:}\hypersetup{urlcolor=blue}\href{https://ui.adsabs.harvard.edu/abs/2012arXiv1201.0490P}{1201.0490}

\bibitem[{{Perley} {et~al.}(2019{\natexlab{\hspace{0pt}a}}){Perley}, {Yan},
  {Andreoni}, {Karambelkar}, {Sharma}, {De}, {Fremling}, \&
  {Kulkarni}}]{Perley19_2019nhs}
{Perley}, D., {Yan}, L., {Andreoni}, I., {et~al.}
  2019{\natexlab{\hspace{0pt}a}}, Transient Name Server Classification Report,
  \hypersetup{urlcolor=blue}\href{https://ui.adsabs.harvard.edu/abs/2019TNSCR1712....1P}{2019-1712,
  1}

\bibitem[{{Perley} {et~al.}(2019{\natexlab{\hspace{0pt}b}}){Perley}, {Yan},
  {Lunnan}, {Gal-Yam}, {Schulze}, \& {Yaron}}]{Perley19}
{Perley}, D., {Yan}, L., {Lunnan}, R., {et~al.} 2019{\natexlab{\hspace{0pt}b}},
  Transient Name Server Classification Report,
  \hypersetup{urlcolor=blue}\href{https://ui.adsabs.harvard.edu/abs/2019TNSCR2829....1P}{2019-2829,
  1}

\bibitem[{{Perley} {et~al.}(2019{\natexlab{\hspace{0pt}c}}){Perley}, {Yan},
  {Gal-Yam}, {Schulze}, {Taggart}, {Bruch}, \& {Sollerman}}]{Perley19_19neq}
{Perley}, D.~A., {Yan}, L., {Gal-Yam}, A., {et~al.}
  2019{\natexlab{\hspace{0pt}c}}, Transient Name Server AstroNote,
  \hypersetup{urlcolor=blue}\href{https://ui.adsabs.harvard.edu/abs/2019TNSAN..79....1P}{79,
  1}

\bibitem[{{Perley} {et~al.}(2016){Perley}, {Quimby}, {Yan}, {Vreeswijk}, {De
  Cia}, {Lunnan}, {Gal-Yam}, {Yaron}, {Filippenko}, {Graham}, {Laher}, \&
  {Nugent}}]{Perley16}
{Perley}, D.~A., {Quimby}, R.~M., {Yan}, L., {et~al.} 2016,
  \hypersetup{urlcolor=magenta}\href{https://dx.doi.org/10.3847/0004-637X/830/1/13}{\apj},
  \hypersetup{urlcolor=blue}\href{https://ui.adsabs.harvard.edu/abs/2016ApJ...830...13P}{830,
  13}

\bibitem[{{Prajs} {et~al.}(2017){Prajs}, {Sullivan}, {Smith}, {Levan},
  {Karpenka}, {Edwards}, {Walker}, {Wolf}, {Balland}, {Carlberg}, {Howell},
  {Lidman}, {Pain}, {Pritchet}, \& {Ruhlmann-Kleider}}]{Prajs17}
{Prajs}, S., {Sullivan}, M., {Smith}, M., {et~al.} 2017,
  \hypersetup{urlcolor=magenta}\href{https://dx.doi.org/10.1093/mnras/stw1942}{\mnras},
  \hypersetup{urlcolor=blue}\href{https://ui.adsabs.harvard.edu/abs/2017MNRAS.464.3568P}{464,
  3568}

\bibitem[{{Prentice} {et~al.}(2019){Prentice}, {Maguire}, {Skillen}, {Magee},
  \& {Clark}}]{Prentice19}
{Prentice}, S.~J., {Maguire}, K., {Skillen}, K., {Magee}, M.~R., \& {Clark}, P.
  2019, Transient Name Server Classification Report,
  \hypersetup{urlcolor=blue}\href{https://ui.adsabs.harvard.edu/abs/2019TNSCR2339....1P}{2019-2339,
  1}

\bibitem[{{Quimby} {et~al.}(2007){Quimby}, {Aldering}, {Wheeler},
  {H{\"o}flich}, {Akerlof}, \& {Rykoff}}]{Quimby07}
{Quimby}, R.~M., {Aldering}, G., {Wheeler}, J.~C., {et~al.} 2007,
  \hypersetup{urlcolor=magenta}\href{https://dx.doi.org/10.1086/522862}{\apjl},
  \hypersetup{urlcolor=blue}\href{https://ui.adsabs.harvard.edu/abs/2007ApJ...668L..99Q}{668,
  L99}

\bibitem[{{Quimby} {et~al.}(2011){Quimby}, {Kulkarni}, {Kasliwal}, {Gal-Yam},
  {Arcavi}, {Sullivan}, {Nugent}, {Thomas}, {Howell}, {Nakar}, {Bildsten},
  {Theissen}, {Law}, {Dekany}, {Rahmer}, {Hale}, {Smith}, {Ofek}, {Zolkower},
  {Velur}, {Walters}, {Henning}, {Bui}, {McKenna}, {Poznanski}, {Cenko}, \&
  {Levitan}}]{Quimby11}
{Quimby}, R.~M., {Kulkarni}, S.~R., {Kasliwal}, M.~M., {et~al.} 2011,
  \hypersetup{urlcolor=magenta}\href{https://dx.doi.org/10.1038/nature10095}{\nat},
  \hypersetup{urlcolor=blue}\href{https://ui.adsabs.harvard.edu/abs/2011Natur.474..487Q}{474,
  487}

\bibitem[{{Quimby} {et~al.}(2018){Quimby}, {De Cia}, {Gal-Yam}, {Leloudas},
  {Lunnan}, {Perley}, {Vreeswijk}, {Yan}, {Bloom}, {Cenko}, {Cooke}, {Ellis},
  {Filippenko}, {Kasliwal}, {Kleiser}, {Kulkarni}, {Matheson}, {Nugent}, {Pan},
  {Silverman}, {Sternberg}, {Sullivan}, \& {Yaron}}]{Quimby18}
{Quimby}, R.~M., {De Cia}, A., {Gal-Yam}, A., {et~al.} 2018,
  \hypersetup{urlcolor=magenta}\href{https://dx.doi.org/10.3847/1538-4357/aaac2f}{\apj},
  \hypersetup{urlcolor=blue}\href{https://ui.adsabs.harvard.edu/abs/2018ApJ...855....2Q}{855,
  2}

\bibitem[{{Roy} {et~al.}(2016){Roy}, {Sollerman}, {Silverman}, {Pastorello},
  {Fransson}, {Drake}, {Taddia}, {Fremling}, {Kankare}, {Kumar}, {Cappellaro},
  {Bose}, {Benetti}, {Filippenko}, {Valenti}, {Nyholm}, {Ergon}, {Sutaria},
  {Kumar}, {Pand ey}, {Nicholl}, {Garcia-{\'A}lvarez}, {Tomasella},
  {Karamehmetoglu}, \& {Migotto}}]{Roy16}
{Roy}, R., {Sollerman}, J., {Silverman}, J.~M., {et~al.} 2016,
  \hypersetup{urlcolor=magenta}\href{https://dx.doi.org/10.1051/0004-6361/201527947}{\aap},
  \hypersetup{urlcolor=blue}\href{https://ui.adsabs.harvard.edu/abs/2016A&A...596A..67R}{596,
  A67}

\bibitem[{{S{\'a}nchez-S{\'a}ez} {et~al.}(2020){S{\'a}nchez-S{\'a}ez}, {Reyes},
  {Valenzuela}, {F{\"o}rster}, {Eyheramendy}, {Elorrieta}, {Bauer},
  {Cabrera-Vives}, {Est{\'e}vez}, {Catelan}, {Pignata}, {Huijse}, {De Cicco},
  {Ar{\'e}valo}, {Carrasco-Davis}, {Abril}, {Kurtev}, {Borissova}, {Arredondo},
  {Castillo-Navarrete}, {Rodriguez}, {Ruz-Mieres}, {Moya},
  {Sabatini-Gacit{\'u}a}, \& {Sep{\'u}lveda-Cobo}}]{Sanchez20}
{S{\'a}nchez-S{\'a}ez}, P., {Reyes}, I., {Valenzuela}, C., {et~al.} 2020, arXiv
  e-prints,
  \hypersetup{urlcolor=magenta}\href{https://arxiv.org/abs/2008.03311}{arXiv}{:}\hypersetup{urlcolor=blue}\href{https://ui.adsabs.harvard.edu/abs/2020arXiv200803311S}{2008.03311}

\bibitem[{{Schlafly} \& {Finkbeiner}(2011)}]{Schlafly11}
{Schlafly}, E.~F., \& {Finkbeiner}, D.~P. 2011,
  \hypersetup{urlcolor=magenta}\href{https://dx.doi.org/10.1088/0004-637X/737/2/103}{\apj},
  \hypersetup{urlcolor=blue}\href{https://ui.adsabs.harvard.edu/abs/2011ApJ...737..103S}{737,
  103}

\bibitem[{{Schulze} {et~al.}(2018){Schulze}, {Kr{\"u}hler}, {Leloudas},
  {Gorosabel}, {Mehner}, {Buchner}, {Kim}, {Ibar}, {Amor{\'\i}n},
  {Herrero-Illana}, {Anderson}, {Bauer}, {Christensen}, {de Pasquale}, {de
  Ugarte Postigo}, {Gallazzi}, {Hjorth}, {Morrell}, {Malesani}, {Sparre},
  {Stalder}, {Stark}, {Th{\"o}ne}, \& {Wheeler}}]{Schulze18}
{Schulze}, S., {Kr{\"u}hler}, T., {Leloudas}, G., {et~al.} 2018,
  \hypersetup{urlcolor=magenta}\href{https://dx.doi.org/10.1093/mnras/stx2352}{\mnras},
  \hypersetup{urlcolor=blue}\href{https://ui.adsabs.harvard.edu/abs/2018MNRAS.473.1258S}{473,
  1258}

\bibitem[{{Short} {et~al.}(2019){Short}, {Nicholl}, {Muller}, {Angus}, \&
  {Yaron}}]{Short19_enz}
{Short}, P., {Nicholl}, M., {Muller}, T., {Angus}, C., \& {Yaron}, O. 2019,
  Transient Name Server Classification Report,
  \hypersetup{urlcolor=blue}\href{https://ui.adsabs.harvard.edu/abs/2019TNSCR.772....1S}{2019-772,
  1}

\bibitem[{{Tachibana} \& {Miller}(2018)}]{Tachibana18}
{Tachibana}, Y., \& {Miller}, A.~A. 2018,
  \hypersetup{urlcolor=magenta}\href{https://dx.doi.org/10.1088/1538-3873/aae3d9}{\pasp},
  \hypersetup{urlcolor=blue}\href{https://ui.adsabs.harvard.edu/abs/2018PASP..130l8001T}{130,
  128001}

\bibitem[{{van der Walt} {et~al.}(2011){van der Walt}, {Colbert}, \&
  {Varoquaux}}]{numpy}
{van der Walt}, S., {Colbert}, S.~C., \& {Varoquaux}, G. 2011,
  \hypersetup{urlcolor=magenta}\href{https://dx.doi.org/10.1109/MCSE.2011.37}{CSE},
  \hypersetup{urlcolor=blue}\href{https://ui.adsabs.harvard.edu/abs/2011CSE....13b..22V}{13,
  22}

\bibitem[{{van Velzen} {et~al.}(2020){van Velzen}, {Gezari}, {Hammerstein},
  {Roth}, {Frederick}, {Ward}, {Hung}, {Cenko}, {Stein}, {Perley}, {Taggart},
  {Sollerman}, {Andreoni}, {Bellm}, {Brinnel}, {De}, {Dekany}, {Feeney},
  {Foley}, {Fremling}, {Giomi}, {Golkhou}, {Ho}, {Kasliwal}, {Kilpatrick},
  {Kulkarni}, {Kupfer}, {Laher}, {Mahabal}, {Masci}, {Nordin}, {Riddle},
  {Rusholme}, {Sharma}, {van Santen}, {Shupe}, \& {Soumagnac}}]{Velzen20}
{van Velzen}, S., {Gezari}, S., {Hammerstein}, E., {et~al.} 2020, arXiv
  e-prints,
  \hypersetup{urlcolor=magenta}\href{https://arxiv.org/abs/2001.01409}{arXiv}{:}\hypersetup{urlcolor=blue}\href{https://ui.adsabs.harvard.edu/abs/2020arXiv200101409V}{2001.01409}

\bibitem[{{Villar} {et~al.}(2018){Villar}, {Nicholl}, \& {Berger}}]{Villar18}
{Villar}, V.~A., {Nicholl}, M., \& {Berger}, E. 2018,
  \hypersetup{urlcolor=magenta}\href{https://dx.doi.org/10.3847/1538-4357/aaee6a}{\apj},
  \hypersetup{urlcolor=blue}\href{https://ui.adsabs.harvard.edu/abs/2018ApJ...869..166V}{869,
  166}

\bibitem[{{Villar} {et~al.}(2019){Villar}, {Berger}, {Miller}, {Chornock},
  {Rest}, {Jones}, {Drout}, {Foley}, {Kirshner}, {Lunnan}, {Magnier},
  {Milisavljevic}, {Sanders}, \& {Scolnic}}]{Villar19}
{Villar}, V.~A., {Berger}, E., {Miller}, G., {et~al.} 2019,
  \hypersetup{urlcolor=magenta}\href{https://dx.doi.org/10.3847/1538-4357/ab418c}{\apj},
  \hypersetup{urlcolor=blue}\href{https://ui.adsabs.harvard.edu/abs/2019ApJ...884...83V}{884,
  83}

\bibitem[{{Villar} {et~al.}(2020){Villar}, {Hosseinzadeh}, {Berger},
  {Ntampaka}, {Jones}, {Challis}, {Chornock}, {Drout}, {Foley}, {Kirshner},
  {Lunnan}, {Margutti}, {Milisavljevic}, {Sanders}, {Pan}, {Rest}, {Scolnic},
  {Magnier}, {Metcalfe}, {Wainscoat}, \& {Waters}}]{Villar20}
{Villar}, V.~A., {Hosseinzadeh}, G., {Berger}, E., {et~al.} 2020, arXiv
  e-prints,
  \hypersetup{urlcolor=magenta}\href{https://arxiv.org/abs/2008.04921}{arXiv}{:}\hypersetup{urlcolor=blue}\href{https://ui.adsabs.harvard.edu/abs/2020arXiv200804921V}{2008.04921}

\bibitem[{{Vreeswijk} {et~al.}(2014){Vreeswijk}, {Savaglio}, {Gal-Yam}, {De
  Cia}, {Quimby}, {Sullivan}, {Cenko}, {Perley}, {Filippenko}, {Clubb},
  {Taddia}, {Sollerman}, {Leloudas}, {Arcavi}, {Rubin}, {Kasliwal}, {Cao},
  {Yaron}, {Tal}, {Ofek}, {Capone}, {Kutyrev}, {Toy}, {Nugent}, {Laher},
  {Surace}, \& {Kulkarni}}]{Vreeswijk14}
{Vreeswijk}, P.~M., {Savaglio}, S., {Gal-Yam}, A., {et~al.} 2014,
  \hypersetup{urlcolor=magenta}\href{https://dx.doi.org/10.1088/0004-637X/797/1/24}{\apj},
  \hypersetup{urlcolor=blue}\href{https://ui.adsabs.harvard.edu/abs/2014ApJ...797...24V}{797,
  24}

\bibitem[{{Vreeswijk} {et~al.}(2017){Vreeswijk}, {Leloudas}, {Gal-Yam}, {De
  Cia}, {Perley}, {Quimby}, {Waldman}, {Sullivan}, {Yan}, {Ofek}, {Fremling},
  {Taddia}, {Sollerman}, {Valenti}, {Arcavi}, {Howell}, {Filippenko}, {Cenko},
  {Yaron}, {Kasliwal}, {Cao}, {Ben-Ami}, {Horesh}, {Rubin}, {Lunnan}, {Nugent},
  {Laher}, {Rebbapragada}, {Wo{\'z}niak}, \& {Kulkarni}}]{Vreeswijk17}
{Vreeswijk}, P.~M., {Leloudas}, G., {Gal-Yam}, A., {et~al.} 2017,
  \hypersetup{urlcolor=magenta}\href{https://dx.doi.org/10.3847/1538-4357/835/1/58}{\apj},
  \hypersetup{urlcolor=blue}\href{https://ui.adsabs.harvard.edu/abs/2017ApJ...835...58V}{835,
  58}

\bibitem[{{Whitesides} {et~al.}(2017){Whitesides}, {Lunnan}, {Kasliwal},
  {Perley}, {Corsi}, {Cenko}, {Blagorodnova}, {Cao}, {Cook}, {Doran},
  {Frederiks}, {Fremling}, {Hurley}, {Karamehmetoglu}, {Kulkarni}, {Leloudas},
  {Masci}, {Nugent}, {Ritter}, {Rubin}, {Savchenko}, {Sollerman}, {Svinkin},
  {Taddia}, {Vreeswijk}, \& {Wozniak}}]{Whitesides17}
{Whitesides}, L., {Lunnan}, R., {Kasliwal}, M.~M., {et~al.} 2017,
  \hypersetup{urlcolor=magenta}\href{https://dx.doi.org/10.3847/1538-4357/aa99de}{\apj},
  \hypersetup{urlcolor=blue}\href{https://ui.adsabs.harvard.edu/abs/2017ApJ...851..107W}{851,
  107}

\bibitem[{{Yan} {et~al.}(2019{\natexlab{\hspace{0pt}a}}){Yan}, {Chen},
  {Perley}, {Schulze}, {Taggart}, \& {Gal-Yam}}]{Yan19_2019sgg}
{Yan}, L., {Chen}, Z., {Perley}, D., {et~al.} 2019{\natexlab{\hspace{0pt}a}},
  Transient Name Server Classification Report,
  \hypersetup{urlcolor=blue}\href{https://ui.adsabs.harvard.edu/abs/2019TNSCR2041....1Y}{2019-2041,
  1}

\bibitem[{{Yan} {et~al.}(2019{\natexlab{\hspace{0pt}b}}){Yan}, {Perley},
  {Lunnan}, {Schulze}, {Gal-Yam}, {Taggart}, {Yaron}, \&
  {Velzen}}]{Yan19_slsne}
{Yan}, L., {Perley}, D., {Lunnan}, R., {et~al.} 2019{\natexlab{\hspace{0pt}b}},
  Transient Name Server AstroNote,
  \hypersetup{urlcolor=blue}\href{https://ui.adsabs.harvard.edu/abs/2019TNSAN..45....1Y}{45,
  1}

\bibitem[{{Yan} {et~al.}(2015){Yan}, {Quimby}, {Ofek}, {Gal-Yam}, {Mazzali},
  {Perley}, {Vreeswijk}, {Leloudas}, {De Cia}, {Masci}, {Cenko}, {Cao},
  {Kulkarni}, {Nugent}, {Rebbapragada}, {Wo{\'z}niak}, \& {Yaron}}]{Yan15}
{Yan}, L., {Quimby}, R., {Ofek}, E., {et~al.} 2015,
  \hypersetup{urlcolor=magenta}\href{https://dx.doi.org/10.1088/0004-637X/814/2/108}{\apj},
  \hypersetup{urlcolor=blue}\href{https://ui.adsabs.harvard.edu/abs/2015ApJ...814..108Y}{814,
  108}

\bibitem[{{Yan} {et~al.}(2017){Yan}, {Lunnan}, {Perley}, {Gal-Yam}, {Yaron},
  {Roy}, {Quimby}, {Sollerman}, {Fremling}, {Leloudas}, {Cenko}, {Vreeswijk},
  {Graham}, {Howell}, {De Cia}, {Ofek}, {Nugent}, {Kulkarni}, {Hosseinzadeh},
  {Masci}, {McCully}, {Rebbapragada}, \& {Wo{\'z}niak}}]{Yan17}
{Yan}, L., {Lunnan}, R., {Perley}, D.~A., {et~al.} 2017,
  \hypersetup{urlcolor=magenta}\href{https://dx.doi.org/10.3847/1538-4357/aa8993}{\apj},
  \hypersetup{urlcolor=blue}\href{https://ui.adsabs.harvard.edu/abs/2017ApJ...848....6Y}{848,
  6}

\bibitem[{{Yan} {et~al.}(2020){Yan}, {Perley}, {Schulze}, {Lunnan},
  {Sollerman}, {De}, {Chen}, {Fremling}, {Gal-Yam}, {Taggart}, {Chen},
  {Andreoni}, {Bellm}, {Cunningham}, {Dekany}, {Duev}, {Fransson}, {Laher},
  {Hankins}, {Ho}, {Jencson}, {Kaye}, {Kulkarni}, {Kasliwal}, {Golkhou},
  {Graham}, {Masci}, {Miller}, {Neill}, {Ofek}, {Porter}, {Mr{\'o}z}, {Reiley},
  {Riddle}, {Rigault}, {Rusholme}, {Shupe}, {Soumagnac}, {Smith}, {Tartaglia},
  {Yao}, \& {Yaron}}]{Yan20}
{Yan}, L., {Perley}, D., {Schulze}, S., {et~al.} 2020, arXiv e-prints,
  \hypersetup{urlcolor=magenta}\href{https://arxiv.org/abs/2006.13758}{arXiv}{:}\hypersetup{urlcolor=blue}\href{https://ui.adsabs.harvard.edu/abs/2020arXiv200613758Y}{2006.13758}

\bibitem[{{Young}(2016)}]{Young16}
{Young}, D. 2016, Transient Name Server Classification Report,
  \hypersetup{urlcolor=blue}\href{https://ui.adsabs.harvard.edu/abs/2016TNSCR..68....1Y}{2016-68,
  1}

\end{thebibliography}

\appendix
\setcounter{figure}{0}
\setcounter{table}{0}

We show in Table~A.\ref{tab:SLSNe} the sample of all the SLSN-I used for this classifier, sorted by redshift.

\startlongtable
\begin{deluxetable*}{ccc|ccc|ccc}
    \tablecaption{Type-I SLSNe \label{tab:SLSNe}}
    \tablehead{\colhead{Name} & \colhead{Redshift}  & \colhead{Reference} & \colhead{Name} & \colhead{Redshift}  & \colhead{Reference} & \colhead{Name} & \colhead{Redshift}  & \colhead{Reference}}
    \startdata
	SN2017egm	&	0.0307	&	49			&	PS15cjz		&	0.2200	&	2				&	DES17C3gyp	&	0.4700	&	2			\\
	PTF11hrq	&	0.0571	&	21			&	SN2016wi	&	0.2240	&	27				&	DES14C1rhg	&	0.4810	&	2			\\
	SN2018hti	&	0.0600	&	62      	&	SN2018gft	&	0.2300	&	56				&	SN2019itq	&	0.4810	&	this work	\\
	SN2019unb	&	0.0635	&	1			&	SN2010gx	&	0.2301	&	30				&	SN2016aj	&	0.4850	&	60			\\
	SN2018bgv	&	0.0795	&	23			&	SN2018ffj	&	0.2340	&	this work		&	SN2019kwq	&	0.5000	&	53			\\
	SN2012aa	&	0.0830	&	48			&	SN2018gkz	&	0.2400	&	29				&	PTF09atu	&	0.5015	&	10			\\
	SN2019hge	&	0.0866	&	61			&	SN2011kf	&	0.2450	&	28				&	PS114bj		&	0.5125	&	4			\\
	SN2017gci	&	0.0900	&	47			&	iPTF16bad	&	0.2467	&	27				&	SN2019otl	&	0.5140	&	this work	\\
	SN2010md	&	0.0987	&	10			&	SN2019enz	&	0.2550	&	$26^{\dagger}$	&	PS112bqf	&	0.5220	&	4			\\
	SN2016eay	&	0.1013	&	46			&	LSQ12dlf	&	0.2550	&	25				&	PS111ap		&	0.5240	&	4			\\
	PTF12hni	&	0.1056	&	30			&	LSQ14mo		&	0.2560	&	24				&	DES16C3dmp	&	0.5620	&	2			\\
	PTF12dam	&	0.1070	&	41			&	PTF09cnd	&	0.2584	&	10				&	DES15S1nog	&	0.5650	&	2			\\
	SN2019neq	&	0.1075	&	45			&	SN2019dlr	&	0.2600	&	53				&	SN2019sgg	&	0.5726	&	54			\\
	SN2018kyt	&	0.1080	&	44			&	SN2019hno	&	0.2600	&	53				&	SN2019kwu	&	0.6000	&	53			\\
	SN2017ens	&	0.1086	&	43			&	SN2018fd	&	0.2630	&	this work		&	DES14X3taz	&	0.6080	&	2			\\
	SN2015bn	&	0.1136	&	42			&	SN2013dg	&	0.2650	&	25				&	PS110bzj	&	0.6500	&	4			\\
	PTF10nmn	&	0.1237	&	$10,21$		&	SN2018lfd	&	0.2700	&	55				&	SN2013hy	&	0.6630	&	9,2			\\
	SN2007bi	&	0.1279	&	41			&	iPTF13bjz	&	0.2712	&	30				&	SN2019fiy	&	0.6700	&	53			\\
	SN2017dwh	&	0.1300	&	40			&	SN2018bym	&	0.2740	&	23				&	PS112zn		&	0.6740	&	52			\\
	SN2018avk	&	0.1320	&	23			&	SN2011ep	&	0.2800	&	16				&	DES17X1blv	&	0.6900	&	2			\\
	SN2020exj	&	0.1330	&	59			&	SN2005ap	&	0.2832	&	22				&	DES16C3cv	&	0.7270	&	2			\\
	SN2019lsq	&	0.1400	&	39			&	PTF10uhf	&	0.2879	&	21				&	PS111bdn	&	0.7380	&	4			\\
	SN2018ffs	&	0.1420	&	this work	&	SN2016inl	&	0.2980	&	this work		&	iPTF13ajg	&	0.7403	&	8			\\
	SN2011ke	&	0.1429	&	21			&	MLS121104	&	0.3030	&	52				&	SNLS07D3bs	&	0.7570	&	51			\\
	SN2019bgu	&	0.1480	&	58			&	SN2019eot	&	0.3057	&	20				&	DES15X3hm	&	0.8600	&	2			\\
	SN2019cdt	&	0.1530	&	38			&	SN2017beq	&	0.3100	&	19				&	DES14X2byo	&	0.8680	&	2			\\
	LSQ14an		&	0.1630	&	37			&	PS112cil	&	0.3200	&	4				&	PS113gt		&	0.8840	&	4			\\
	SN2019ujb	&	0.1647	&	this work	&	SN2019cwu	&	0.3200	&	53				&	PS110awh	&	0.9084	&	7			\\
	SN2019obk	&	0.1656	&	61			&	PTF12mxx	&	0.3296	&	10				&	DES17X1amf	&	0.9200	&	2			\\
	SN2018ibb	&	0.1660	&	57			&	iPTF13ehe	&	0.3434	&	18				&	DES16C3ggu	&	0.9490	&	2			\\
	SN2019pvs	&	0.1670	&	this work	&	SN2019sgh	&	0.3440	&	this work		&	PS110ky		&	0.9558	&	7			\\
	PTF10bfz	&	0.1701	&	10			&	LSQ14bdq	&	0.3450	&	17				&	PS111aib	&	0.9970	&	4			\\
	SN2012il	&	0.1750	&	28			&	SN2018lfe	&	0.3500	&	63      		&	DES16C2aix	&	1.0680	&	2			\\
	PTF12gty	&	0.1768	&	21			&	SN2019kwt	&	0.3562	&	53				&	PS110ahf	&	1.1000	&	4			\\
	CSS160710	&	0.1800	&	36			&	PTF10bjp	&	0.3584	&	10				&	DES15X1noe	&	1.1880	&	2			\\
	SN2019gfm	&	0.1816	&	35			&	LSQ14fxj	&	0.3600	&	16				&	SCP06F6		&	1.1890	&	6			\\
	SN2009cb	&	0.1864	&	21			&	SN2019zbv	&	0.3700	&	this work		&	PS110pm		&	1.2060	&	5			\\
	SN2009jh	&	0.1867	&	$10,21$		&	SN2006oz	&	0.3760	&	15				&	PS111tt		&	1.2830	&	4			\\
	iPTF16asu	&	0.1870	&	34			&	SN2019zeu	&	0.3900	&	this work		&	DES14C1fi	&	1.3020	&	2			\\
	SN2019nhs	&	0.1900	&	33			&	DES15C3hav	&	0.3920	&	2				&	PS111afv	&	1.4070	&	4			\\
	SN2018cxa	&	0.1900	&	this work	&	iPTF13cjq	&	0.3962	&	30				&	SNLS07d2bv	&	1.5000	&	3			\\
	SN2010hy	&	0.1901	&	$10,30$		&	SN2019kcy	&	0.4000	&	53				&	DES14S2qri	&	1.5000	&	2			\\
	SN2011kg	&	0.1924	&	10			&	iPTF13bdl	&	0.4030	&	30				&	PS113or		&	1.5200	&	4			\\
	SN2019kws	&	0.1977	&	53,61		&	SN2019cca	&	0.4103	&	14				&	PS111bam	&	1.5650	&	4			\\
	SN2019xaq	&	0.2000	&	this work	&	iPTF16eh	&	0.4270	&	13				&	PS112bmy	&	1.5720	&	4			\\
	SN2016ard	&	0.2025	&	32			&	CSS130912	&	0.4305	&	$11,12$			&	SNLS06d4eu	&	1.5881	&	3			\\
	PTF10aagc	&	0.2060	&	10			&	PTF10vqv	&	0.4518	&	10				&	DES16C2nm	&	1.9980	&	2			\\
	SN2016els	&	0.2170	&	31			&	CSS140925	&	0.4600	&	16				&	SN2213		&	2.0500	&	50			\\
	\enddata
    \tablecomments{All the SLSN-I used to train our classifier. Note there are more SLSNe candidates in the literature, but we keep only the unambiguous ones to avoid polluting the sample. 1:\cite{Prentice19}; 2:\cite{Angus19}; 3:\cite{Howell13}; 4:\cite{Lunnan18}; 5:\cite{McCrum15}; 6:\cite{Quimby11}; 7:\cite{Chomiuk11}; 8:\cite{Vreeswijk14}; 9:\cite{Papadopoulos15}; 10:\cite{Perley16}; 11:\cite{Vreeswijk17}; 12:\cite{Liu18}; 13:\cite{Lunnan18_iPTF16eh}; 14:\cite{Perley19}; 15:\cite{Leloudas12}; 16:\cite{Schulze18}; 17:\cite{Nicholl15_LSQ14bdq}; 18:\cite{Yan15}; 19:\cite{Kasliwal19}; 20:\cite{Fremling19_2019eot}; 21:\cite{Quimby18}; 22:\cite{Quimby07}; 23:\cite{Lunnan19_four}; 24:\cite{Chen17}; 25:\cite{Nicholl14}; 26:\cite{Short19_enz}; 27:\cite{Yan17}; 28:\cite{Inserra13}; 29:\cite{Fremling18}; 30:\cite{Cia18}; 31:\cite{Fraser16}; 32:\cite{Blanchard18}; 33:\cite{Perley19_2019nhs}; 34:\cite{Whitesides17}; 35:\cite{Chen19}; 36:\cite{Drake09}; 37:\cite{Inserra17}; 38:\cite{Fremling19_2019cdt}; 39:\cite{Fremling19_2019lsq}; 40:\cite{Blanchard19}; 41:\cite{Nicholl13}; 42:\cite{Nicholl16_15bn}; 43:\cite{Chen18}; 44:\cite{Fremling19_2018kyt}; 45:\cite{Perley19_19neq}; 46:\cite{Nicholl17_16apd}; 47:\cite{Lyman17}; 48:\cite{Roy16}; 49:\cite{Nicholl17_17egm}; 50:\cite{Cooke12}; 51:\cite{Prajs17}; 52:\cite{Lunnan14}; 53:\cite{Yan19_slsne}; 54:\cite{Yan19_2019sgg}; 55:\cite{Fremling19_2018lfd}; 56:\cite{Fremling18_2018gft}; 57:\cite{Fremling18_2018ibb}; 58:\cite{Fremling19_2019bgu}; 59:\cite{Dahiwale20}; 60:\cite{Young16}; 61:\cite{Yan20}; 62:\cite{Lin20}; 63: Yin et al., in prep.}
    \tablenotetext{\dagger}{We find that a redshift of $z = 0.255$ is a better match to the SNe spectral features than the $z = 0.22$ reported in \cite{Short19_enz}.}
\end{deluxetable*}

\end{document}